\documentclass[longauth]{aa} 

\usepackage{courier}
\usepackage{multirow}
\usepackage{multicol}
\usepackage{booktabs}
\usepackage{graphicx}
\usepackage{lscape}
\usepackage{cancel}

\usepackage{txfonts}
\usepackage{xcolor}
\usepackage{lineno}
\usepackage{subcaption}

\usepackage[normalem]{ulem}

\usepackage{hyperref}

\begin{document} 


\newcommand{\fixme}[1]{{\color{red} #1}}
\newcommand{\barbara}[1]{{\color{black} #1}}
\newcommand{\nikhil}[1]{{\color{black} #1}}
\newcommand{\mattia}[1]{{\color{black} #1}}
\newcommand{\marko}[1]{{\color{black} #1}}
\newcommand{\silvia}[1]{{\color{black} #1}}
\newcommand{\stefano}[1]{{\color{black} #1}}
\newcommand{\ld}[1]{{\textcolor{black}{#1}}}
\newcommand{\khunsang}[1]{{\textcolor{black}{#1}}}
\newcommand{\ignore}[1]{}

\definecolor{tangerine}{rgb}{1.0, 0.6, 0.4}

    \title{Can we constrain the binary neutron star merger time from kilonova light curves?} 


   \author{Noah Jamsin
          \inst{1,2,3}
          \and
          Marion Pillas
          \inst{1,2}
          \and
          Marko Risti\'c  
           \inst{4,5}
          \and
          Laurence E. H. Datrier
          \inst{6}
          \and
          Daniel Brethauer
          \inst{7,8}
          \and
          Stefano Ascenzi
          \inst{9,10,11}\thanks{These authors coordinated the project. Corresponding author: barbara.patricelli@pi.infn.it}
          \and
          Mattia Bulla
          \inst{12,13,14}\footnotemark[1]
          \and
          Barbara Patricelli
          \inst{15,16,17}\footnotemark[1]
          \and
          Silvia Piranomonte
          \inst{17}\footnotemark[1]
          \and
          Henrik Rose
          \inst{18}
          \and
          Nikhil Sarin
          \inst{19,20}
          \and
          Thibeau Wouters
          \inst{21,22}
          \and
          Khun Sang Phukon
          \inst{23}
          \and
          Alex Jacquesson
          \inst{1}
          \and
          Samuele Ronchini
          \inst{9,10}
          \and
          Alberto Revilla-Pe\~{n}a
          \inst{24,25}
          \and          
          Tyler Barna
          \inst{26}
          \and          
          Michael W. Coughlin
          \inst{26}
          \and          
          Tim Dietrich
          \inst{18,27}
          \and          
          Richard O’Shaughnessy
          \inst{28}
          \and          
          Fabio Ragosta
          \inst{29,30}         
          }

   \institute{
        IJCLab, Université Paris-Saclay, CNRS/IN2P3, rue Georges Clémenceau, 91405 Orsay, France 
        \and 
        Institut d’Astrophysique de Paris, Sorbonne Université, CNRS, Bd Arago, 75014 Paris France
        \and 
        STAR institute, Université de Liège, All du Six Août, B-4000 Liège Belgique 
        \and
        Los Alamos National Laboratory, LANL P.O. Box 1663, Los Alamos, NM 87545
        \and
        UC San Diego, 9500 Gilman Dr. La Jolla, CA 92093
        \and
        The Nicholas and Lee Begovich Center for Gravitational-Wave Physics and Astronomy, California State University, Fullerton, CA 92831, USA
        \and
        Department of Astronomy, University of California, Berkeley, CA 94720-3411, USA
        \and
        Berkeley Center for Multi-messenger Research on Astrophysical Transients and Outreach (Multi-RAPTOR), University of California, Berkeley, CA 94720-3411, USA
        \and
        Gran Sasso Science Institute (GSSI), Viale F. Crispi 7, L’Aquila (AQ), I-67100, Italy
        \and
        INFN - Laboratori Nazionali del Gran Sasso, L’Aquila (AQ), I-67100, Italy
        \and
        INAF - Osservatorio Astronomico di Brera, via E. Bianchi 46, I23807 Merate (LC), Italy
        \and
        Department of Physics and Earth Science, University of Ferrara, via Saragat 1, I-44122 Ferrara, Italy
        \and
        INFN - Sezione di Ferrara, via Saragat 1, I-44122 Ferrara, Italy 
        \and
        INAF - Osservatorio Astronomico d’Abruzzo, via Mentore Maggini snc, 64100 Teramo, Italy
        \and
        Physics Department, University of Pisa, Largo B. Pontecorvo 3, I-56127 Pisa, Italy
        \and
        INFN - Pisa, Largo B. Pontecorvo 3, I-56127 Pisa, Italy
        \and
        INAF - Osservatorio Astronomico di Roma, Via Frascati 33, I-00078 Monte Porzio Catone (Rome), Italy
        \and
        Universität Potsdam, Institut für Physik und Astronomie, Karl-Liebknecht-Str. 24/25, 14476 Potsdam, Germany
        \and
        Kavli Institute for Cosmology, University of Cambridge, Madingley Road, Cambridge CB3 0HA UK
        \and
        Institute of Astronomy, Unversity of Cambridge, Madingley Road, Cambridge CB3 0HA UK
        \and
        Institute for Gravitational and Subatomic Physics (GRASP), Utrecht University, Princetonplein 1, 3584 CC Utrecht, The Netherlands
        \and
        Nikhef, Science Park 105, 1098 XG Amsterdam, The Netherlands
        \and
        School of Physics and Astronomy and Institute for Gravitational Wave Astronomy, University of Birmingham, Edgbaston, Birmingham, B15 2TT
        \and
        Institut de Ciències del Cosmos (ICCUB), Universitat de Barcelona (UB), Martí i Franquès, 1, 08028 Barcelona, Spain
        \and
        Departament de Física Quàntica i Astrofísica (FQA), Universitat de Barcelona (UB), Martí i Franquès, 1, 08028 Barcelona, Spain
        \and
        School of Physics and Astronomy, University of Minnesota, Minneapolis, MN 55455, USA
        \and
        Max Planck Institute for Gravitational Physics (Albert Einstein Institute), Am M\"uhlenberg 1, Potsdam 14476, Germany
        \and
        Rochester Institute of Technology, 1 Lomb Memorial Dr, Rochester, NY 14623-5603
        \and
        Dip. di Fisica "E. Pancini" -- Universit\`a degli Studi di Napoli Federico II, Via Cinthia 21, Edificio 6, 80126 Napoli
        \and
        INAF - Istituto Nazionale di Astrofisica - Osservatorio Astronomico di Capodimonte, Salita Moiariello 16 – 80131, Napoli, Italy
        }


\date{}

 
  \abstract
   {\barbara{The joint observation of gravitational waves (GWs) from a binary neutron star (BNS) merger (GW170817), together with a short Gamma-Ray Burst (GRB 170817A) and a kilonova (AT2017gfo), marked the birth of multimessenger astronomy with GWs, and clearly demonstrated the huge informative power of multimessenger observations. In the future, more BNS merger and kilonova detections are expected, thanks to the increasing sensitivity of GW detectors and the advent of new electromagnetic facilities such as the Vera C. Rubin Observatory. However, the association between a BNS merger and a kilonova might not be trivial: the detection of a kilonova might not coincide with the onset of the EM emission, making it difficult to clearly define a temporal window to assess the coincidence with a GW event.}} 
   {\barbara{In this work we investigate the possibility to constrain the BNS merger time from kilonova light curves, taking into account different observing conditions and with focus on the Vera C. Rubin Observatory.}}
   {\barbara{We used different kilonova models and different inference frameworks to simulate and fit kilonova light curves. We explored different possible cadences and different wavelength coverages for observations, to investigate how the different observing conditions can affect the capability to recover the BNS merger time. The tests performed also include real data related to AT2017gfo, plus a few GRBs that have been associated to possible kilonova candidates.}}
   {\barbara{We find that, under realistic observing conditions with the Vera C. Rubin Observatory, the BNS merger time can be recovered with a precision of 5.5 days (90 \% CI) with a best-case scenario with an accuracy of 0.18 day and a precision of 0.74 day (90 \% CI). Results can slightly improve (by $\sim$ 5 \%)  if the redshift of the source is available. Upper limits near the merger time can also improve the results, reducing the width of the posterior by $\sim60$ \%. The BNS merger time estimate could also benefit from additional EM follow-up observations with other telescopes. }}
   {}


   \maketitle
%

\section{Introduction} 


\barbara{On August 17, 2017, Advanced LIGO \citep{2015CQGra..32g4001L} and Advanced Virgo \citep{2015CQGra..32b4001A} observed for the first time a gravitational wave (GW) signal from a binary neutron star (BNS) inspiral (GW170817, \citealp{2017PhRvL.119p1101A}). Less than two seconds later, a short gamma-ray burst (sGRB), GRB 170817A, was detected by 
the Gamma-ray Burst Monitor (GBM) onboard Fermi \citep{2017ApJ...848L..14G}, and the spectrometer anti-coincidence shield (SPI-ACS) onboard INTEGRAL \citep{2017ApJ...848L..15S}. This joint observation provided the first direct evidence that at least a fraction of sGRBs have a BNS system as progenitor, and marked the birth of multi-messenger astronomy with GWs. The electromagnetic (EM) follow-up campaign that followed this multi-messenger observation led to the discovery of an optical transient (SSS17a/AT 2017gfo)
in the host galaxy NGC 4993, also detected in ultraviolet and infrared
wavelengths (see \citealp{2017ApJ...848L..12A} and references therein). Photometric and spectroscopic observations of the source allowed to identify it as a kilonova \citep{2017Natur.551...67P,2017Natur.551...75S}: this supports the hypothesis that BNS mergers are sites of r-process nucleosynthesis of heavy elements. A second GW signal from a BNS merger has been observed in 2019, during the third LIGO-Virgo-KAGRA (LVK) observing run: GW190425 \citep{2020ApJ...892L...3A}. Despite an extensive EM follow-up campaign, no EM counterpart was found in association with this GW event (see, e.g., \citealp{2019ApJ...880L...4H}), mainly due to the poorly constrained sky localisation of the event (of the order of 8000 deg$^2$). No other significant BNS merger has been reported since then \citep{2025arXiv250818082T,2026arXiv260527225T}. 
}
\silvia{The Advanced LIGO, Advanced Virgo, and KAGRA \citep{2019CQGra..36p5008A} detectors concluded the fourth observing run (O4) in November 2025 and are now in a period of upgrades and commissioning. Even before this transition, synergy between LVK and the Vera C. Rubin Observatory was already underway: while Rubin was still in its early-operations phase, prior to the start of sustained LSST observations, it carried out Target-of-Opportunity follow-up of gravitational-wave candidates reported during the final part of O4, such as the sub-solar-mass neutron-star merger candidate S251112cm, for which Rubin observations began about two days later \citep{macbride2026too}. This collaboration is expected to deepen further: the Vera C. Rubin Observatory officially began its ten year Legacy Survey of Space and Time (LSST) on 2026 June 30,  
and a new, LVK six-month observing run, Intermediate Run 1 (IR1), is expected to start in January 2027, during which GW interferometers will operate in closer synergy with Rubin's now fully operational LSST survey.}

\barbara{
In the future, our potential of achieving multi-messenger detections can increase using different, complementary approaches. In fact, aside from EM follow-up of candidate GW signals, other kind of searches can be performed, such as externally triggered searches. In this last scenario, 
an EM/neutrino transient is detected, and then GW data are analysed offline in detail to look for a possible GW signal in temporal and spatial coincidence with the observed non-GW signal (e.g., a GRB). The scope of these targeted searches is to enhance our ability to detect GW signals in coincidence with non-GW transients with respect to standard, all-sky GW searches. Pipelines used by the LIGO-Virgo-KAGRA collaboration for such searches are X-pipeline \citep{2010NJPh...12e3034S,2012PhRvD..86b2003W}, which performs a search for generic GW signals, and PyGRB \citep{2011PhRvD..83h4002H,2014PhRvD..90l2004W}, that looks for GW signals produced by BNS or NS-BH mergers.

In addition to this, there are online pipelines that perform low-latency searches of coincidences between EM events or neutrino event candidates and GW triggers. Specifically, the Rapid, on-source VOEvent Coincident Monitor (RAVEN) pipeline searches for and analyses coincidences between GW candidates and non-GW events such as, e.g., Gamma-Ray Bursts (GRBs) and neutrinos in low-latency (see \citealp{Urban2016,Piotrzkowski2022}). RAVEN typically analyses coincidences in real-time, listening to alerts from GRB/low energy neutrino experiments, submitting these to the Gravitational-Wave Candidate Event Database (GraceDB)\footnote{https://gracedb.ligo.org/}, and then checking whether there are coincident GW candidates already in GraceDB and vice-versa. If a coincidence is found, the significance of the multi-messenger event is determined by computing the overlap between the skymaps of the two events, and generating a resulting joint false alarm rate (FAR). 

One crucial aspect of all the above described multi-messenger searches (including EM follow-up of GW events) is the definition of time window for two events to be considered as coincident. For GRBs this is quite straightfoward: the GRB trigger time typically corresponds to the onset of the GRB emission, and theoretical models predict the possible time offset of the GRB emission with respect to GW signal; for instance with PyGRB the search for a GW signal is performed in a time window between -5 s and +1 s from the trigger time of short GRBs (see, e.g., \citealp{2022ApJ...928..186A}). The definition of this time window can be much more complicated for other transients such as, e.g., kilonovae: their emission evolves over timescales of days, and the first observation might not coincide with the onset of the EM emission, making it difficult to clearly define a temporal window for coincidence and consequently to assess the significance of the association with a GW event. 
The aim of this work is to investigate if --- and with which precision --- this time window can be constrained, based on estimates of the BNS merger time that can be inferred from kilonova light curves obtained under several, realistic observing conditions with the Vera C. Rubin Observatory (e.g., different cadence of observations, different filters, etc) opening a new era of kilonova discoveries.}

\silvia{The paper is organised as follows: Section~\ref{sec:simulations} presents the kilonova models and light-curve inference frameworks used to simulate and fit kilonova light curves. Section~\ref{sec:results} describes the constraints on the merger time obtained from the fitting of simulated kilonova light curves and real kilonova candidates, including an assessment of the robustness of these constraints to early-time emission mechanisms not captured by standard r-process heating models. Finally, Sec.~\ref{sec:discussion} contains a discussion of these results and their implications for a GW-targeted search around kilonova candidates seen by Vera C. Rubin.
}

\section{Simulating kilonova data}
\label{sec:simulations}
\subsection{Kilonova models}
\label{sec:kn_models}

\stefano{Kilonova models can be generally divided among three classes: (i) semi-analytic energy-balance models \citep{Metzger2017}, (ii) semi-analytic radiation-diffusion models \citep{Wollaeger2018, Ricigliano2024}, and (iii) Monte Carlo radiative transfer models \citep{Kasen2013,Tanaka2013,Wollaeger2018, Bulla:2019,Shingles2023}. The first group is inspired by Arnett's approach for supernova modelling \citep{Arnett1980, Arnett1982}, and consists in solving the energy balance equation within the ejecta, where the transient luminosity represent the energy-loss term. The second group is rooted in the supernova model by \citet{PintoEastman2000}, which solves the radiation diffusion equation in an optically thick medium, obtaining the luminosity as the radiation flux at the photosphere. Both approaches assume spherical geometry of the ejecta and grey opacity. Furthermore, they can self-consistently provide a bolometric luminosity, while a spectrum (generally a black-body) needs to be assumed to obtain the luminosity in different filters.

The third approach solves the radiative transfer equation within the ejecta with a Monte Carlo method, which consists in describing the transport as a sequence of stochastic events representing the interaction of the radiation with the matter. This approach allows to make use of a frequency-dependent opacity, and, in principle, can be applied to a non-spherical geometry with multiple ejecta components, whose radiative interaction can be treated self-consistently. Additionally, these models calculate self-consistently not only the bolometric lightcurve, but also the spectrum and the polarization of the transient. Unlike the first two approaches, the high computational cost of Monte Carlo radiative transfer models prevents their direct use in inference frameworks (see Sec. \ref{sec:inferenceframework}); to overcome this limitation, surrogate models that interpolate the lightcurves from a pre-computed grid of models are employed instead (\emph{e.g.} \citealp{Lukosiute2022, Peng2024, Saha2024})}

Another crucial aspect differentiating kilonova models is how key microphysical quantities are treated across the ejecta. In some models, particularly in the semi-analytic ones, nuclear heating rates, thermalization efficiencies, and r-process opacities are prescribed as global quantities and applied uniformly throughout the ejecta (e.g. by assuming constant efficiencies and ``grey'' opacities). Other models, particularly those employing detailed radiative transfer calculations, compute these properties locally, allowing them to depend on the density, temperature, velocity, and composition of each ejecta region. Because these properties directly control the energy deposited in the ejecta and the rate at which radiation escapes, differences between global and local prescriptions can lead to substantial variations in the predicted spectra and light curves \citep{Bulla2023,Sarin2024b,Brethauer2024}. 

\stefano{In this work, we make use of both the second and third categories of models. For the semi-analytic radiation-diffusion approach, we employ the \texttt{two\_component\_kilonova\_with\_cocoon} model from \texttt{Redback}~\citep{Sarin2024} that adds a shock-cooling emission from a cocoon \citep{2018ApJ...855..103P} to a two-component kilonova model. Concerning Monte Carlo radiative transfer models, we use instead predictions from three different codes: \textsc{sedona} \citep{Kasen2006}, \textsc{possis} \citep{Bulla:2019,Bulla2023} and \textsc{SuperNu} \citep{SuperNu}. Grids of models computed with {\sc sedona} are those presented in \citet{Kasen2017} and in \citep{Brethauer2026}. The first grid, which we will refer to as \texttt{Ka2017} in the rest of the paper, includes one-dimensional kilonova models characterised by different combinations of three free parameters: the ejecta mass $m_{\rm ej}\in[0.001,0.1]\,M_\odot$, averaged velocity $v_{\rm ej}\in[0.03,0.3]$ c and lanthanide-fraction $X_{\rm lan}\in[10^{-9},10^{-1}]$. The second grid, which we will refer to as {\tt Brethauer26}, is similarly one-dimensional with kilonova models parametrised by three free parameters: the ejecta mass $m_{\rm ej}\in[0.001,0.1]\,M_\odot$, averaged velocity $v_{\rm ej}\in[0.05,0.3]$~c and lanthanide-fraction $X_{\rm lan}\in[10^{-9},10^{-1}]$. However, the {\tt Brethauer26} grid differs from {\tt Ka2017} in that it includes non-thermal ionization from high-energy electrons emitted via radioactive decay of r-process material that can change the ionisation state of the ejecta instead of assuming local thermodynamic equilibrium (see \citealt{Brethauer2026} for more details). 

Among the models computed by {\sc possis}, we choose two grids of axially-symmetric two-components models: one grid computed with the first version of the code \citep{Bulla:2019,Dietrich:2020}, and one with a more recent version featuring improved sets of heating rates, thermalisation efficiencies and opacities \citep{Bulla2023,Koehn2026}. The former, referred to as {\tt Bu2019lm}, varies the mass of the dynamical and disk-wind ejecta components, $m_{\rm ej,d}\in[0.001,0.02]\,M_\odot$ and $m_{\rm ej,w}\in[0.01,0.13]\,M_\odot$, and the half-opening angle of the lanthanide-rich component distributed around the merger plane, $\phi\in[15,75]^\circ$. The latter, referred to as {\tt Bu2026\_MLP}, varies the ejecta mass, averaged velocity and averaged $Y_{\rm e}$ of both components: $m_{\rm ej,d}\in[10^{-4},10^{-1.3}]\,M_\odot$, $v_{\rm ej,d}\in[0.12,0.35]$ c, $Y_{\rm e,d}\in[0.15,0.35]$, $m_{\rm ej,w}\in[10^{-4},10^{-0.55}]\,M_\odot$, $v_{\rm ej,w}\in[0.05,0.15]$ c and $Y_{\rm e,w}\in[0.2,0.4]$. Both grids add the viewing angle $\theta$ as an extra parameter, varying from a face-on ($\theta=0^\circ$) to a edge-on ($\theta=90^\circ$) case.
}

\marko{The last grid of kilonova models used is that developed using \textsc{SuperNu} \citep{SuperNu} in \cite{Wollaeger2018} and expanded in \cite{Korobkin2021} and \cite{Ristic-LC}, which we will refer to as the ``Los Alamos National Laboratory" (LANL) suite of models. 
Similar to {\tt Bu2019lm} and {\tt Bu2026\_MLP}, these assume a two-component, axisymmetric (two-dimensional) kilonova model consisting of lanthanide-rich, equatorial dynamical ejecta and lanthanide-poor, axial wind ejecta. Each component is assumed to be homologously expanding and parameterized by a mass and velocity such that ($m_{\rm{d}}$, $v_{\rm{d}}$) and ($m_{\rm{w}}$, $v_{\rm{w}}$) describe the dynamical and wind components' masses and averaged velocities, respectively. The two components are evolved simultaneously in any given simulation (as opposed to, e.g., evolving two separate one-dimensional components and summing their flux); in spatial cells where the two components overlap, their contribution to the total cell heating rate is weighted by mass and assumed to be uniformly mixed. The {\tt LANL} model ejecta parameter ranges are presented in Table~\ref{tab:LANL_priors} and also correspond to matching prior bounds during inference.
}

\subsection{Inference framework}
\label{sec:inferenceframework}


In this work, we used four different frameworks to perform inference: \textsc{nmma}, \textsc{Redback}, \textsc{nessai/bilby} and \textsc{rift}. We provide some info of each framework in this section, but refer the reader to the relevant publications for further details.

The first framework is the Nuclear-physics and Multi-Messenger Astronomy (\textsc{nmma}) framework \citep{Pang:2023,Dietrich:2020,Rose:2026}, a Bayesian analysis suite built on top of \textsc{Bilby} \citep{Ashton:2019,Romero-Shaw:2020owr} that enables the joint interpretation of GW signals, kilonova and GRB-afterglow light curves, and nuclear-physics information. Rather than evaluating the radiative-transfer simulations directly at each likelihood call, \textsc{nmma} relies on fast machine-learning surrogate models trained on grids of precomputed light curves from Monte Carlo radiative-transfer codes such as \textsc{possis} \citep{Bulla:2019,Bulla2023} and \textsc{sedona} \citep{Kasen2006,Kasen2017}; each surrogate predicts the apparent magnitude in a given band as a function of the ejecta parameters, viewing angle, and time since merger $t_0$. Posterior distributions of the source parameters are obtained through nested sampling \citep{Skilling:2004pqw,Skilling:2006gxv} (using, e.g., \textsc{pymultinest} \citep{Feroz:2008xx,PyMultinest} or \textsc{dynesty} \citep{Speagle:2019ivv}). 
Moreover, in \textsc{nmma} we also use \textsc{pocoMC} \citep{Karamanis:2022ksp}, a sequential Monte Carlo sampler \citep{del2006sequential,chopin2020introduction} that uses normalizing flows to precondition the parameter space during sampling, and which has already been used in GW parameter estimation \citep{Karamanis:2022alw,Williams:2025szm}. 
The likelihood includes an additional systematic-uncertainty term $\sigma_{\rm sys}$ that broadens the per-point Gaussian likelihood to absorb the mismatch between the surrogate models and the true emission (see Sec.~\ref{sec:170817}). \textsc{nmma} has recently been interfaced with \textsc{Fiesta} \citep{Koehn:2025b}, a \textsc{jax}-based package that provides GPU-accelerated, high-precision surrogates for kilonova and GRB-afterglow light curves; this drastically reduces the computational cost of combined multi-messenger analyses and enables the use of gradient-based samplers.

We also use {\sc Redback} framework~\citep{Sarin2024}, an electromagnetic transient modelling and inference framework, which also uses \textsc{Bilby}~\citep{Ashton:2019} for inference on CPUs with a JAX-native extension~{\sc Redback-JAX}\footnote{\url{https://github.com/nikhil-sarin/redback-jax}}, which uses {\sc BlackJAX}~\citep{2024arXiv240210797C} as the inference backend. This framework has a smaller library of surrogate models for kilonovae compared to \textsc{nmma}, in particular a smaller subset of \textsc{possis} radiative transfer models~\citep{Bulla:2019} and radiative-transfer simulations with {\sc sedona}~\citep{Kasen2017} via {\sc kilonovanet}~\citep{Lukosiute2022}. However, there is a larger library of semi-analytical kilonova models built on different assumptions, such as the inclusion of shock-cooling emission from a cocoon \citep{2018ApJ...855..103P}, free neutrons \citep{Metzger2015,2015ApJ...815...82L} and magnetar engines \citep{Sarin2022}, heating-rates \citep{2012MNRAS.426.1940K, Sarin2024b} and thermalisation assumptions \citep{2016ApJ...829..110B}. We only use these latter class of models within {\sc Redback} for our analysis here as this allows for an easier test of the inferred parameters across a more diverse range of modelling assumptions than allowed by more detailed radiative transfer simulations and the surrogates they are built on. For GRB afterglows {\sc Redback} includes a range of implementations from other popular afterglow packages such as afterglowpy, vegas, jetsimpy, as well as an inbuilt library of afterglow models including inverse Compton, synchrotron-self absorption, and pre-deceleration physics. {\sc Redback} provides a library of likelihoods ranging from the standard assumption of a Gaussian, a Gaussian with a constant systematic model or data error $\sigma_{\rm sys}$ similar to {\sc nmma}, a GP covariance noise source or a Gaussian mixture model. Posteriors are estimated via {\sc dynesty} sampler~\citep{Speagle:2019ivv} wrapped via {\sc bilby}. For our analysis with {\sc redback}, we employ the standard Gaussian likelihood for kilonova model in the flux density domain, assuming independent Gaussian uncertainties on the measured flux in each band. We employ the \texttt{two\_component\_kilonova\_with\_cocoon} semi-analytic model. The total emission is constructed as the sum of three ejecta components, each characterised independently by an ejecta mass $m_{{\rm ej},i}$, expansion velocity $v_{{\rm ej},i}$, grey opacity $\kappa_i$, and temperature floor $T_{{\rm floor},i}$, with $i=1,2,3$.

\ld{A third, light-weight inference code is based on the 2017 one-dimensional radiative transfer models developed by \cite{Kasen2017}, and serves as another check of modelling systematics on the recovery of parameters. To get expected light curves, bandpass filters are convolved with the time-resolved spectra. Those spectra have a 0.1 day time resolution, but in order to smooth out the numerical noise present in the curve, light curves are smoothed with a rolling window of 0.5 day past their peak brightness. We also discard data points beyond $t=12$\,days post-merger, or truncate light curves where models become unreliable (marked by a sudden, sharp decline in brightness). A Gaussian Process is trained and optimised on these processed light curves for each band, using the \textsc{george} python package. The light curves with $X_{\rm lan} = 10^{-9}$ are discarded from the training dataset due to the large gap in the parameter grid between $X_{\rm lan} = 10^{-9}$ and $X_{\rm lan} = 10^{-5}$. We use a combination of two square exponential kernels, one acting on all dimensions of the data, and one acting on only the time dimension to take into account both the lightcurve peaks and their slow decay. Final kilonova light curves are made up of two component light curves, red ($X_{\rm lan} \geq 10^{-2}$) and blue ($X_{\rm lan} \leq 10^{-2}$, $v_{\rm ej}^{\rm blue} > v_{\rm ej}^{\rm red}$). These components are added together post-simulation and therefore do not take into account any interactions between the two; this model also ignores the impact of viewing angles.}
\ld{The inference is carried out using \textsc{nessai} as it is implemented in \textsc{bilby} with a fixed $d_{L}$ and allowing all ejecta parameters and $t_{0}$ to vary.}

\marko{Finally, we use the Rapid parameter inference on gravitational wave sources via Iterative FiTting (\textsc{rift}) algorithm as an independent parameter inference framework separate from \textsc{Bilby}. \textsc{rift} is an extremely flexible inference engine, capable of running on both CPUs \citep{RIFT} and GPUs \citep{RIFT-GPU}, primarily applied to gravitational wave inference \citep{Jan2020, Wagner2024, RuizRocha2025, Henshaw2025, Ristic-GWEM}, but frequently used for broader astrophysical contexts \citep[largely kilonova inference, e.g.][]{Ristic-LC, Kedia2023, Ristic-nucleosynthesis, Ristic-spectra, Peng2024, Ristic-LGRB}. At a very high level, \textsc{rift} leverages the calculation of a marginalized intrinsic parameter likelihood by integrating the total likelihood over extrinsic, or ``nuisance'', parameters through direct, adaptive Monte Carlo integration which prioritizes regions of high likelihood. The posterior distributions over the intrinsic parameters, in our case the kilonova model inputs, are calculated through a separate Monte Carlo integration step. These two steps are repeated iteratively until convergence is confirmed through comparison of the one-dimensional marginal distributions' empirical cumulative distributions across iterations; see \cite{Pankow2015} for further information on the Monte Carlo integration algorithm employed in \textsc{rift}. As with the previously described inference approaches, we do not directly evaluate the LANL radiative transfer model predictions. Instead, we use an emulator trained on the active learning model grid presented in \cite{Ristic-LC}; this simulation grid was intentionally designed to reduce emulator uncertainty during iterative grid construction, and is thus optimized for light-curve inference. We use the neural network light-curve emulator presented in \cite{Peng2024} which generates vector light-curve predictions for a given set of simulation input parameters, using a separate neural network for each of the \textit{grizyJHK} bands we consider in our inference. Our log-likelihood is constructed as

\begin{equation}
    \ln \mathcal{L}(\vec{x}) = \sum_{t, b} \frac{(\hat{y}_{t, b}(\vec{x}) - d_{t, b})^2}{\sigma_{t, b}^2 + \sigma_{\rm sys}^2},
\end{equation}

where $\hat{y}_{(t, b)}(\vec{x})$ is the neural network surrogate's prediction for the AB magnitude in band $b$ at time $t$ for a given set of input parameters $\vec{x} = (m_d, v_d, m_w, v_w, \theta)$, $y_{t, b}(\vec{x})$ the observed data (e.g. from AT2017gfo) in band $b$ at time $t$ with associated observation uncertainty $\sigma_{t, b}$, and the systematic uncertainty $\sigma_{\rm sys}$ which we use as a catch-all term to encapsulate the broader family of uncertainty contributions. We apply a Gaussian prior on $\theta$ with $\mu = 20$ and $\sigma = 5$ based on radio observations of AT2017gfo \citep{Mooley2018}.}





\section{Results}
\label{sec:results}

In the following we explore the recovery of the merger time from kilonova lightcurves from various kilonova cases and under different observing scenario. In Sec.~\ref{sec:kn-generation}, we start with simulated kilonova lightcurves, generated with 
\texttt{Bu2026\_MLP}, in order to assess the performances of the merger time recovery over a kilonova population under various observational constraints. We perform the Bayesian inference of the source properties with {\sc nmma}, inferring the merger time from a freely sampled timeshift parameter that provides an additional degree of freedom to fit the surrogates to these synthetic data. Specifically, we start by an optimistic observing scenario and the use of the same model for lightcurve generation and recovery in Sec.~\ref{sec:self-optimistic} (called TEST 1a), we then move to a more realistic LSST observing scenario in Sec.~\ref{sec:self-LSST} (called TEST 1b), besides, still under this realistic configuration, we test the use of different models for the lightcurve generation and the merger time inference in Sec.~\ref{sec:cross-LSST} (called TEST 1c). Furthermore, we move to real case examples with observed kilonova candidates: AT2017gfo in Sec.~\ref{sec:170817} (called TEST 2a) and a sample of kilonovae associated with GRBs in Sec.~\ref{sec:GRB} (called TEST 2b). Finally, we investigate the impact of early-time blue emission from free neutrons on parameter estimation in Sec.~\ref{sec:early-time}.

\begin{figure*}
    \centering
    \includegraphics[width=0.9\linewidth]{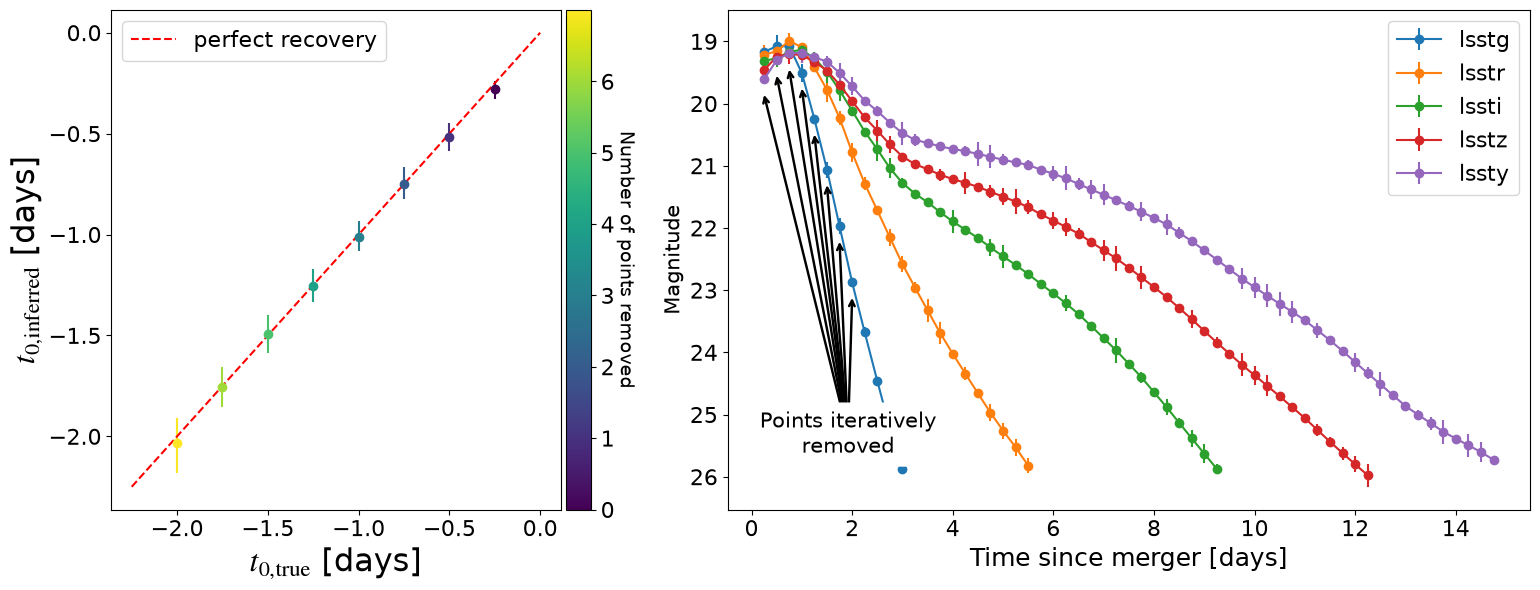}
    \caption{Illustration of the iterative $t_0$ recovery procedure when removing early detections: true vs inferred $t_0$ distribution for one lightcurve of the grid, the error bars represent the $1\sigma$ credible interval colored according to the number of removed detections (left). The corresponding lightcurve in the $grizy$ bands is shown on the right.}
    \label{fig:ts-recovery-example-Bu19}
\end{figure*}

\subsection{Generation of synthetic kilonova lightcurves}
\label{sec:kn-generation}


Synthetic kilonova lightcurves allows us to explore a broad population, sample a wide range of physical parameters, and avoid restricting our analysis to the few events that have been detected to date. To generate these synthetic light curves, we use the {\sc nmma} framework and focus on 
the \texttt{Bu2026\_MLP} model. We start by generating a BNS population from distributions presented in Table~\ref{tab:NMMAinjectionprior-generation}. In particular, we use fitting formulae implemented in {\sc nmma} \citep{Dietrich:2020, formuladyn} to convert the binary masses to ejecta masses, assuming an equation of state (EOS) from \cite{EOS_paper} (EOS labelled 4818). Other properties not easily related to the BNS properties, such as the electron fraction are chosen following a uniform distribution across the model parameter space. Here, potential correlations between the ejecta masses and velocities are not taken into account. Although correlations between these quantities may arise from the underlying merger dynamics, we treat them as independent phenomenological parameters, as commonly done when considering sampling kilonova radiative-transfer model grids \citep{Wollaeger_2021}. 

\begin{table}[th]
\centering
\begin{tabular}{llcl}
\hline
\hline
\textbf{Parameter} & \textbf{Distribution} & \textbf{Range / Bounds} & Use \\ \hline
$M_1$ & Uniform & ${1, 2.25}~~ [M_\odot]$ & Generation \\
$M_2$ & Uniform & ${1, 2.25}~~ [M_\odot]$ & Generation\\
$\log_{10}M^{\rm{dyn}}_{\rm{ej}}$ & Uniform  & ${-4, -1.302}$ & Analysis\\
$\log_{10}M^{\rm{wind}}_{\rm{ej}}$ & Uniform  & ${-4, -0.56}$ & Analysis\\
$v_\mathrm{ej}^\mathrm{dyn}$ & Uniform & ${0.12 - 0.35}\,\,[c]$ & Both \\
$v_\mathrm{ej}^\mathrm{wind}$ & Uniform  & ${0.05 - 0.15}\,\,[c]$ & Both\\ 
$Y_e^\mathrm{dyn}$ & Uniform  & ${0.15 - 0.35}$ & Both \\
$Y_e^\mathrm{wind}$ & Uniform & ${0.2 - 0.4}$ & Both \\
$\theta$ & Sine & ${0 - \displaystyle \pi/2} \,\, [\text{rad}]$ & Both\\
$D_L$ & $\rm{Uniform}_{cv}$ & $10 - 200\,\,[\text{Mpc}]$ & Generation\\
$D_L$ & Uniform & $10 - 300\,\,[\text{Mpc}]$ & Analysis\\
$\sigma_{\rm sys}$ & Uniform & ${0.3-2}$ & Analysis \\

\hline
\hline
\end{tabular}
\caption{Priors used to generate and analyse synthetic kilonova lightcurves with \texttt{Bu2026\_MLP}. $\rm{Uniform}_{cv}$ refers to a uniform co-moving volume prior and  $\sigma_{\rm sys}$ to the model systematic uncertainty.}

\label{tab:NMMAinjectionprior-generation}
\end{table}

We then use these parameters to generate corresponding multi-band lightcurves. These lightcurves are then injected into simulated observations, reproducing realistic survey conditions by specifying the observing filters, cadence, and photometric uncertainties. This procedure enables us to assess the kilonova detectability and the recovery of its properties across a wide range of intrinsic properties and observational configurations. 
The \textsc{nmma} \texttt{Bu2026\_MLP} surrogate is trained with \textsc{fiesta} \citep{Koehn:2025b}, predicting fluxes across a broad spectrum and enabling to construct light curves in LSST filters.

Different observing scenario (cadence, filters, noise) have been tested.
We first evaluate the performances of the merger time recovery overcoming observational constraints. To do so, we build an optimistic observing scenario: four detection points per day (0.25 day between each detection) in $grizy$ bands, no observing noise with a measurement error between 0 and 0.2 mag, drawn from a uniform distribution and assuming a telescope sensitivity of 26 mag. The second scenario, more realistic, relies on \texttt{rubin\_sim}, the Rubin Observatory simulation package, to produce light curves as seen by LSST. We use the Rubin Observatory Deep Drilling Fields (DDF) baseline simulation \citep{lsst_mag, jones_2026_20607766}. The DDF survey consists of single pointings of about 10 square degrees each, which receive intensive observations on a regular basis. Each DDF will reach at least 1 magnitude deeper than the “Wide Fast Deep” (WFD) coadded depths, which is the bulk of the LSST survey. This baseline simulation allows us to find so-called \textit{valid sources}, i.e. sky locations that are visited at least $N_{\min}$ times within a given time window of $\Delta t$ days following a user-specified time, for which the telescope's observational conditions are extracted from the simulation in the form of the $5\sigma$ depth magnitude (magnitude threshold at which a signal is detected with an SNR 5$\sigma$ above the background level) in each filter for each observation. Those sources can then be used to inject kilonova detections with {\sc nmma} at the respective observation times and filters. 

\begin{figure*}
    \centering
    \includegraphics[width=0.95\linewidth]{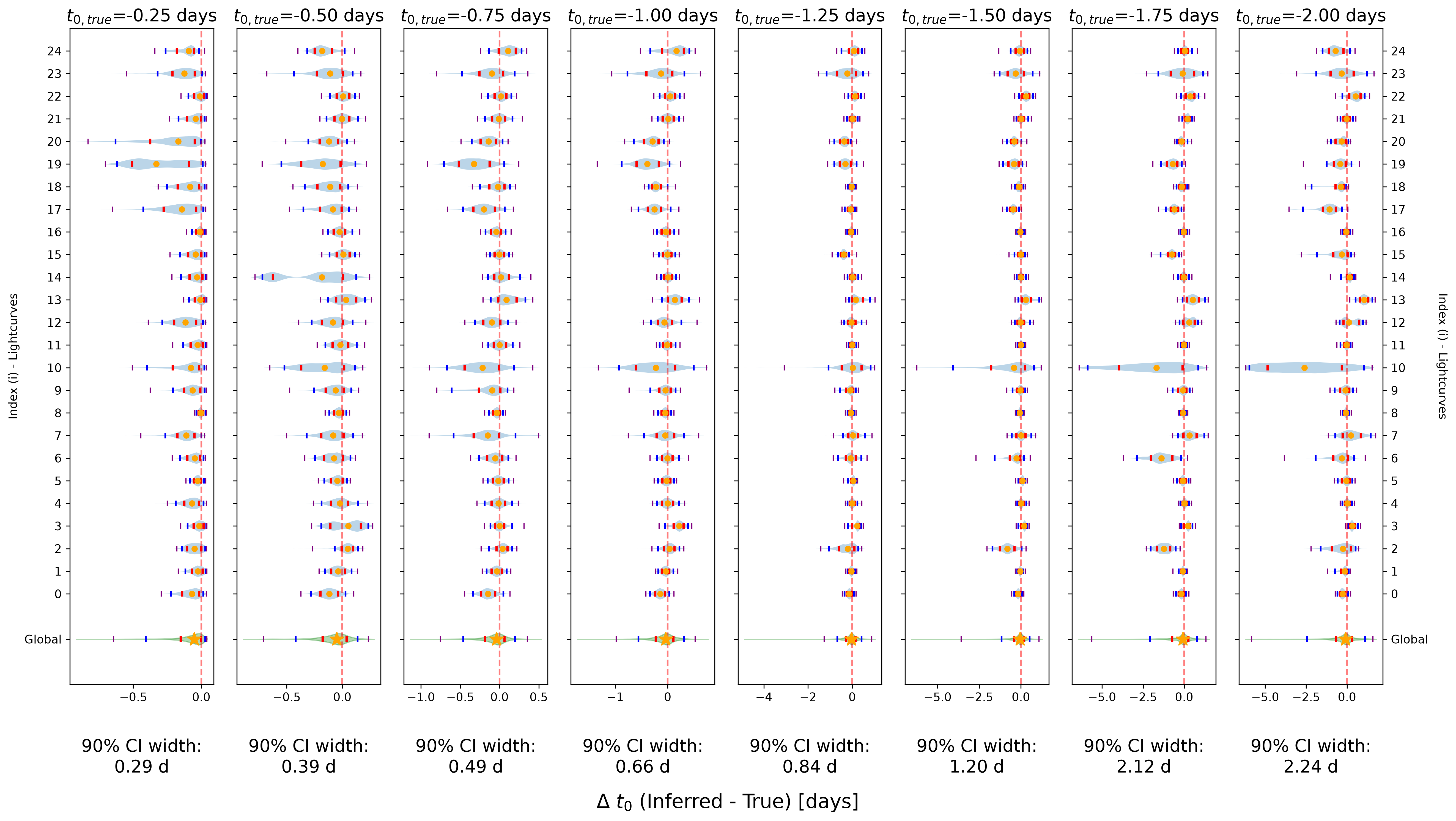}
    \caption{Violin plots of the shifted posterior distributions of the $t_0$ for the optimistic configuration with \texttt{Bu2026\_MLP}.
    Each column corresponds to a case where we remove early observations. The true time delay between the first detection and the merger time is given in the top x axis.
    The orange points indicate the median of the distribution. The 1, 2 and 3$\sigma$ credible intervals are shown by the red, blue and purple lines, respectively, for each violin. The red dotted line indicates the true value, and the green violin is the merged distribution. The corresponding median 90\% credible interval is marked below each corresponding column.}
    \label{fig:violin-optimistic-Bu26}
\end{figure*}

The priors used to analyze the lightcurve are presented in Table~\ref{tab:NMMAinjectionprior-generation}. 
In addition to observation uncertainties, to account for potential modelling inaccuracies, {\sc nmma} includes a systematic uncertainty term $\sigma_{\rm sys}$ that treats the true magnitude as a Gaussian-distributed quantity centred on the model prediction. The width of this distribution quantifies the level of systematic uncertainty. $\sigma_{\rm sys}$ is therefore also included in the analysis prior.
Results using both observing scenario are described below: Sec.~\ref{sec:self-optimistic} presents the optimistic case, injecting synthetic lightcurves with 
\texttt{Bu2026\_MLP} and performing the parameter inference using the same kilonova grid. Section~\ref{sec:self-LSST} presents similar tests, this time using the realistic LSST observing scenario. Finally, Sec.~\ref{sec:cross-LSST} stills uses the realistic observing configuration. However, we generate synthetic lightcurves using one kilonova model while performing the Bayesian parameter estimation with a different model. This cross-model analysis evaluates the ability of each model to recover the properties of kilonovae described by alternative theoretical predictions, rather than only fitting data generated from the same model, which can lead to biases. Such an approach is particularly important given the limited number of observed kilonovae so far, which prevents us from establishing a definitive description of their emission. Therefore, we account for the possibility that one theoretical model provides a closer representation of real kilonovae than the others and assess the robustness of our results against modelling uncertainties. 

For each test, we infer the posterior distribution of the merger time using the priors presented in Table~\ref{tab:NMMAinjectionprior-generation}. This priors are mostly similar to the ones used for the lightcurve generation. This ensures the same parameter space coverage, although the diversity of kilonovae is not fully known and we might encounter kilonova scenarios that are not encompassed by our model parameter space. 



\begin{figure*}
    \centering
    \includegraphics[width=0.9\linewidth]{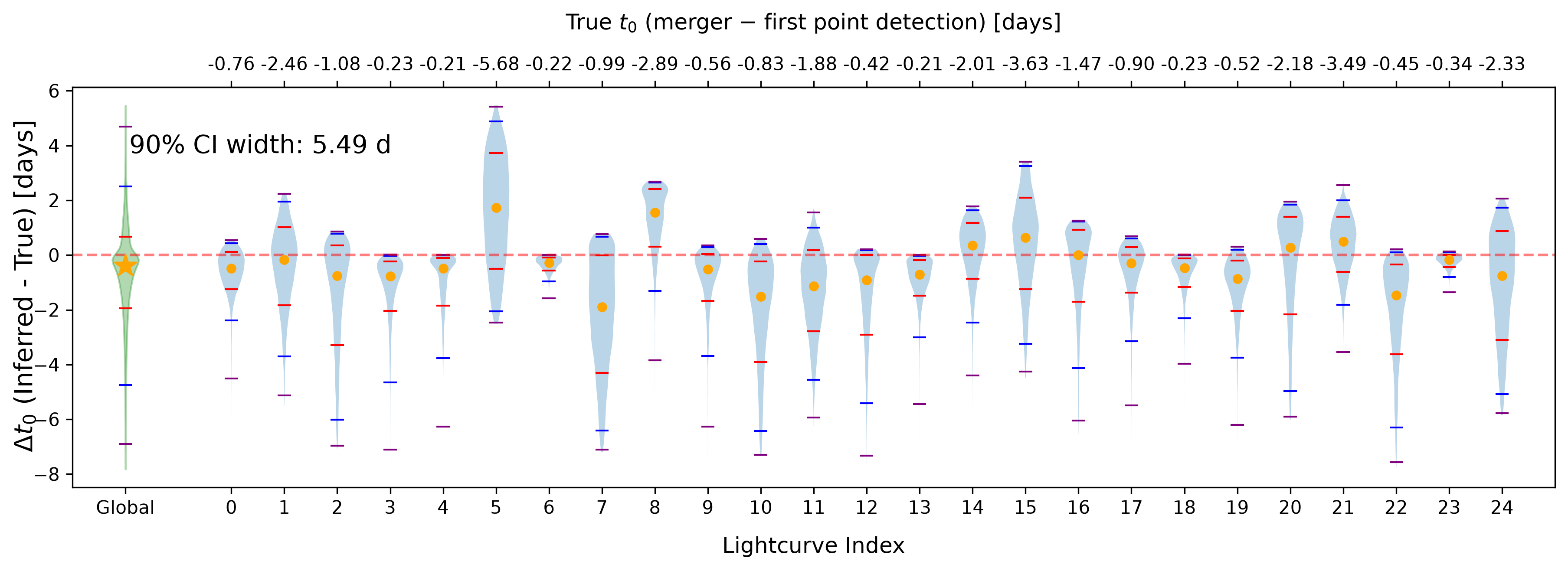} 
    
    \caption{Violin plots of the shifted posterior distributions of the $t_0$ for the realistic LSST configuration with 
    \texttt{Bu2026\_MLP}. 
    The true time delay between the first detection and the merger time is given in the top x axis. The orange points indicate the median of the distribution. The 1, 2 and 3$\sigma$ credible intervals are shown by the red, blue and purple lines, respectively, for each violin. The red dotted line indicates the true value, and the green violin is the marginalized distribution. The median 90\% credible interval is marked below each corresponding column.}
    \label{fig:Bu19_26_LSST_self}
\end{figure*}

We present the results in the form of violin plots displaying the distribution of the shifted posterior $\Delta \theta = \theta_\mathrm{inferred} - \theta_\mathrm{true} $ (here, $\theta$ being $t_0$ the timeshift between the first detection and the merger time) for each individual lightcurve, as well as the merged posterior distribution, which is defined as
\begin{equation}
    P(\theta | t_{0,\mathrm{true}}) =\frac{1}{N_\mathrm{LC}} \sum_{i=1}^{N_\mathrm{LC}} P(\theta|\mathrm{LC}_i,t_{0,\mathrm{true}}),
    \label{eq:marginalised}
\end{equation}
where $\theta$ is the merger time here. The merged posterior distribution shown in each section provides an estimate of the general behaviour of the inference across all light curves, capturing the systematic trends and overall spread of the residuals at the population level, independently of the variability between individual events.

\subsubsection{TEST 1a - Optimistic scenario}
\label{sec:self-optimistic}
We generate a grid of 25 synthetic lightcurves based on 
\texttt{Bu2026\_MLP}. This number was chosen not to build a realistic kilonova population, but rather to assess the performance of the $t_0$ recovery while keeping a reasonable computational cost. Within {\sc nmma}, the intrinsic lightcurve is converted into detection points. To assess how accurately the merger time can be recovered as a function of the first detection epoch, we perform an iterative analysis of each synthetic lightcurve. We first analyse the complete lightcurve and infer the merger time using all available observations. We then progressively remove the earliest detection and repeat the inference, effectively simulating increasingly later discoveries of the transient. By repeating this process until only a limited number of observations remain, we quantify how the precision and accuracy of the merger time estimate evolve as information from the early phase of the lightcurve is lost.
The procedure is illustrated in Fig.~\ref{fig:ts-recovery-example-Bu19} for the 
\texttt{Bu2026\_MLP} model. For this specific lightcurve, we can see on the left panel that the true merger time is well within the recovered merger time 1$\sigma$ posterior distribution (represented by the error bars) and the size of the error bars increases as more early detection points are removed (light-colored points in the left panel).

The final results
are shown in
Fig.~\ref{fig:violin-optimistic-Bu26},
respectively. Each column corresponds to a case where we remove early observations, increasing progressively the timeshift between the first detection and the merger from 0.25 to 2 days with a step of 0.25 days. 
The true $t_0$ (dotted red line) is well recovered for most of the sample, with the posterior median (orange star) lying close to the true value. The median 90\% credible interval width grows steadily as more detections are removed while keeping a width contained between a few hours in case of an early detection up to a bit above 2 days for a later detection of the kilonova. This widening of the credible interval underscores how strongly early kilonova detections constrain $t_0$.
The last two columns shows that the light curve labelled 10 clearly stands out due to the much larger width of its posterior. This important widening of the posterior is due to the faintness of the synthetic KN and therefore the overall low amount of detections despite the high observing cadence and telescope sensitivity. Indeed, when removing the first five detection epochs of this light curve, only three multi-band detections are left. 
\ignore{For a subset of lightcurves, we fail to recover the true $t_0$ value. Two main reasons explain this behavior: surrogate model predictions at very early times are less reliable, due to the rapid evolution of several parameters (such as the opacity), which makes them harder to model accurately \citep{Kasen2017}. Secondly, the non-monotonic morphologies (i.e., multiple distinct peaks) of synthetic lightcurves can mislead the inference toward incorrect $t_0$ posterior values.}


\ignore{Finally, median values of the $t_0$ posterior size over the whole kilonova sample are summarized in Table~\ref{tab:table-median-optimistic}.
This table shows that the 
\texttt{Bu2026\_MLP} model is in good agreement with comparable 1, 2 and 3$\sigma$ credible interval widths for both early and late detection, with credible interval widths increasing as we shift toward a later detection.
\begin{table}
    \centering
    \ignore{\begin{tabular}{ccc}
    \hline
    \hline
       $t_0$ posterior & \texttt{Bu2019lm} & \texttt{Bu2026\_MLP} \\
       size (median) & $t_{0,true}=-0.28$ day & $t_{0,true}=-0.28$ day \\
        \hline 
        \hline 
        $1\sigma$ & 0.3019 & 0.4066 \\
        $2\sigma$ & 0.6854 & 0.7976 \\
        $3\sigma$ & 1.1706 & 1.1987 \\
        \hline
        \hline
         & $t_{0,true}=-1.28$ day & $t_{0,true}=-1.28$ day \\
        \hline\hline
        $1\sigma$ & 0.9416 & 0.7299 \\
        $2\sigma$ & 1.7973 & 1.5721 \\
        $3\sigma$ & 2.7469 & 2.7156 \\
        \hline
    \end{tabular}}
    \begin{tabular}{cccc}
    \hline
    \hline
        \multicolumn{4}{c}{\texttt{Bu2026\_MLP}~~~~ $t_{0,true}=-0.28$ day}\\
        & 16th percentile & median & 84th percentile\\
        \hline 
        $1\sigma$ & 0.3135 & 0.4066 & 0.6045 \\
        $2\sigma$ & 0.6064 & 0.7976 & 1.2016 \\
        $3\sigma$ & 0.9164 & 1.1987 & 1.6873 \\
        \hline
        \multicolumn{4}{c}{\texttt{Bu2026\_MLP}~~~~ $t_{0,true}=-1.28$ day}\\
        \hline
        $1\sigma$ & 0.6219 & 0.7299 & 1.6185 \\
        $2\sigma$ & 1.3001 & 1.5721 & 2.9555 \\
        $3\sigma$ & 2.1364 & 2.7156 & 4.4227 \\
        \hline\hline
    \end{tabular}
    \caption{16th percentile, median and 84th percentile values of the $t_0$ posterior size over the whole lightcurve grid for 
    \texttt{Bu2026\_MLP} considering optimistic observing scenario. Values shown for two different time delays between the merger time and the first detection: -0.28 and -1.28 days.}
    \label{tab:table-median-optimistic}
\end{table}}

\subsubsection{TEST 1b - Self-grid Injection+Recovery LSST}
\label{sec:self-LSST}

To illustrate how kilonovae will be observed by LSST, we inject synthetic lightcurves into simulated observations of the DDF survey. We use the same KN parameters as in Sec.~\ref{sec:self-optimistic} and we generate a grid of 25 light curves using the 
\texttt{Bu2026\_MLP} model injected in LSST survey. 
The resulting simulated observations include both photometric detections and upper limits, reflecting realistic survey conditions.
Unlike Sec.~\ref{sec:self-optimistic}, we do not artificially remove the earliest observations to investigate the merger time recovery. Instead, LSST observing cadence often naturally produces a time shift between the merger and the first detection, depending on the survey schedule and the kilonova brightness. We therefore analyse each simulated LSST lightcurve as observed, using its corresponding detection history to assess the recovery of the merger time under realistic observing conditions. Results are shown as violin plots in Fig.~\ref{fig:Bu19_26_LSST_self}.

The inference 
recovers well the true merger time at least within the $3\sigma$ credible intervals. 
As expected due to lower total amount of detections, most of the light curves have large posterior distribution widths of the order of several days as indicated by the median 90\% credible interval width of 5.49 days. While the $1\sigma$ posterior size is smaller, the merger time could not be reliably inferred for a subset of light curves. This is due to insufficient or poorly constraining observational coverage. The main causes include sparse detections (often only one or two epochs), observations in a single filter or without measurable colour evolution, detections occurring only several days after the merger, and limited temporal sampling with multiple observations obtained at the same epoch. In a few cases, large photometric uncertainties or intrinsically complex light-curve morphologies (e.g., variability or bimodal brightness peak) also prevented a robust determination of the merger time.

\ignore{
\begin{table}[h]
    \centering
    \ignore{
    \begin{tabular}{ccc}
    \hline
    \hline
       $t_0$ posterior & \texttt{Bu2019lm} & \texttt{Bu2026\_MLP} \\
       size (median) & [days] & [days] \\
        \hline 
        \hline 
        $1\sigma$ & 4.8145 & 3.7958 \\
        $2\sigma$ & 8.0060 & 7.2312 \\
        $3\sigma$ & 9.6850 & 9.4549 \\
        \hline
    \end{tabular}}
    \begin{tabular}{cccc}
    \hline
    \hline
    \multicolumn{4}{c}{\texttt{Bu2026\_MLP}~~~~ $t_{0}$ posterior size [days]}\\
         & 16th percentile & median & 84th percentile\\
        \hline
        $1\sigma$ & 2.3045 & 3.7958 & 5.4240 \\
        $2\sigma$ & 4.7379 & 7.2312 & 8.6708 \\
        $3\sigma$ & 6.7445 & 9.5145 & 9.8976 \\
        \hline
        \hline
    \end{tabular}
    \caption{16th percentile, median and 84th percentile values of the $t_0$ posterior size in days over the whole lightcurve grid for 
    \texttt{Bu2026\_MLP}, considering LSST simulated observing scenario.}
    \label{tab:table-median-LSST-self}
\end{table}}

\begin{figure}
    \centering
\includegraphics[width=\linewidth]{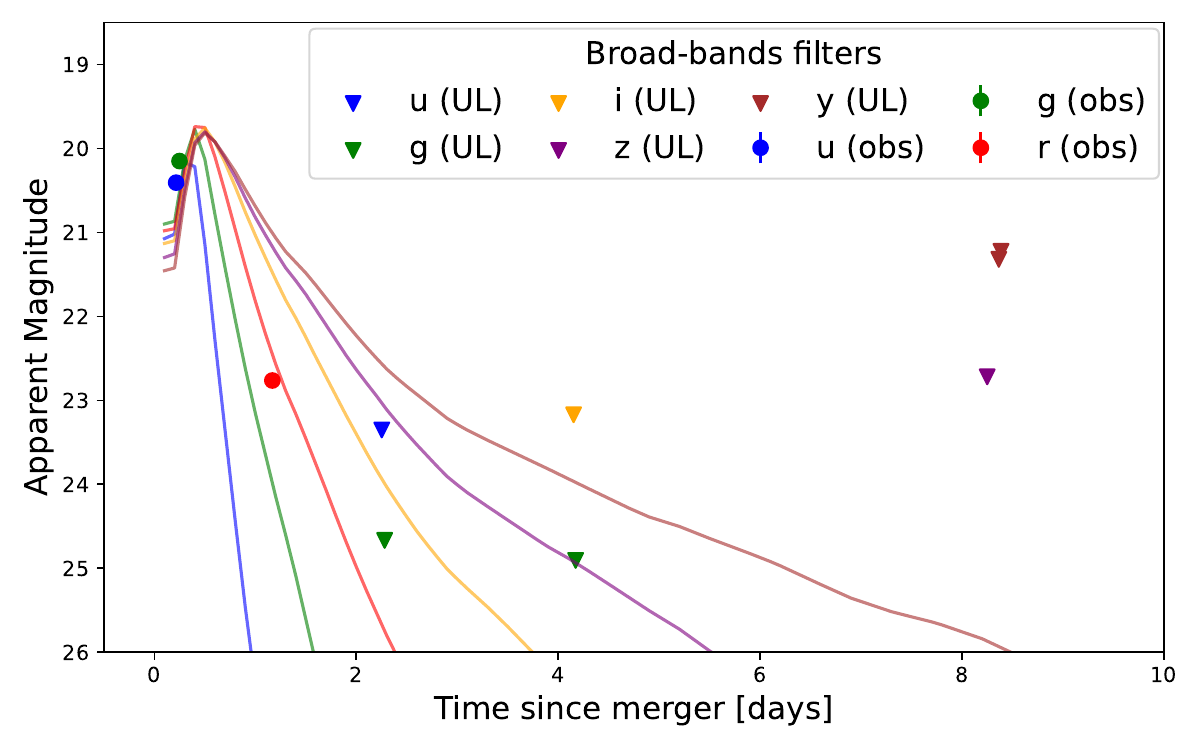}
    \caption{Lightcurve labeled 6 generated with \texttt{Bu2026\_MLP} and injected in LSST DDF survey. The injected parameters are: $\log_{10}M^{\rm{dyn}}_{\rm{ej}}=-3.704$, $\log_{10}M^{\rm{wind}}_{\rm{ej}}=-1.297$, $\theta=0.515 ~\rm{rad}$, $D_L=198.98~$Mpc. Detection are shown as points and upper limits as inverted triangles.}
    \label{fig:LC1}
\end{figure}

The light curves labeled 6 and 23 represent some of the most promising case for the LSST. We show the photometric dataset for the light curve 6 together with \texttt{Bu2026\_MLP} bestfit in Fig.~\ref{fig:LC1}. Despite the low total number of detection points and a large observational uncertainty in one detection in $lsstu$ band, they are are multi-band, allowing us to study the color evolution and in the rising phase, further helping to constrain the merger time as already stated in Sec~\ref{sec:self-optimistic}.

\begin{figure*}
    \centering
    \includegraphics[width=0.9\linewidth]{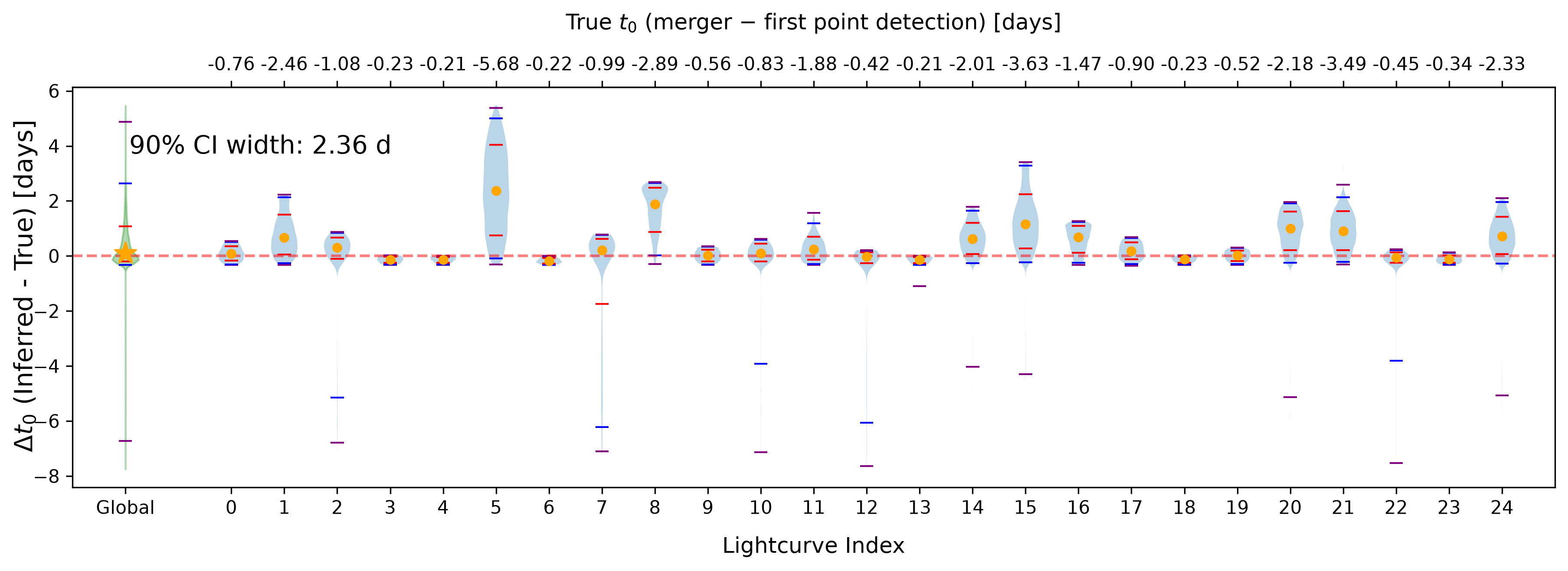}
    \caption{
    Same as Fig~\ref{fig:Bu19_26_LSST_self} but with upper limits 3 hours before merger time.
    }
    \label{fig:bu19_lsst_ul}
\end{figure*}

\begin{figure*}
    \centering
    \includegraphics[width=0.9\linewidth]{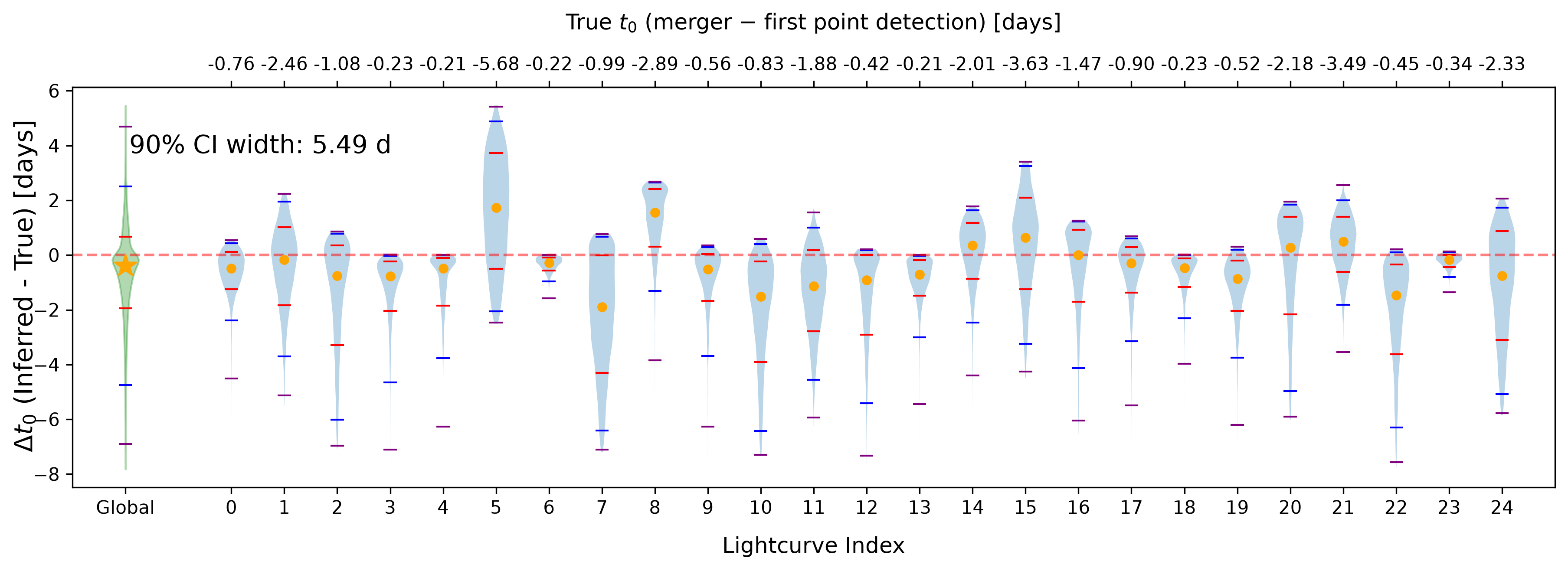}
    
    \caption{Same as Fig. \ref{fig:Bu19_26_LSST_self}, but with a fixed luminosity distance for the inference.}
    \label{fig:Bu19_26_LSST_self_Dl}
\end{figure*}

\ignore{\paragraph{Quantifying the information gain --} 
To directly quantify the extent to which the inference constrains the merger time search space and thereby provide a quantitative assessment of recovery performance in terms of precision, we employ the Kullback-Leibler (KL) divergence, also known as the relative entropy, which expresses the information gain of the inference in units of bits \citep{kl-div}. Unlike the posterior width alone, the KL divergence measures the reduction of uncertainty relative to the prior, providing a quantitative measure of the precision with which the merger time is recovered. For a uniform prior of width $W$ and assuming a normally distributed posterior of standard deviation $\sigma$, the information gain in bits is given as 
\begin{equation}
    D_\mathrm{KL} = \log_2(W) - \frac{1}{2}\log_2(2 \pi e\sigma^2).
    \label{eq:kldiv}
\end{equation}
As the result is expressed in bits, an information gain of 3 indicates that the parameter space is reduced by a factor $2^3=8$. Fig. \ref{fig:kldiv} shows the relative entropy as a function of the total number of detections and timeshift between the merger time and the first observation (detection or UL) for both \texttt{Bu2019lm} and \texttt{Bu2026\_MLP}. As expected, the information gain increases for lower timeshifts, indicating that mergers detected earlier in their evolution are recovered with greater precision. This is consistent with the conclusions drawn from the optimistic scenario study, where earlier detections were already found to yield tighter posteriors, underlining the strong impact of an early observation to constrain the merger time. Similarly, $D_\mathrm{KL}$ tends to be higher for larger number of detections, indicating that a more sampled lightcurve produces more precise posterior. In particular, lightcurve labeled 1 for \texttt{Bu2019lm}, already mentioned as the most promising for LSST, stands out as the best-recovered case, combining a high number of detections with a small timeshift (UL in the rising phase), and consequently reaches the highest information gain of the sample. In particular, the information gain starts to be significant for timeshift smaller than 2 days and a number of detections above 7.

\begin{figure}
    \centering
    \includegraphics[width=0.9\linewidth]{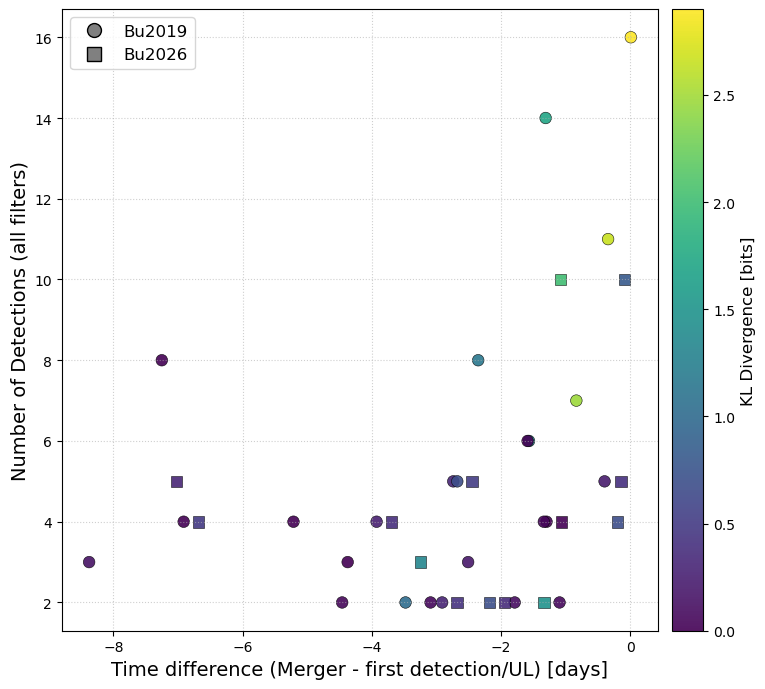}
    \caption{Evolution of the Kullback-Leibler divergence as a function of the timeshift between the merger time and the first observation (detection or UL) and the total number of detections across all filters (excluding upper limits) for the realistic LSST configuration with \texttt{Bu2019lm} and \texttt{Bu2026\_MLP}. The circles represent the lightcurves generated with \texttt{Bu2019lm} and the squares, those generated with \texttt{Bu2026\_MLP}.}
    \label{fig:kldiv}
\end{figure}}

\paragraph{Exploring the importance of upper limits --} 
\ignore{We isolate the synthetic kilonovae for which the merger time recovery failed or was correctly estimated but with large uncertainties ($>~2~$ days)}
In the context of wide-field surveys such as the LSST DDF, it is expected to have fields observed every two days. Knowing this, one could expect to have upper limits prior to the explosion of the kilonova. To explore this scenario, we fit again the light curves adding a fake upper limit 3 hours before the merger time in two neighbouring filters (e.g. $g$ and $r$) to mimic \href{https://www.lsst.org/scientists/survey-design}{LSST filter rotation strategy}. Results are shown in Fig.~\ref{fig:bu19_lsst_ul}. 
\ignore{The histograms compare the widths of the minimum credible intervals (1, 2 or 3$\sigma$) containing the true merger time for each light-curve in the sample under the standard and additional fake upper limit configurations.}
We can see that an upper limit before the merger time provides stronger constraints, largely reducing the width of the posterior for lightcurves with sparse detections. Still, for some lightcurves such as the 5 and 7, this upper limit does not improve the width of the posterior, due to a short number of detections (between 2 and 3) shortly spaced and without any measurable color evolution for light curve 7 combined to
an overly late detection of the kilonova occurring several days after the merger for light curve 5. Still, for light curve 7, the addition of upper limits still changed the probability distribution of the posterior to concentrate it around the true merger time, effectively dividing the width of the $1\sigma$ credible interval by more than 2.  On average, an upper limit before the merger time narrows the posterior width by 60\%, corresponding to a shortening of several days, as written in Fig~\ref{fig:bu19_lsst_ul}.

\ignore{\begin{table}[h]
    \centering
    \begin{tabular}{ccc}
    \hline
    \hline
       $t_0$ posterior & \texttt{Bu2019lm} & Width \\
       size (median) & (UL) & reduction\\
        \hline 
        \hline 
        $1\sigma$ & 1.6101 & 66.56\% \\
        $2\sigma$ & 2.5425 & 68.24\% \\
        $3\sigma$ & 3.2099 & 66.86\% \\
        \hline
    \end{tabular}
    \caption{Median values of the $t_0$ posterior size over the whole lightcurve grid, for \texttt{Bu2019lm} with an upper limit before the merger and the corresponding reduction (in percent) compared to the lightcurve without the upper limit, considering LSST simulated observing scenario.}
    \label{tab:table-median-LSST-UL}
\end{table}}

\paragraph{Exploring the importance of redshift estimation --} 
Realistically, if the source is identified as a kilonova, a redshift might be estimated (from host galaxy identification, for example). Thus, we re-analyzed the lightcurves with a fixed luminosity distance. 
The results are shown in Fig. \ref{fig:Bu19_26_LSST_self_Dl}.
\ignore{For some lightcurves (such as the 7 or 23 for \texttt{Bu2019lm} or the 12 or 18 for \texttt{Bu2026\_MLP}), fixing the distance considerably reduces the width of the $3\sigma$ credible interval by between $\sim30$ to 90\%. Still, apart from those cases, for most lightcurves the difference is below 5\%, indicating an already accurate estimation of the distance from the inference of the lightcurve only.}
For most of the lightcurves, the difference compared to Fig.~\ref{fig:Bu19_26_LSST_self} is below 5\%, indicating an already accurate estimation of the distance from the inference of the lightcurve only with the 90\% credible interval width indicated on both Fig~\ref{fig:Bu19_26_LSST_self} and Fig~\ref{fig:Bu19_26_LSST_self_Dl} being the same.
\ignore{However, in certain cases, the difference can be substantial, with a reduction of the interval by a factor of around 2. This has been observed in sparse lightcurves with between 2 and 4 detections in the fading phase. This led to an overestimation (resp. underestimation) of the distance, which, in turn, caused a bias of the timeshift towards greater (resp. smaller) values. }

\begin{figure*}
    \centering
    \includegraphics[width=0.9\linewidth]{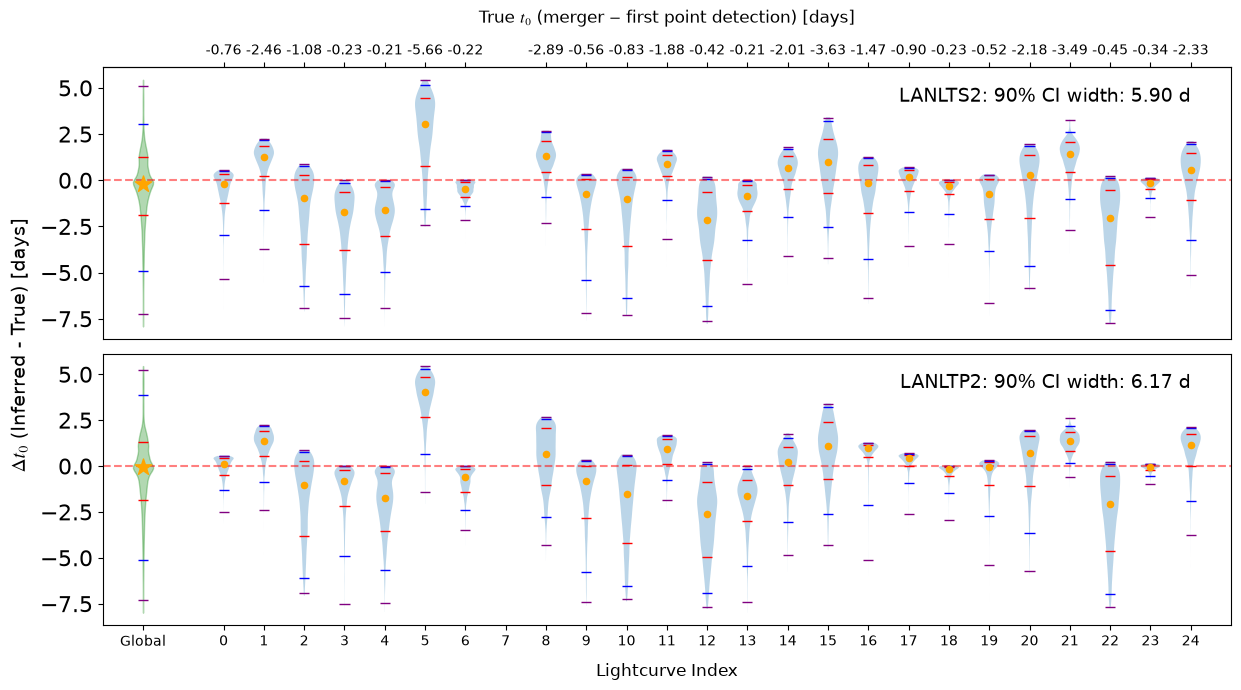}
    
    \caption{Violin plots of the shifted posterior distributions of the $t_0$ for the realistic LSST configuration genrated with \texttt{LANLTS2} (top) and \texttt{LANLTP2} (bottom) inferred with \texttt{Bu2026\_MLP}. The true time delay between the first detection and the merger time is given in the top x axis. The orange points indicate the median of the distribution. The 1, 2 and 3$\sigma$ credible intervals are shown by the red, blue and purple lines, respectively, for each violin. The red dotted line indicates the true value, and the green violin is the marginalized distribution. The median 90\% credible interval for both models is indicated on the figure.}
    \label{fig:Bu19_LSST_cross_LANL}
\end{figure*}

\subsubsection{TEST 1c - Cross-grid Injection+Recovery LSST}
\label{sec:cross-LSST}
Finally, using the same model for lightcurve generation and inference, can lead to biases. In principle, we expect that one model should be able to fit the lightcurve generated by another model if it is built upon the same kilonova theoretical hypotheses (e.g. two two-component models). Therefore, to check the ability of a model to recover the parameters of lightcurves generated with a different model, we perform a cross-grid injection-inference using the LSST configuration described in Sec.~\ref{sec:kn-generation}. We use two models, we continue to use \texttt{Bu2026\_MLP} for inference and \texttt{LANLTS2} and \texttt{LANLTP2} based on SuperNu for light curve generation. The two LANL models differ by their geometry: \texttt{LANLTS2} includes a red toroidal dynamical ejecta and a bluer spherically symmetric wind ejecta, whereas \texttt{LANLTP2} includes the same dynamical ejecta but uses a peanut-shaped wind ejecta instead. 

The violin plots in Fig.~\ref{fig:Bu19_LSST_cross_LANL} display the
results for \texttt{LANLTS2} (top) and \texttt{LANLTP2} (bottom). From this figure we can see that light curve 7 is missing. Owing to the different treatment of the microphysics such as the opacities or heating rates between \texttt{Bu2026\_MLP} and the \texttt{LANL} models, hence affecting the brightness of the predicted light curves, resulting in a fail to satisfy the LSST detection criteria in the case of light curve 7. For both models, the recovered posteriors remain broadly consistent with the injected values, with median offsets close to zero for most lightcurves. Compared to the self-consistent case (median 90\% CI width of 5.49 days), the median 90\% CI width increases slightly to 5.90 days for \texttt{LANLTS2} and 6.17 days for \texttt{LANLTP2}. This modest broadening points to a small bias in the recovery when the inference model differs from the injection model.
\ignore{, whereas Fig.~\ref{fig:bu19_lsst_comparison} illustrates the difference between self-recovered analysis and cross-inference for the models considered. From this last figure, we can see that the results of the inference with the 3 models are consistent as expected. In some cases, the posterior width is less large when doing cross-inference with a median 3 sigma width smaller for both \texttt{LANL} models than \texttt{Bu2019lm}. Conversely, some posterior are significantly larger when performing inference with a \texttt{LANL} model; this can be justified by, first, a very sparse lightcurve and secondly, a pronounced non-uniform peak (as in lightcurve 1 showed in Fig.~\ref{fig:LC1}) which, as mentioned previously, can reduce the robustness of the inference of the merger epoch. Additionally, we see that the recovery with \texttt{LANLTS2} provides better estimates than with \texttt{LANLTP2}. This is explainable by the different geometries of the LANL models, as the \texttt{LANLTS2} geometry resembles more closely the geometry of \texttt{Bu2019lm} which includes the same dynamical and wind ejecta geometries as well as a supplementary blue polar dynamical ejecta.}

\ignore{\begin{figure*}
    \centering
    \includegraphics[width=0.8\linewidth]{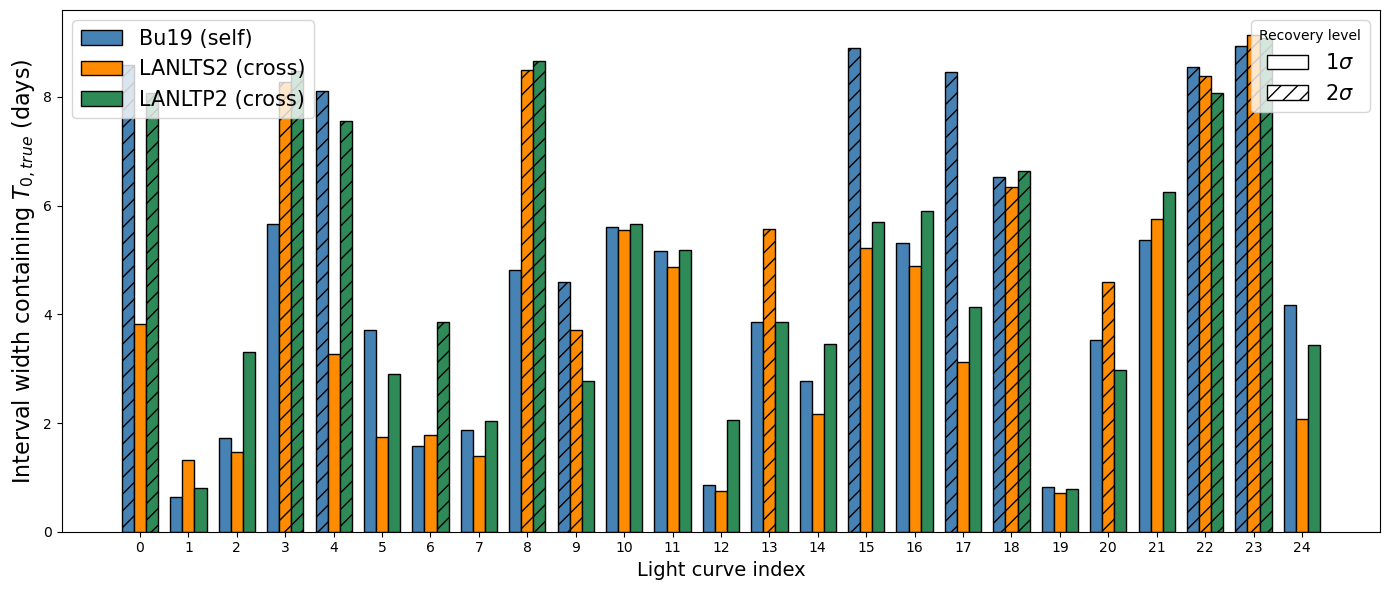}
    \caption{Credible interval containing the true $t_0$ for the realistic LSST configuration gener9ated with \texttt{Bu2019lm}. The blue, orange and green bars show, respectively, inference with \texttt{Bu2019lm}, \texttt{LANLTS2} and \texttt{LANLTP2}. The plain and hatched bars indicate, respectively, a true $t_0$ at recovered within the 1 and 2$\sigma$ credible interval.}
    \label{fig:bu19_lsst_comparison}
\end{figure*}}

\subsection{TEST 2a - GW170817/AT2017gfo}
\label{sec:170817}
\begin{figure*}
    \centering
    \includegraphics[width=0.85\linewidth]{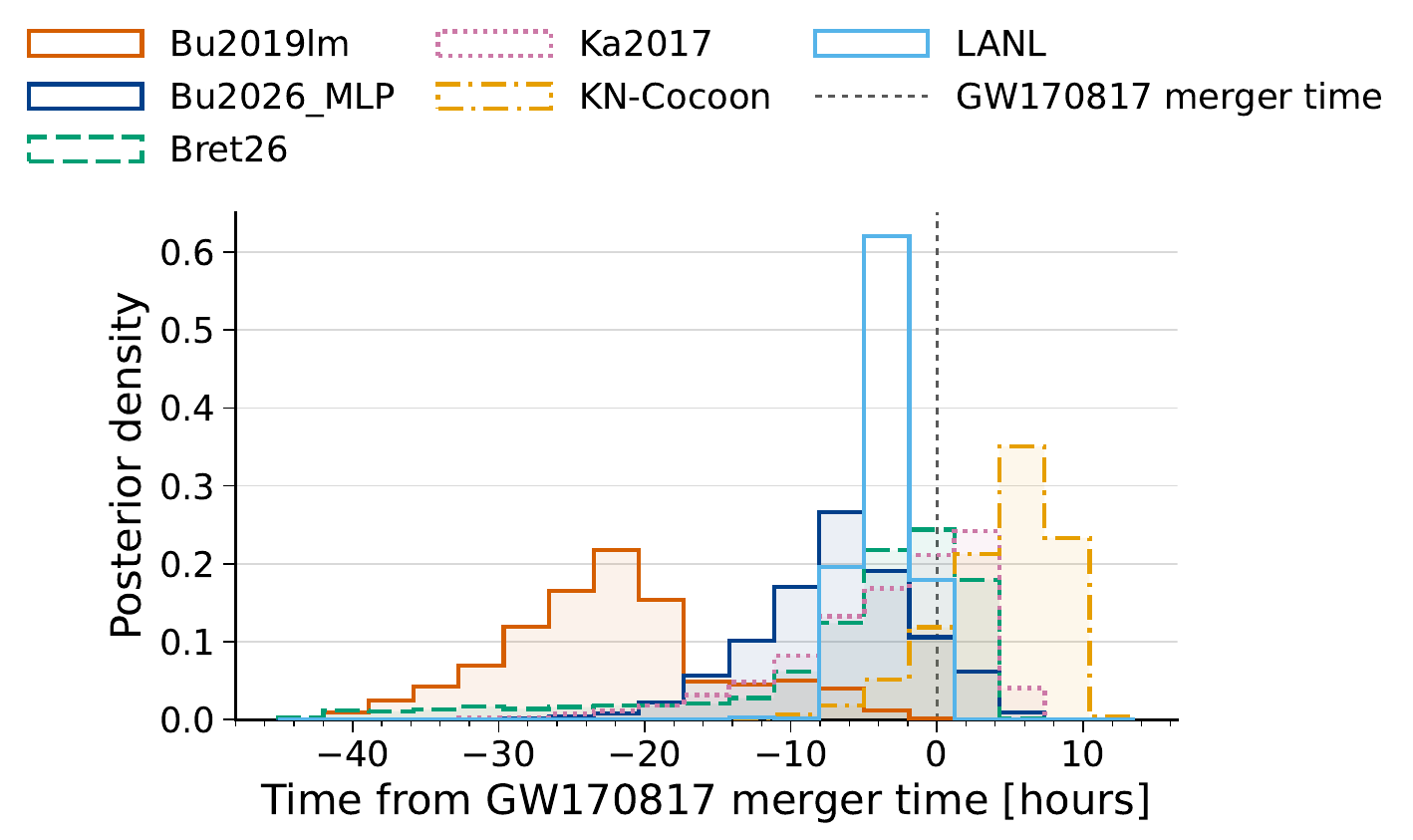}
    \caption{AT2017gfo merger time posterior distributions inferred with all models and inference frameworks considered in this work.}
    \label{fig:170817-all-results}
\end{figure*}

To get rid of the model biases, we also assess the performances of the merger time recovery using real kilonova candidates. This also allows us to investigate a scenario with a lightcurve enriched by various optical telescope observations, which is likely to be the case following a kilonova detection by LSST. AT2017gfo remains the only confirmed kilonova with an unambiguous association to a BNS merger. It is therefore natural to use it as a toy model for investigating constraints on the merger time $t_0$. The latter is precisely known from the coincident GW signal GW170817, providing a excellent benchmark for testing how well merger time can be inferred from standalone kilonova observations.
Through this section we use different inference frameworks - {\sc nmma}, {\sc Redback}, {\sc nessai/bilby} and {\sc rift} (see Sec.~\ref{sec:inferenceframework}) - and different kilonova models - {\tt Bu2019lm}, {\tt Bu2026\_MLP}, {\tt Ka2017}, {\tt LANL}, {\tt Brethauer26} and the semi-analytic \texttt{two\_component\_kilonova\_with\_cocoon} (see Sec.~\ref{sec:kn_models}) - and we compare the constraints obtained on the merger time. Here we assume a fixed luminosity distance of 40.7 Mpc, inferred from the host galaxy redshift measurement~\citep{Palmese_2017}. Our AT2017gfo dataset is composed of photometric measurements over a period of $\sim$ 30 days in $u\,g \,r \,i\, z\, y\, J\,H\,K$ bands corrected for galactic extinction \citep{Schlafly_2011}. $T_0$ posterior distributions are shown in a single figure (Fig.~\ref{fig:170817-all-results}) summarizing the results obtained with all models and inference frameworks.

\subsubsection{Inference with \textsc{nmma} and \texttt{Bu2019lm/Bu2026\_MLP}}\label{subsec:NMMA_AT2017gfo}
We analyze AT2017gfo lightcurve with \texttt{Bu2019lm} and \texttt{Bu2026\_MLP} models. 
The priors used are described in Appendix~\ref{sec:priors}, Table \ref{tab:NMMA2017prior}. The priors broadly cover the model parameter space in order to be as generic as possible, not assuming anything about the source (for instance an on-axis scenario), except its luminosity distance. Here, we analyze AT2017gfo twice, once leaving $\sigma_{\rm sys}$ as a free parameter, as in Sec.~\ref{sec:kn-generation}, and once with a fixed $\sigma_{\rm sys}$ = 1 mag. The prior on the merger time spans uniformly over the two days before the first detection.


Fig.~\ref{fig:170817-all-results} shows the merger time posterior distributions with $\sigma_{\rm sys}=1$ mag using \texttt{Bu2019lm} (orange curve) and \texttt{Bu2026\_MLP} (dark blue curve) and the best fit light curves in $ugrizyJHK$ generated with \texttt{Bu2026\_MLP} is displayed in Appendix~\ref{sec:bestfitlc}, Fig.~\ref{fig:nmma_lc}. The true merger time is at the edge of \texttt{Bu2019lm} posterior distribution. \texttt{Bu2026\_MLP} model gives a correct estimation of AT2017gfo merger time (black dashed line in the histogram) within the 2$\sigma$ interval (corresponding to a posterior size of approximatively 21 hours). 
The difference in the recovery of $t_0$ between the \texttt{Bu2019lm} and the \texttt{Bu2026\_MLP} models likely reflects differences between the versions of \textsc{possis} used to generate the two grids. 
In particular, \texttt{Bu2026\_MLP} was produced with a more recent version of the code \citep{Bulla2023}, which incorporates updated heating rates, thermalization efficiencies, and opacity treatments compared with those adopted for the \texttt{Bu2019lm} grid \citep{Bulla:2019}. At early times, the opacity prescription used in \texttt{Bu2019lm} tends to overestimate the opacity, delaying photon escape and producing a longer rise to maximum light. When $t_0$ is allowed to vary, the fit can compensate for this delayed model peak by shifting the inferred merger time to earlier epochs. In contrast, the opacities used in \texttt{Bu2026\_MLP} are lower and allow radiation to escape more readily, making the light curves peak shortly after the merger. This more rapid evolution reduces the time shift required to reproduce the observed peak and leads to a more accurate recovery of $t_0$.

We show the posterior distributions using a free $\sigma_{\rm sys}$ in Fig.~\ref{fig:Bulla-nmma-freesigmasys} (Appendix~\ref{appendix-AT2017gfo}).
We also found an inferred $\sigma_{\rm sys}$ lower than 1 mag and a $t_0$ deviating from the true value. A similar effect is observed in \textsc{Redback} results below: decreasing ($\sigma_{\rm sys}$) results in a slight shift of the posterior distributions, while noticeably narrowing them. 
A distinctive feature of kilonova emission is its hours-to-days long rise, reaching peak luminosity on timescales of roughly one to two days after the merger, in contrast to, for example, GRB afterglows, which typically peak much earlier. This is followed by a rapid fading phase. As a result, the slope of the rising lightcurve is highly sensitive to the merger time. Detections during this phase - which were not available for AT2017gfo, whose first detection occurred $\sim$11 hours after the merger - can significantly improve constraints on the merger time. 
Given the good performance of \texttt{Bu2026}, we now focus on the \texttt{Bu2019lm} case, for which the recovered merger time shows a larger deviation from the true value. To assess whether detections during the rising phase can improve the merger-time inference, we first fit the AT2017gfo light curve while fixing the merger time to the value independently measured from the GW signal.
We then inject fake detection points from the resulting best-fit lightcurve, in $ugrizyJHK$ filters around 0.25 day post merger 
and repeat the Bayesian inference with {\sc nmma} treating the merger time as a free parameter again. AT2017gfo photometric lightcurve with synthetic detection points, using \texttt{Bu2019lm} best-fit, is shown in Appendix~\ref{sec:bestfitlc}, Fig.~\ref{fig:2017-fake-detection}.

\begin{figure}
    \centering
    \includegraphics[width=\linewidth]{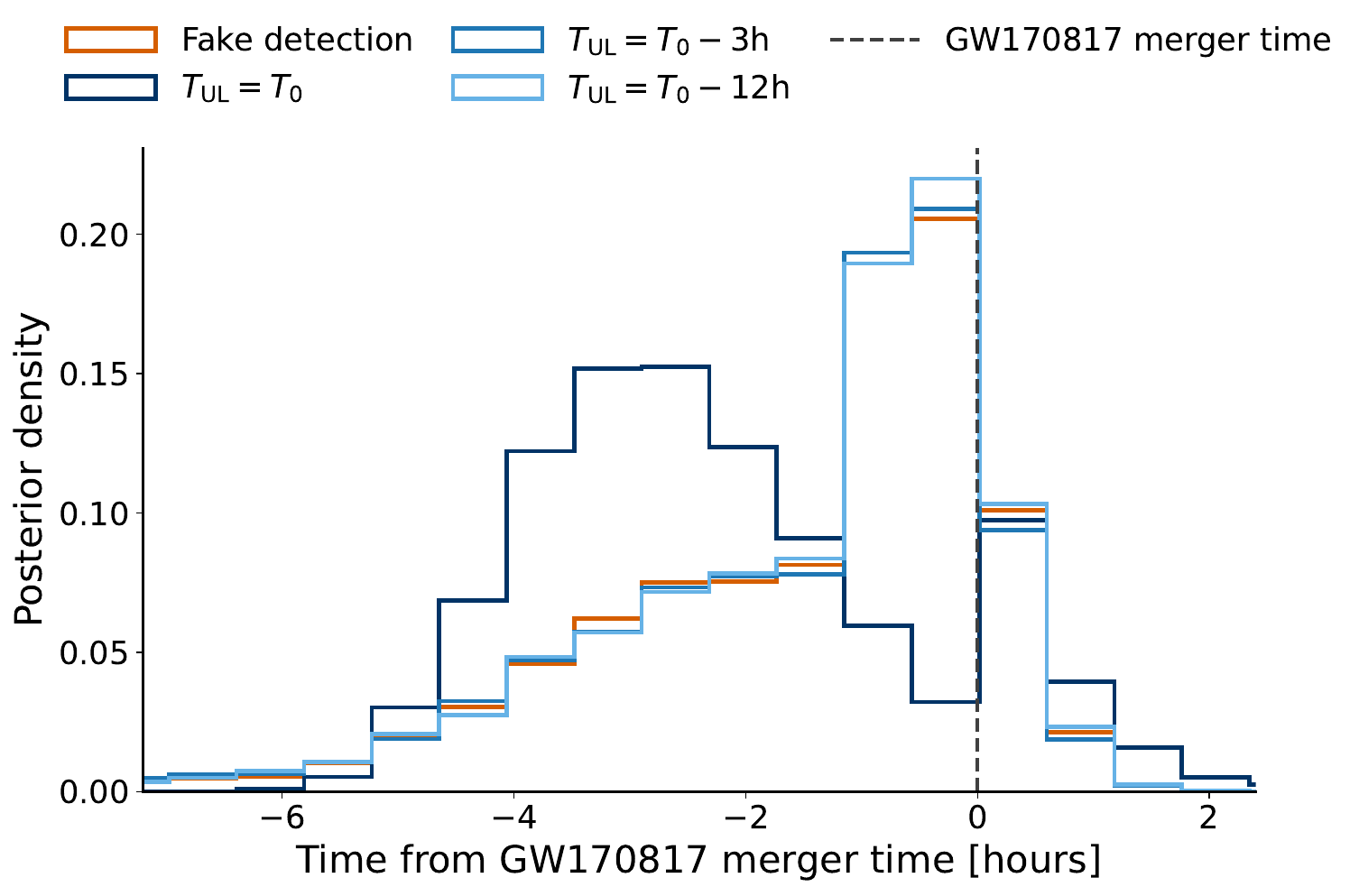}
    \caption{AT2017gfo merger time posterior distribution inferred with {\sc nmma} using \texttt{Bu2019lm} when adding a fake detection point in the rising phase or upper limits before the first detection.}
    \label{fig:fake-det-UL-nmma-AT2017gfo}
\end{figure}

Updated posterior distribution of the merger time is presented in Fig.~\ref{fig:fake-det-UL-nmma-AT2017gfo}. Adding these fake detection points significantly improve the constraints. GW170817 merger time is now at the edge of the $1\sigma$ posterior distribution and the size of the uncertainties is reduced, reaching 0.15 days for \texttt{Bu2019lm}. 



Finally, in the context of wide-field surveys such as Vera C. Rubin, observations are expected to be obtained nightly, with fields revisited approximately every two days. Given this cadence, most kilonovae are unlikely to be detected during their rising phase. And even with one detection both in the rising and fading phases, it would still be challenging to identify the source as a kilonova, since the observed fading rate over the sampled interval could be too shallow to provide a distinctive classification signature. Thus, we try here a third configuration without a fake detection in the rising phase but with upper limit near the merger time, assuming the last visit occurred in the day before similarly to what was done in Sec.~\ref{sec:self-LSST}. 
We use the same methodology as with the fake detection to add upper limits in the photometry dataset, resulting in additional fake 23 mag upper limit (i) at the merger time $t_0$, (ii) 0.1 and (iii) 0.5 days before the merger in $ugrizyJHK$ bands. Posterior distributions of the merger time using \texttt{Bu2019lm} 
are presented in Fig.~\ref{fig:fake-det-UL-nmma-AT2017gfo} and Table~\ref{tab:table-UL-AT2017gfo} summarizes the size of the posterior distribution for each upper limit scenario. As for the fake detection GW170817 merger time is now at the edge of the $1\sigma$ posterior distribution and fully contained by the $2\sigma$ level when the upper limit deviates from the true merger time. 
Finally, to verify that our result does not depend on the choice of sampler, we repeated the \texttt{Bu2019lm} analysis using the nested sampler \texttt{PyMultiNest}, which explores the posterior in a fundamentally different way from pocoMC. Results are shown in Appendix~\ref{appendix-AT2017gfo}, Fig.~\ref{fig:Bu19lm-AT2017gfo-pymultinest}. In particular, in the fake-detection case, the merger time constraint is consistent (and slightly improved), reaching a $1\sigma$ posterior size of 0.03 days.

\begin{table}[h]
    \centering
    \begin{tabular}{cc}
    \hline
    \hline
    Fake observation epoch & \texttt{Bu2019lm} - $t_0$ $1\sigma$ posterior size \\
     $[$days$]$ & [days] \\
    \hline
    \hline
        $T_{0,GW}+0.25$ (detection) & 0.15 ($T_{0,GW}$ at the edge) \\
        $T_{0,GW}$ (UL) & 0.16 ($\ni$ $T_{0,GW}$) \\
        $T_{0,GW}-0.1$ (UL) & 0.16 ($T_{0,GW}$ at the edge) \\
        $T_{0,GW}-0.5$ (UL) & 0.15 ($T_{0,GW}$ at the edge) \\
        \hline
        \hline
    \end{tabular}
    \caption{Size of merger time posterior distribution for AT2017gfo using fake upper limit before and at the merger time, for \texttt{Bu2019lm}.
    $T_{0,GW}$ is GW170817 merger time here.}
    \label{tab:table-UL-AT2017gfo}
\end{table}


\subsubsection{Inference with \textsc{nmma} and Brethauer26}
We repeat the analysis on the extinction-corrected AT2017gfo photometry using surrogate models trained on the {\tt Brethauer26} grid from FIESTA \citep{Koehn:2025b} using the priors detailed in Table \ref{tab:Bret26_priors} (Appendix~\ref{sec:priors}) with $\sigma_{\rm sys}$ = 0.5 and 1.

We find that for a single component with a $\sigma_{\rm sys}$ = 0.5 (1), the posterior distribution of the ejecta parameters are $v_{\rm kin}$ = $0.16^{+0.03}_{-0.01}  \,(0.15^{+0.03}_{-0.02})c$, $\log_{10}$ ($m_{\rm ej}$/M$_\odot$) = $-1.39^{+0.03}_{-0.03}\,(-1.37^{+0.07}_{-0.08})$,  and $\log_{10}$X$_{\rm lan}$ = $-3.64^{+0.07}_{-0.10}\,(-3.70^{+0.17}_{-1.15}$). Given our setup, $t_{\rm0, true} \sim$ -0.23 days which is contained within the 1$\sigma$ error bars of our models with $t_0 = -0.21^{+0.07}_{-0.15}\,(-0.34^{+0.18}_{-0.39})$. The larger systematic uncertainty results in a wider posterior for all of the parameters, but are consistent within 1$\sigma$. The posterior for $t_0$ is shown in Fig.~\ref{fig:170817-all-results}. We then show the light curve of the best-fitting model with $\sigma_{\rm sys}$ = 1 in Appendix~\ref{sec:bestfitlc}, Figure \ref{fig:nmma_lc_Bret26}. Fitting with a two-component model may lead to improved recovery of $t_0$ and better-fitting light curves.

\subsubsection{Inference with \textsc{george} and \texttt{Ka17}}

We carry out a similar test using \texttt{Ka2017} to infer $t_0$ from the AT2017gfo photometry. This model is a two-component model (red and blue) where components are added together, ignoring the geometry of the ejecta and interactions between the two components. The test is carried out for two systematic uncertainties on the models, $\sigma_{\rm{sys}}=1.0$ and $\sigma_{\rm{sys}}=0.5$.
\ld{The priors used are described in Appendix~\ref{sec:priors},  Table~\ref{tab:Ka2017_priors}. We keep the priors on $t_0$ and $D_{L}$ consistent with the priors described in Table~\ref{tab:NMMA2017prior}. We note that this inference is done without taking into accounts the upper limits in the observed AT2017gfo photometry that are accounted for in Section~\ref{subsec:NMMA_AT2017gfo}. Posteriors on $t_{0}$ are presented in Fig.~\ref{fig:170817-all-results} for $\sigma_{\rm{sys}}=1.0$ (pink curve) and in Fig.~\ref{fig:Ka2017-sigmasys0.5}, (Appendix~\ref{appendix-AT2017gfo}) for $\sigma_{\rm{sys}}=0.5$. The true $t_0$ is in the recovery window for both cases. Where $\sigma_{\rm sys} = 1.0$, the width of the recovery window for the 1$\sigma$ posterior is $0.50$\, days ($\ni t_{0}$), and where $\sigma_{\rm sys} = 0.5$, the width is $0.22$\, days ($\ni t_0$). Fig.~\ref{fig:bestfit_Ka17_george} (Appendix~\ref{sec:bestfitlc}) shows the generated $ugrizJHK$ light curves for the best fit ejecta parameters.}


\subsubsection{Inference with \textsc{rift} and \texttt{LANL} models}
We use the emulators presented in \cite{Peng2024}, which are based on the simulation data presented in \cite{Ristic-LC}, with the {\sc rift} parameter inference engine to infer the $t_0$ posterior for AT2017gfo. \emph{This emulator and model parameter grid (but not the underlying kilonova model) are entirely separate from the {\tt LANLTP2} model grid implemented in {\tt NMMA} and should not be assumed equivalent.} Table~\ref{tab:LANL_priors} (Appendix~\ref{sec:priors}) highlights our inference parameters and their associated domain bounds. Notably, our ejecta components have fixed composition (loosely described by $Y_{\rm e, d} = 0.05$ and $Y_{\rm e, w}
 = 0.27$) as opposed to the \textsc{possis} and \textsc{Sedona} models considered in this work. Thus, our inference is limited to the variation of ejecta parameters ($m_{\rm d}, v_{\rm d}, m_{\rm w}, v_{\rm w}$), viewing angle $\theta$, systematic uncertainty $\sigma_{\rm{sys}}$ (for the analysis in this section, we fix $\sigma_{\rm{sys}} = 1$, but generally leave it as a free parameter as in Appendix~\ref{appendix-AT2017gfo}), which we treat as a catch-all systematic term for light-curve fitting, and $t_0$. Our luminosity distance is fixed to $40$ Mpc, and we perform inference using the AdaptiveVolume sampler in \textsc{rift}.

Figure~\ref{fig:RIFT_LANL_AT2017gfo} (Appendix~\ref{sec:bestfitlc}) shows our best-fit light curves using the AT2017gfo dataset presented in the KN Open Catalog, which encompasses a richer dataset than considered in the previous two analyses. Since our \texttt{LANL} models are post-processed using the Vera C. Rubin observatory's \textit{grizy} filters, we do not match our models to observations from a specific telescope. Instead, we fit to the full dataset, and account for any systematic bandpass differences through our $\sigma_{\rm{sys}}$ parameter (here fixed to $\sigma_{\rm{sys}} = 1$). We find overall good agreement between our models and data, with the largest discrepancies occurring for all bands at the $0.5$ day observations, the \textit{g}-band data (1-3 days) and the \textit{J}-band observations at ${\sim}10$ days.

Fig.~\ref{fig:170817-all-results} shows the recovered $t_0$ posterior using the \texttt{LANL} models (light blue curve). We get a moderately sharp posterior located at $t_0 = -0.17$ days, with narrower tails extending out to slightly larger and smaller $t_0$ values compared to other analyses. We do recover the true GW170817 merger time, same as the \textsc{george} and \texttt{Ka2017} analysis and \textsc{nmma} using \texttt{Bu2026\_MLP}, and our $t_0$ recovery is less discrepant from the true merger time as compared to the \textsc{nmma} recovery using \texttt{Bu2019lm} in the analyses which do not consider additional artificial data during inference (Fig.~\ref{fig:2017-fake-detection}, Appendix~\ref{sec:bestfitlc}). 

One of the most striking features in Fig.~\ref{fig:170817-all-results} is the sharpness of the \texttt{LANL} posterior compared to the others; notably, this sharpness is present in both the fixed and free $\sigma_{\rm{sys}}$ analyses (Figs. \ref{fig:RIFT_LANL_AT2017gfo} and \ref{fig:LANL-freesigma-sys}). For the fixed $\sigma_{\rm{sys}}$ analysis, the true merger time does not land in the region of highest posterior density, and in our fixed $\sigma_{\rm{sys}}$ inference in Appendix~\ref{appendix-AT2017gfo}, the sharply-peaked posterior does not recover $t_0$ altogether. In both cases, our inference has a clear systematic offset which favors an earlier $t_0$. The \texttt{Bu2019lm} and \texttt{Bu2026\_MLP} posteriors reduce in width and increase in height by about a factor of two in Figure \ref{fig:Bulla-nmma-freesigmasys} when $\sigma_{\rm{sys}}$ is treated as a free parameter in those analyses. The results with free $\sigma{\rm{sys}}$ imply the possibility of a connection between the flexibility in $\sigma_{\rm{sys}}$ and the recovery of $t_0$ to some significance level. While both the \texttt{LANL} and \texttt{Bu2026\_MLP} analyses employ multi-layer perceptron emulators, the underlying kilonova models and interplay between emulator systematic uncertainty and inference methodologies introduce non-trivial complexities in posterior results. These sensitivities to systematic uncertainty have been shown before \citep{Heinzel21}, and the results of Figure~\ref{fig:170817-all-results} reinforce that a careful treatment of systematic uncertainty is required for any one of these analyses. Though in this work we present a comparison of $t_0$ recovery across inference frameworks for completeness, a fully detailed comparison of more generalized inference results using standardized simulation, emulation, and inference circumstances is necessary for a field-wide understanding of the implications of each of these different groups' inference pipelines.



\subsubsection{Inference with \textsc{Redback}}
We analyse the same extinction-corrected AT2017gfo photometry as the analyses above (113 detections and three limiting magnitudes in $ugrizyJHK$) with
\textsc{Redback}~\citep{Sarin2024}. We use the
\texttt{two\_component\_kilonova\_with\_cocoon} model implemented in \textsc{Redback} following~\cite{Sarin2024b} via \textsc{Redback} v1.18.0. We fix the luminosity
distance to 40.7~Mpc, corresponding to $z=0.00912$ in the Planck18
cosmology~\citep{Planck2018}. \textsc{Redback} intrinsically requires a redshift and not solely a luminosity distance to capture the effect of k-corrections, although this is minimal at the redshift of AT2017gfo. Our model combines two independently parametrised, r-process-powered ejecta components with
the shock-cooling cocoon prescription of~\citet{2018ApJ...855..103P}. Models such as these have been broadly successful in interpreting the observations of AT2017gfo~\citep[e.g.][]{Kasliwal2017,Arcavi2018}.

As with the analyses above, we ignore the merger time as known from the gravitational-wave observation and instead sample this with a uniform prior relative to the first observation, $t_0-t_{\rm first}\sim\mathcal{U}(-2,0)$. The remaining priors are
the default \textsc{Redback} priors for this model. Our likelihood includes the standard Gaussian likelihood for detections and non-detections as implemented in \textsc{Redback} on the Flux-space data transformed self-consistently from the input magnitude data. However, for a data point, $i$, the corresponding uncertainty, $\sigma_{i}$ is the intrinsic error on the data alongside a $\sigma_{\rm sys}=1$ magnitude uncertainty added in quadrature, consistent with the
systematic uncertainty adopted for the corresponding analyses above. We sample the posterior via a two stage approach. First, we estimate the max posterior via \textsc{scipy}~\citep{2020SciPy-NMeth} optimization followed by an ensemble of $64$ \textsc{emcee}~\citep{2013PASP..125..306F} walkers initialized around in a narrow Gaussian around the max posterior point. The ensemble is evolved for 5000 steps, of which the first 1500 are discarded as burn-in.

For our fiducial choice of $\sigma_{\rm sys}=1$ mag, we infer a merger-time offset relative to the first observation of
$t_0-t_{\rm first}=-0.260^{+0.124}_{-0.197}\ {\rm days}$, where the quoted uncertainties correspond to the $68\%$ credible interval. The central 95\% credible interval is $[-0.696,-0.068]$ days. The $t_0$ posterior distribution is shown in Fig.~\ref{fig:170817-all-results} (yellow curve). The independently measured GW170817 merger time corresponds to $t_0-t_{\rm first}=-0.4715$ days and lies within the $95\%$ credible interval. To explore the sensitivity of our result to the choice of systematic uncertainty, we repeat the full inference with $\sigma_{\rm sys}=0.5$ mag while keeping the data, physical priors, and sampler settings unchanged. This provides a constraint on $t_0-t_{\rm first}=-0.243^{+0.083}_{-0.113}\ {\rm days}$,
with a central 95\% credible interval of $[-0.488,-0.103]$ days. Reducing $\sigma_{\rm sys}$ narrows the central 68\% interval from 0.321 to 0.196 days and shifts the posterior median later by only 0.017 days. The known GW merger time remains inside the central 95\% interval but marginally less consistent due to the more precise posterior further away from the true value. Given the approximate nature of the semi-analytic model and underlying uncertainties in kilonova modelling, we adopt the more conservative 1-mag result as our primary constraint with the test with a smaller systematic error exploring the sensitivity to this choice.

For the primary analysis, the first ejecta component favours a comparatively low opacity, $\kappa_1=4.37^{+2.12}_{-1.33}\ {\rm cm^2\,g^{-1}}$, whereas the opacity of the second component is substantially less constrained, $\kappa_2=16.3^{+8.6}_{-9.6}\ {\rm cm^2\,g^{-1}}$. The corresponding velocity constraints are $v_1=0.128^{+0.033}_{-0.019}c$ and $v_2=0.231^{+0.109}_{-0.095}c$. The cocoon contribution favours a shocked ejecta fraction of $f_{\rm shock}=0.760^{+0.166}_{-0.178}$ and $\cos\theta_{\rm cocoon}=0.911^{+0.043}_{-0.032}$, corresponding to a characteristic cocoon angle of approximately $24^{\circ}$. The cocoon shock time is inferred to be $t_{\rm shock}=8.37^{+1.20}_{-2.06}$ s. These constraints should be interpreted as conditional on the adopted semi-analytic prescription and its default priors, rather than as model-independent measurements of the ejecta.

The parameter posteriors obtained with $\sigma_{\rm sys}=0.5$ mag broadly overlap those from the fiducial analysis, but are generally narrower. The most noticeable changes are towards a smaller first-component velocity and opacity, a lower first-component temperature floor, and smaller values of $f_{\rm shock}$ and $\eta$. Several posteriors are broad or show support near a prior boundary, particularly $v_1$ and the second-component temperature floor. This behaviour, together with correlations between $t_0$ and the parameters controlling the diffusion and cooling timescales, indicates that the merger-time precision remains sensitive to the adopted physical parametrization. In Fig.~\ref{fig:redback-lightcurve} (Appendix~\ref{sec:bestfitlc}), we show the AT2017gfo photometry together with the posterior-predictive light-curve $68\%$ credible interval. The $t_0$ posterior distribution is shown in Fig.~\ref{fig:170817-all-results}.The model captures the rapid fading of the optical emission and the slower evolution of the redder and near-infrared bands across the first 10 days. The width of the credible interval reflects both parameter uncertainty and the degeneracy in the inferred onset time.





\subsection{TEST 2b - GRBs}
\label{sec:GRB}
Over the last decades, a few kilonova candidates have been observed following a gamma-ray burst (GRB) detections (\emph{e.g.} GRB130603B \citealt{Tanvir2013, Berger2013}; GRB050709 \citealt{Jin2016}; GRB060614 \citealt{Yang2015}; see also \citealt{Gompertz2018, Ascenzi2019, Rastinejad_2021, Rastinejad_2025} for catalogs). Although they have not been unequivocally linked to a neutron star merger through a GW emission, they remain an interesting playground to test the inference of the merger time assuming that the GRB emission has occurred at most a few seconds after the merger. Here we restrict our study on kilonova candidates presented in \cite{Rastinejad_2025}. They are summarized in Table~\ref{tab:otherKNe} and additional details can be found about these events in \cite{Rastinejad_2025}.

\begin{table}[ht!]
\centering
\begin{tabular}{cccc}
\hline
\hline
\textbf{GRB} & \textbf{T$_0$ [mjd]} & \textbf{Redshift $z$} & \textbf{Analyzed [Y/N]} \\ \hline
\hline
050709 & 53560.942 & 0.161 & N \\
060614 & 53900.530 & 0.125 & Y \\
130603B & 56456.659 & 0.356 & N \\
160821B & 57621.937 & 0.162 & Y \\
200522A & 58991.487 & 0.554 & N \\
211211A & 59559.549 & 0.076 & Y\\
230307A & 60010.656 & 0.065 & Y \\
\hline
\end{tabular}
\caption{KN candidates associated with GRB emissions. AT2017gfo is not shown here as its analysis is discussed in Sec.~\ref{sec:170817}. The last column tells if the event has been included in the sample analyzed in this work (Y) or not (N).}
\label{tab:otherKNe}
\end{table}

Both an optical afterglow and a kilonova component have been observed for the whole sample. Therefore we use the afterglow subtracted lightcurves provided in \cite{Rastinejad_2025} to evaluate the constraints placed on the merger time using the kilonova candidate alone, under the assumption that delay between $t_0$ and the GRB prompt emission start time is of the order of seconds (but see \citep{2025A&A...703A.101G} for a discussion of scenarios in which the afterglow may precede the prompt emission). The kilonova-only lightcurve is obtained by fitting the X-ray and radio data and subtracting it from each observed flux to produce an "afterglow-subtracted" lightcurve. The afterglow model used describe a synchrotron emission from a forward shock, incorporating the effects of Inverse Compton cooling. In our analysis, we do not include GRB 130603B, GRB 200522A and GRB 050709 due to their very poor optical dataset (only one or two optical detections). We use {\sc nmma} framework and as previously, we analyze each lightcurve with \texttt{Bu2019lm} and \texttt{Bu2026\_MLP} and fixing the luminosity distance for each event thanks to their redshift estimate using \texttt{Planck18} cosmology \citep{Planck2018}. Although we would likely not have recovered GRB 060614's $t_0$ in a generic search, since the first detection occurred approximately 4.84 days after the prompt emission, we also extend the timeshift to earlier values to include the true start time and assess the recovery performance. 
As in previous sections, results are displayed as violin plots and summary tables.

\begin{table}
\centering
\begin{tabular}{c c c}
\hline
\hline
GRB & Model & Posterior size \\
& & containing $t_0$ [hours] \\
\hline
\hline
\multirow{2}{*}{GRB\,230307A}
  & \texttt{Bu2019lm} & 39.8 ($3\sigma$) \\
  & \texttt{Bu2026\_MLP} & 38.0 ($2\sigma$) \\
\hline
\multirow{2}{*}{GRB\,211211A}
  & \texttt{Bu2019lm} & 36.6 ($2\sigma$)  \\
  & \texttt{Bu2026\_MLP} & 8.7 ($1\sigma$)  \\
\hline
\multirow{2}{*}{GRB\,160821B}
  & \texttt{Bu2019lm} & 33.9 ($1\sigma$) \\
  & \texttt{Bu2026\_MLP} & 18.9 ($1\sigma$) \\
\hline
\multirow{2}{*}{GRB\,060614}
  & \texttt{Bu2019lm} & 126.0 ($1\sigma$) \\
  & \texttt{Bu2026\_MLP} & 133.2 ($2\sigma$) \\
\hline
\hline
\end{tabular}
\caption{Size of the merger time posterior credible intervals containing the true merger time $t_0$ for the four GRB.}
\label{tab:grb_sigma_1mag}
\end{table}

\begin{figure}
    \centering
    \includegraphics[width=0.9\linewidth]{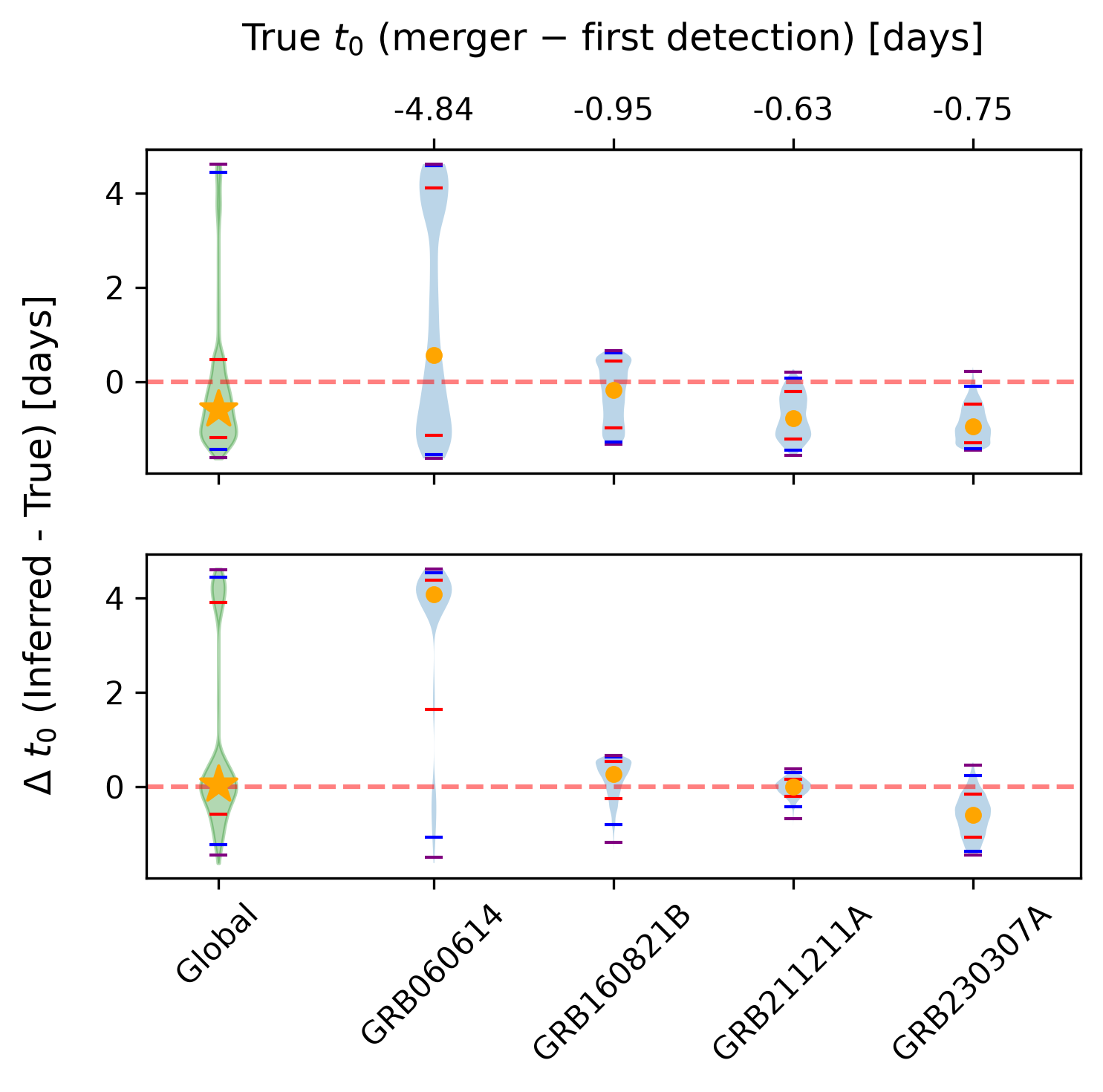}
    \caption{Violin plots of the shifted posterior distributions of the $t_0$ for the inference of the other kilonova candidates with \texttt{Bu2019lm} (top) and \texttt{Bu2026\_MLP} (bottom). The orange points indicate the median of the distribution. The 1, 2 and 3$\sigma$ credible intervals are shown by the red, blue and purple lines, respectively, for each violin. The red dotted line indicates the true value, and the green violin is the marginalized distribution.}
    \label{fig:GRB-violin}
\end{figure}

Results are provided in Fig.~\ref{fig:GRB-violin} and Table~\ref{tab:grb_sigma_1mag} for \texttt{Bu2019lm} (top) and \texttt{Bu2026\_MLP} (bottom). Although we analyze the GRBs twice, using a fixed $\sigma_{\rm sys}=1$ and leaving it as a free parameter in the prior, we only show results with a fixed systematic uncertainty consistently with the AT2017gfo results presented above. Overall, both models recover GRB $t_0$ consistently, although with different levels of precision. 
The \texttt{Bu2019lm} model recovers the $t_0$ within the 1$\sigma$ credible interval for GRB 160821B and within the 2 and 3$\sigma$ credible interval for GRBs 211211A and 230307A respectively. The true $t_0$ sits just outside the 3$\sigma$ edge (0.5h away) for GRB 230307A with \texttt{Bu2019lm}. In contrast, the \texttt{Bu2026\_MLP} model recovers $t_0$ within the 1$\sigma$ credible interval for GRBs 160821B and 211211A, and within the 2$\sigma$ credible interval for GRB 230307A. Overall the size posterior credible internal is smaller with \texttt{Bu2026\_MLP} than with \texttt{Bu2019lm}. Moreover, GRB 160821B's and GRB 211211A's $t_0$ 1$\sigma$ posterior distribution sizes are small, reaching 18.9 and 8.7 hours respectively with \texttt{Bu2026\_MLP}. The merger time associated to these three GRBs is quite well recovered as they all have detections near the kilonova brightness peak.
On the contrary, the very large posterior size for GRB 060614 and its bimodality with both \texttt{Bu2019lm} and \texttt{Bu2026\_MLP} is mainly due to the limited constraining power of the available dataset. In particular, the first detections occur only at $\gtrsim$ 4 days after $t_0$, while the early  observations are relatively shallow upper limits, leaving the overall lightcurve behaviour and therefore the merger time only weakly constrained. Finally, assuming an on-axis afterglow scenario, we reanalyzed the kilonova lightcurves to infer $t_0$. Results were consistent with those previously described treating the inclination as a free parameter. Still, it should be noted that the lightcurves used for these estimates are not representative of the expected orphan kilonova lightcurves as i) the GRB+KN events we considered occurred at much larger distances; (ii) depending on the brightness of the afterglow, the uncertainties on the lightcurve evolution are quite large, especially because the number of observations can be limited in this case.

\subsection{TEST 3 - Robustness to Early-Time Emission}
\label{sec:early-time}
While the $\beta$-decay of $r$-process material powers the light curves in the simulated kilonovae presented so far, there are other sources of energy that have been predicted that could enhance early-time ($t \leq 1$ day) emission such as free neutron decay (\citealt{Metzger2015,Brethauer26b}) jet-cocoon emission \citep{Gottlieb2020}, and central engines \citep{Sarin2022, Ai2025}. The additional energy source will alter the light curve at early times, and can impact estimates of key parameters. 

We assess the systematic uncertainties of early time emission from free neutrons on $t_0$ estimates by using {\sc nmma} to fit a model with $t_{0,true}$ = $8\times10^{-4}$ days, M = $3\times10^{-3}$ M$_\odot$, $v_c = 0.3c$, $v_{\rm max}$ = 0.7$c$, and $X_{\rm lan} = 0.1$ that contains $10^{-4}$ M$_\odot$ of free neutrons and a model without the free neutrons (for more details and light curves, see \citealt{Brethauer26b}). We fit the $g, r, i, z,$ and $y$ band light curve models placed at a distance of 40.7 Mpc using {\sc nmma} fitting routines generated from both the \texttt{Ka2017}, \texttt{Bu2019lm}, and \texttt{Bu2026\_MLP} to isolate systematic uncertainties intrinsic to the early time emission. We then run the same tests, but fix the distance to 40.7 Mpc to examine the scenario where a host galaxy is identified. Due to the construction of the \texttt{Bu2026\_MLP} model, for that fit only we restrict the data points to be $t \geq 0.2$ days. We emphasize that due to systematic uncertainties in cross-grid injection, the utilization of QNLTE in the free neutron models \citep{Brethauer2026} but not the {\sc nmma} fitting routines, and the utilization of all synthetic data from the simulations which translates to unrealistically fast observing cadences, the absolute fitting parameter values will be biased. Rather, the change in values between fitting the model with free neutrons and without are crucial to understanding the systematic uncertainties caused by early time emission.

\begingroup

\renewcommand{\arraystretch}{1.5} 
\begin{table}[h]
    \centering
    \begin{tabular}{c|cc|cc}
    
    \hline
    \hline
    M$_n$ (M$_\odot$) & 0 & 10$^{-4}$ & 0 & 10$^{-4}$\\
          & \multicolumn{2}{c}{Free D$_L$} & \multicolumn{2}{c}{Fixed D$_L$} \\
    \hline
    \hline
         &\multicolumn{4}{c}{\texttt{Ka2017}}   \\
        
    \hline 
    \hline 
       D$_L$ (Mpc) & 40.6$^{+18}_{-4.3}$ & 38.2$^{+12.6}_{-6.6}$ & -- & --\\
        $t_0$ (days) & -0.79$^{+0.14}_{-0.09}$& -0.71$^{+0.06}_{-0.16}$ & -0.84$^{+0.18}_{-0.06}$& -0.71$^{+0.06}_{-0.18}$
        \\
        log$_{10}$($\frac{M_{ej}}{M_\odot}$) & -2.9$^{+0.34}_{-0.07}$ & -2.77$^{+0.24}_{-0.16}$ & -2.92$^{+0.05}_{-0.05}$ & -2.72$^{+0.05}_{-0.04}$\\
        log$_{10}(\frac{v}{c})$ & -1.4 $^{+0.9}_{-0.1}$  & -0.56$^{+0.02}_{-0.54}$ & -1.47$^{+0.90}_{-0.03}$ & -0.56$^{+0.02}_{-0.51}$\\
        log$_{10}$(X$_{\rm lan}$) & -3$^{+0.36}_{-0.97}$ & -4.45$^{+0.3}_{-4.3}$ & -2.87$^{+0.2}_{-0.82}$ & -4.43$^{+0.23}_{-4.4}$\\

        \hline
        \hline
        & \multicolumn{4}{c}{\texttt{Bu2019lm}} \\
             \hline
             \hline
             D$_L$ (Mpc) & 152$^{+20}_{-18}$ & 138$^{+20}_{-16}$ & -- & --\\
             $t_0$ (days) & -1.30$^{+0.19}_{-0.30}$ & -1.31$^{+0.15}_{-0.3}$ & -1.99$^{+0.01}_{-0.01}$& -1.99$^{+0.02}_{-0.01}$\\
             log$_{10}$($\frac{M_{wind}}{M_\odot}$) & -1.93$^{+0.14}_{-0.11}$ & -1.92$^{+0.2}_{-0.1}$ & -2.04$^{+0.03}_{-0.03}$ & -2.04$^{+0.03}_{-0.03}$\\
             log$_{10}$($\frac{M_{dyn}}{M_\odot}$) & -2.83$^{+0.41}_{-0.13}$ & -2.53$^{+0.5}_{-0.4}$ & -1.81$^{+0.04}_{-0.04}$& -1.83$^{+0.04}_{-0.05}$\\
             $\phi$ (degrees) & 20.4$^{+7}_{-4}$ & 17.7$^{+5}_{-2}$ & 30.6$^{+1.9}_{-1.8}$& 25.6$^{+1.6}_{-1.5}$\\
             $\iota$ (radians) & 1.07$^{+0.28}_{-0.37}$& 0.96$^{+0.31}_{-0.32}$ & 1.55$^{+0.02}_{-0.03}$ & 1.54$^{+0.02}_{-0.03}$\\

        \hline
        \hline
        & \multicolumn{4}{c}{\texttt{Bu2026\_MLP}} \\
  
             \hline
             \hline
             D$_L$ (Mpc) & 44$^{+30}_{-19}$ & 28$^{+22}_{-11}$ & -- & --\\
             $t_0$ (days) & -0.73$^{+0.32}_{-0.40}$ & -0.79$^{+0.40}_{-0.49}$ & -0.72$^{+0.33}_{-0.38}$ & -0.84$^{+0.39}_{-0.46}$\\
             log$_{10}$($\frac{M_{wind}}{M_\odot}$) & -2.99$^{+0.53}_{-0.60}$ & -3.07$^{+0.54}_{-0.58}$ & -2.91$^{+0.38}_{-0.67}$ & -2.79$^{+0.42}_{-0.70}$\\
             v$_{wind}$ & 0.1$^{+0.03}_{-0.03}$ & 0.1$^{+0.03}_{-0.03}$ & 0.1$^{+0.03}_{-0.03}$ & 0.11$^{+0.03}_{-0.03}$\\
             Y$_{e, wind}$ & 0.3$^{+0.07}_{-0.06}$ & 0.3$^{+0.07}_{-0.06}$ & 0.3$^{+0.07}_{-0.06}$ & 0.31$^{+0.06}_{-0.07}$\\
             log$_{10}$($\frac{M_{dyn}}{M_\odot}$) & -2.60$^{+0.50}_{-0.72}$ & -2.78$^{+0.46}_{-0.64}$ & -2.68$^{+0.38}_{-0.79}$ & -2.70$^{+0.44}_{-0.87}$\\
             v$_{dyn}$ & 0.23$^{+0.08}_{-0.08}$ & 0.24$^{+0.07}_{-0.08}$ & 0.22$^{+0.09}_{-0.07}$ & 0.25$^{+0.08}_{-0.08}$\\
             Y$_{e, dyn}$ & 0.28$^{+0.05}_{-0.07}$  & 0.26$^{+0.06}_{-0.07}$ & 0.27$^{+0.05}_{-0.06}$ & 0.27$^{+0.05}_{-0.06}$\\
             $\iota$ (radians) & 1.0$^{+0.38}_{-0.48}$ & 1.0$^{+0.4}_{-0.5}$ & 1.01$^{+0.37}_{-0.48}$ & 0.98$^{+0.45}_{-0.51}$\\
             
        \end{tabular}

    \caption{Fitting results for a fiducial model with and without 10$^{-4}$ M$_\odot$ of free neutrons in the ejecta placed at 40.7 Mpc and $t_{0, true}$ = $8\times10^{-4}$ days using \texttt{Ka2017}, \texttt{Bu2019lm}, and \texttt{Bu2026\_MLP} models with the distance parameter free (left) and fixed (right).}
    \label{tab:early}
\end{table}
\endgroup

Table \ref{tab:early} shows the results of this exercise, where for the \texttt{Ka2017} models the fit is pushed towards higher velocities and lower X$_{\rm lan}$ to compensate for the additional blue emission from the free neutrons. The large uncertainty in the velocity and X$_{\rm lan}$ is a reflection of a degeneracy between the two parameters with the posteriors forming a bimodal distribution that favors a $v\sim0.1c$ and X$_{\rm lan} \sim 10^{-9}$ model at a subdominant peak and a $v\sim0.3c$ and X$_{\rm lan} \sim 10^{-4}$ model for the neutron-rich case. Similarly, the neutron-free model shows a bimodal posterior also at $v\sim0.3c$ and X$_{\rm lan} \sim 10^{-4}$ but the second peak has shifted to $v\sim0.1c$ and X$_{\rm lan} \sim 10^{-2.5}$. Despite the significant shift in the estimates of the ejecta properties, $t_0$ is still consistent within 1$\sigma$ suggesting that the uncertainty is shifted into the ejecta properties instead of the $t_0$. Notably, the distance that {\sc nmma} predicts for the \texttt{Ka2017} models when the parameter is left free is consistent with the distance that the light curves were placed.

For the \texttt{Bu2019lm} models when the distance is allowed to vary, the only change to estimated ejecta properties is the dynamical ejecta mass from $\sim 1.6\times10^{-3}$ M$_\odot$ to $\sim 3\times10^{-3}$ M$_\odot$, though each estimate is approximately 1$\sigma$ away from each other. This may be a result that the velocity of the ejecta is not a free parameter in the \texttt{Bu2019lm} grid, and so to achieve the faster decay that comes from the presence of free neutrons the model grid favored minimal values of $\phi$ to reduce the amount of high-opacity material the emission would have to travel through. Similarly, the wind ejecta mass may not be impacted by the inclusion of free neutrons due to the fixed, low velocity ($0.02c\leq v \leq 0.1c$) that prevents more rapid decay. Interestingly, the neutron-free and neutron-rich models fit with the \texttt{Bu2019lm} grid place the kilonova at $\sim150$ Mpc, which is nearly a factor of 4 farther than the distance the light curve was injected at and points to brighter emission predicted by the \texttt{Bu2019lm} models. In contrast, when the distance is fixed the inclusion of free neutrons appears to decrease the lanthanide-rich opening angle, $\phi$. This is likely a result of smaller $\phi$ leading to less high-opacity ejecta and therefore bluer emission that fits better to the model with free neutrons. Additionally, the fit is much more constrained overall when the distance is fixed with {\sc nmma} strongly preferring a more massive dynamical component and a less massive wind component. 

The last grid of models, \texttt{Bu2026\_MPL}, did not show any significant change to the fitting parameters between the neutron-rich and neutron-free models, in both cases of when the distance was allowed to vary and when the distance was fixed. This may be a result of the restriction of data to $t > 0.2$ days, where the most prominent feature from free neutrons is the excess absorption seen at optical wavelengths (see \citealt{Brethauer26b} for more details). 

However, while the $t_0$ parameter was not greatly affected in any of the model grids, this may be a result of the model grids not having sufficient temporal resolution at early times for the \texttt{Ka2017} and \texttt{Bu2019lm} models. The \texttt{Ka2017} and \texttt{Bu2019lm} grids have a spectral temporal resolution of 0.1 days and 0.2 days, respectively. Additionally, for the \texttt{Bu2019lm} models, when the distance was fixed the fit preferred $t_0$ values at the edge of the prior farthest from the true $t_0$ which could have prevented variance in the fitting estimates. While the \texttt{Bu2026\_MLP} grid has logarithmic temporal resolution, similarly to the light curves being fit, restrictions from {\sc nmma} prevented the use of data at $t \leq 0.2$ days, which could also have influenced the variability of $t_0$ between fits.



\section{Discussion and Conclusions}
\label{sec:discussion}
In this work, we explored the possibility of inferring the BNS merger time directly from kilonova lightcurves. We also assess which factors derived from observation and modeling have the greatest influence on the recovery of $t_0$. Our results shows that the ability to constrain the merger time strongly depends on the temporal coverage, wavelength sampling, and the choice of kilonova model. A clear outcome of this study is the importance of early observations, which carry most of the timing information, while kilonova light curves first observed after peak rapidly lose constraining power on the explosion time. According to studies such as~\citep{stevenson_strategy_2026}, most kilonovae will be detected 1 to 2 days after the merger, under typical LSST observing scenarios. In this context, upper limits obtained before the first detection are also highly valuable, as they effectively bound the merger epoch and reduce the allowed search parameter space.
In the most favorable case, based on the exceptionally well-sampled GW170817 dataset supplemented with a single synthetic observation during the rising phase, the merger time can be recovered with an uncertainty as low as 0.03 days (0.72 hours). 

Under optimistic observing conditions, with a cadence of four observations per day in the (grizy) filters, we recover the true merger time in most of cases and the median $1\sigma$ uncertainty over our synthetic kilonova population increases to approximately 0.15 days (90\% CI: 0.3 days). Similarly, observations obtained with the baseline LSST observing strategy alone yield a best-case $1\sigma$ uncertainty of about 0.4 days (90\% CI: 0.7 days) and when adding upper limits, 0.2 days (90\% CI: 0.3 days). However, for many kilonova synthetic lightcurves, sparse sampling or unfavorable observing conditions can lead to uncertainties of several days, reaching up to about one week in the worst cases.

The inferred merger time also depends on the adopted kilonova model. Different models encode different assumptions about the ejecta properties, opacities, and radiative transfer, which can lead to systematic differences in the recovered merger time. Rather than assuming a single model a priori, one could evaluate the statistical evidence for competing models through log-likelihood model comparison, to select the model describing the dataset the best before interpreting the merger-time posterior. This approach, not presented here, has been tested for GRB 211211A dataset. A log-likelihood of $ln(\mathcal{Z})=-32.9$ for \texttt{Bu2019lm} and $ln(\mathcal{Z})=-28.6$ for \texttt{Bu2026\_MLP} have been found, indicating a preference for the \texttt{Bu2026\_MLP} model, consistent with its improved recovery of the merger time. 
Extending this type of analysis to future events would provide a robust framework for both model selection and uncertainty quantification. From an observational perspective, these results provide practical guidance for future follow-up strategies. Rapid observations are essential, but equally important is obtaining multi-band photometry. At least two filters should be observed whenever possible to capture the kilonova color evolution. At the interface between the optimistic and realistic LSST observing configuration, we can notice a third scenario, the most realistic one, in which wide field facilities such as LSST provide the initial discovery while dedicated follow-up programs enrich the light curve with rapid, multi-band observations. Such an intermediate configuration combines the discovery power of large surveys with the detailed temporal and spectral information required for accurate merger-time reconstruction. This merger time reconstruction from orphan KNe seen by LSST can be used to inform GW analyses in order to narrow the search temporal window. Other tools such as \cite{rochiniTDR} can use the temporal window derived in this work to evaluate the GW detection probability informed by the KN sky position and properties. Finally, similar studies could be performed for neutron-star black hole mergers, as the latter can also produce a kilonova emission under some favourable conditions in terms of masses and spins.



\begin{acknowledgements}
LD was supported by Nicholas and Lee Begovich
and the Dan Black Family, and the National Science Foundation
Awards 2308985 and PHY-2608017. DB is partially supported by a NASA Future Investigators in
NASA Earth and Space Science and Technology (FINESST) award
No. 80NSSC23K1440 and the Hearts to Humanity Eternal Research
Grant through UC Berkeley. MB acknowledges the Department of Physics and Earth Science of the University of Ferrara for the financial support through the FIRD 2025 grant. HR acknowledges funding from the EU Horizon under ERC Starting Grant, no. SMArt-101076369. Views and opinions expressed are those of the authors only and do not necessarily reflect those of the European Union or the European Research Council. Neither the European Union nor the granting authority can be held responsible for them. NS is supported by the Kavli Foundation. TW is supported by the research program of the Netherlands Organization for Scientific Research (NWO) under grant number OCENW.XL21.XL21.038. KSP acknowledges support from STFC grant ST/V005677/1. SR, TB, and SP acknowledge support from the Astrophysics Center for Multi-messenger Studies in Europe (ACME), funded under the European Union’s Horizon Europe Research and Innovation Program, Grant Agreement No. 101131928. AR acknowledges support from 2025 STEP 00204, PID2024-159689NB-C22, RYC2022-035983-I, and CEX2024-001451-M. TD acknowledges funding from the EU Horizon under ERC Starting Grant, no. SMArt-101076369 and through from the German Research Foundation (project number 
543044491). Views and opinions expressed are those of the authors only and do not necessarily reflect those of the European Union or the European Research Council. Neither the European Union nor the granting authority can be held responsible for them. ROS acknowledges support from NSF AST 2206321. High-end computing (HEC) resources supporting this work were provided by the NASA Advanced Supercomputing (NAS) Division
at Ames Research Center. 
\end{acknowledgements}

%
%
\bibliographystyle{aa}
\bibliography{bibliography}

@preamble{ " \newcommand{\noop}[1]{} " }

@ARTICLE{2012MNRAS.426.1940K,
       author = {{Korobkin}, O. and {Rosswog}, S. and {Arcones}, A. and {Winteler}, C.},
        title = "{On the astrophysical robustness of the neutron star merger r-process}",
      journal = {\mnras},
         year = 2012,
        month = nov,
       volume = {426},
       number = {3},
        pages = {1940-1949},
          doi = {10.1111/j.1365-2966.2012.21859.x},
archivePrefix = {arXiv},
       eprint = {1206.2379},
 primaryClass = {astro-ph.SR},
       adsurl = {https://ui.adsabs.harvard.edu/abs/2012MNRAS.426.1940K}
}

@ARTICLE{Arnett1980,
       author = {{Arnett}, W.~D.},
        title = "{Analytic solutions for light curves of supernovae of Type II}",
      journal = {\apj},
         year = 1980,
        month = apr,
       volume = {237},
        pages = {541-549},
          doi = {10.1086/157898},
       adsurl = {https://ui.adsabs.harvard.edu/abs/1980ApJ...237..541A}
}

@ARTICLE{Arnett1982,
       author = {{Arnett}, W.~D.},
        title = "{Type I supernovae. I - Analytic solutions for the early part of the light curve}",
      journal = {\apj},
         year = 1982,
        month = feb,
       volume = {253},
        pages = {785-797},
          doi = {10.1086/159681},
       adsurl = {https://ui.adsabs.harvard.edu/abs/1982ApJ...253..785A}
}

@ARTICLE{Metzger2017,
       author = {{Metzger}, Brian D.},
        title = "{Kilonovae}",
      journal = {Living Reviews in Relativity},
         year = 2017,
        month = dec,
       volume = {20},
       number = {1},
          eid = {3},
        pages = {3},
          doi = {10.1007/s41114-017-0006-z},
archivePrefix = {arXiv},
       eprint = {1610.09381},
 primaryClass = {astro-ph.HE},
       adsurl = {https://ui.adsabs.harvard.edu/abs/2017LRR....20....3M}
}

@article{Kasen2017,
author={Kasen, Daniel and Metzger, Brian
and Barnes, Jennifer and Quataert, Eliot and Ramirez-Ruiz, Enrico},
title={Origin of the heavy elements in binary neutron-star mergers from a gravitational-wave event},
journal={Nature},
year={2017},
month={Nov},
day={01},
volume={551},
number={7678},
pages={80-84},
issn={1476-4687},
doi={10.1038/nature24453},
url={https://doi.org/10.1038/nature24453}
}

@ARTICLE{Peng2024,
       author = {{Peng}, Yinglei and {Risti{\'c}}, Marko and {Kedia}, Atul and {O'Shaughnessy}, Richard and {Fontes}, Christopher J. and {Fryer}, Chris L. and {Korobkin}, Oleg and {Mumpower}, Matthew R. and {Villar}, V. Ashley and {Wollaeger}, Ryan T.},
        title = "{Kilonova light-curve interpolation with neural networks}",
      journal = {Physical Review Research},
         year = 2024,
        month = jul,
       volume = {6},
       number = {3},
          eid = {033078},
        pages = {033078},
          doi = {10.1103/PhysRevResearch.6.033078},
archivePrefix = {arXiv},
       eprint = {2402.05871},
 primaryClass = {astro-ph.HE},
       adsurl = {https://ui.adsabs.harvard.edu/abs/2024PhRvR...6c3078P}
}

@ARTICLE{PintoEastman2000,
       author = {{Pinto}, Philip A. and {Eastman}, Ronald G.},
        title = "{The Physics of Type IA Supernova Light Curves. I. Analytic Results and Time Dependence}",
      journal = {\apj},
         year = 2000,
        month = feb,
       volume = {530},
       number = {2},
        pages = {744-756},
          doi = {10.1086/308376},
       adsurl = {https://ui.adsabs.harvard.edu/abs/2000ApJ...530..744P}
}

@ARTICLE{Ricigliano2024,
       author = {{Ricigliano}, Giacomo and {Perego}, Albino and {Borhanian}, Ssohrab and {Loffredo}, Eleonora and {Kawaguchi}, Kyohei and {Bernuzzi}, Sebastiano and {Lippold}, Lukas Chris},
        title = "{xkn: a semi-analytic framework for the modelling of kilonovae}",
      journal = {\mnras},
         year = 2024,
        month = mar,
       volume = {529},
       number = {1},
        pages = {647-663},
          doi = {10.1093/mnras/stae572},
archivePrefix = {arXiv},
       eprint = {2311.15709},
 primaryClass = {astro-ph.HE},
       adsurl = {https://ui.adsabs.harvard.edu/abs/2024MNRAS.529..647R}
}

@ARTICLE{Saha2024,
       author = {{Saha}, Surojit and {Williams}, Michael J. and {Datrier}, Laurence and {Hayes}, Fergus and {Nicholl}, Matt and {Kong}, Albert K.~H. and {Hendry}, Martin and {Heng}, IK Siong and {Lamb}, Gavin P. and {Lin}, En-Tzu and {Williams}, Daniel},
        title = "{Rapid Generation of Kilonova Light Curves Using Conditional Variational Autoencoder}",
      journal = {\apj},
         year = 2024,
        month = feb,
       volume = {961},
       number = {2},
          eid = {165},
        pages = {165},
          doi = {10.3847/1538-4357/ad02f4},
archivePrefix = {arXiv},
       eprint = {2310.17450},
 primaryClass = {astro-ph.HE},
       adsurl = {https://ui.adsabs.harvard.edu/abs/2024ApJ...961..165S},
}

@ARTICLE{macbride2026too,
       author = {{MacBride}, Sean Patrick and {Jones}, R. Lynne and {Yoachim}, Peter and {Ribeiro}, Tiago and {Guy}, Leanne P. and {Anand}, Shreya and {Howard}, Erin Leigh and {Sullivan}, Ian S. and {Wang}, Daniel L. and {Bellm}, Eric C. and {Armstrong}, Robert and {Wood-Vasey}, W.~M. and {Drlica-Wagner}, Alex and {Herner}, Kenneth and {Narayan}, Gautham and {Acero-Cuellar}, Tatiana and {Bettina Bianco}, Federica and {Andreoni}, Igor and {S{\'a}nchez}, Bruno O. and {Banovetz}, John and {Alexov}, Anastasia and {Dennihy}, Erik and {Blum}, Robert D. and {Utsumi}, Yousuke and {Soares-Santos}, Marcelle and {Lupton}, Robert H. and {Bechtol}, Keith and {Ivezi{\'c}}, {\v{Z}}eljko and {Mortensen}, Kris and {Venegas}, Paulina and {Kang}, Yijung and {Shugart}, Alysha B. and {Khadka}, Narayan and {Kelkar}, Kshitija and {{\v{Z}}ilkov{\'a}}, Danica and {Fanning}, Kevin and {Walter}, Christopher W. and {Weaver}, Chris and {Araneda-Baltierra}, Fabian and {Gomboc}, Andreja and {Tyson}, J. Anthony and {Chandler}, Colin Orion and {Bernardinelli}, Pedro H. and {Singh}, Devanshi},
        title = "{The Rubin Observatory Target-of-Opportunity System in the First Year of Operations}",
      journal = {arXiv e-prints},
         year = 2026,
        month = jun,
          eid = {arXiv:2607.00217},
        pages = {arXiv:2607.00217},
          doi = {10.48550/arXiv.2607.00217},
archivePrefix = {arXiv},
       eprint = {2607.00217},
 primaryClass = {astro-ph.IM},
       adsurl = {https://ui.adsabs.harvard.edu/abs/2026arXiv260700217M}
}

@ARTICLE{2015CQGra..32g4001L,
       author = {{LIGO Scientific Collaboration} and {Aasi}, J. and {Abbott}, B.~P. and {Abbott}, R. and {Abbott}, T. and {Abernathy}, M.~R. and {Ackley}, K. and {Adams}, C. and {Adams}, T. and {Addesso}, P. and {Adhikari}, R.~X. and {Adya}, V. and {Affeldt}, C. and {Aggarwal}, N. and {Aguiar}, O.~D. and {Ain}, A. and {Ajith}, P. and {Alemic}, A. and {Allen}, B. and {Amariutei}, D. and {Anderson}, S.~B. and {Anderson}, W.~G. and {Arai}, K. and {Araya}, M.~C. and {Arceneaux}, C. and {Areeda}, J.~S. and {Ashton}, G. and {Ast}, S. and {Aston}, S.~M. and {Aufmuth}, P. and {Aulbert}, C. and {Aylott}, B.~E. and {Babak}, S. and {Baker}, P.~T. and {Ballmer}, S.~W. and {Barayoga}, J.~C. and {Barbet}, M. and {Barclay}, S. and {Barish}, B.~C. and {Barker}, D. and {Barr}, B. and {Barsotti}, L. and {Bartlett}, J. and {Barton}, M.~A. and {Bartos}, I. and {Bassiri}, R. and {Batch}, J.~C. and {Baune}, C. and {Behnke}, B. and {Bell}, A.~S. and {Bell}, C. and {Benacquista}, M. and {Bergman}, J. and {Bergmann}, G. and {Berry}, C.~P.~L. and {Betzwieser}, J. and {Bhagwat}, S. and {Bhandare}, R. and {Bilenko}, I.~A. and {Billingsley}, G. and {Birch}, J. and {Biscans}, S. and {Biwer}, C. and {Blackburn}, J.~K. and {Blackburn}, L. and {Blair}, C.~D. and {Blair}, D. and {Bock}, O. and {Bodiya}, T.~P. and {Bojtos}, P. and {Bond}, C. and {Bork}, R. and {Born}, M. and {Bose}, Sukanta and {Brady}, P.~R. and {Braginsky}, V.~B. and {Brau}, J.~E. and {Bridges}, D.~O. and {Brinkmann}, M. and {Brooks}, A.~F. and {Brown}, D.~A. and {Brown}, D.~D. and {Brown}, N.~M. and {Buchman}, S. and {Buikema}, A. and {Buonanno}, A. and {Cadonati}, L. and {Calder{\'o}n Bustillo}, J. and {Camp}, J.~B. and {Cannon}, K.~C. and {Cao}, J. and {Capano}, C.~D. and {Caride}, S. and {Caudill}, S. and {Cavagli{\`a}}, M. and {Cepeda}, C. and {Chakraborty}, R. and {Chalermsongsak}, T. and {Chamberlin}, S.~J. and {Chao}, S. and {Charlton}, P. and {Chen}, Y. and {Cho}, H.~S. and {Cho}, M. and {Chow}, J.~H. and {Christensen}, N. and {Chu}, Q. and {Chung}, S. and {Ciani}, G. and {Clara}, F. and {Clark}, J.~A. and {Collette}, C. and {Cominsky}, L. and {Constancio}, Jr., M. and {Cook}, D. and {Corbitt}, T.~R. and {Cornish}, N. and {Corsi}, A. and {Costa}, C.~A. and {Coughlin}, M.~W. and {Countryman}, S. and {Couvares}, P. and {Coward}, D.~M. and {Cowart}, M.~J. and {Coyne}, D.~C. and {Coyne}, R. and {Craig}, K. and {Creighton}, J.~D.~E. and {Creighton}, T.~D. and {Cripe}, J. and {Crowder}, S.~G. and {Cumming}, A. and {Cunningham}, L. and {Cutler}, C. and {Dahl}, K. and {Dal Canton}, T. and {Damjanic}, M. and {Danilishin}, S.~L. and {Danzmann}, K. and {Dartez}, L. and {Dave}, I. and {Daveloza}, H. and {Davies}, G.~S. and {Daw}, E.~J. and {DeBra}, D. and {Del Pozzo}, W. and {Denker}, T. and {Dent}, T. and {Dergachev}, V. and {DeRosa}, R.~T. and {DeSalvo}, R. and {Dhurandhar}, S. and {D́{\i}az}, M. and {Di Palma}, I. and {Dojcinoski}, G. and {Dominguez}, E. and {Donovan}, F. and {Dooley}, K.~L. and {Doravari}, S. and {Douglas}, R. and {Downes}, T.~P. and {Driggers}, J.~C. and {Du}, Z. and {Dwyer}, S. and {Eberle}, T. and {Edo}, T. and {Edwards}, M. and {Edwards}, M. and {Effler}, A. and {Eggenstein}, H.-B. and {Ehrens}, P. and {Eichholz}, J. and {Eikenberry}, S.~S. and {Essick}, R. and {Etzel}, T. and {Evans}, M. and {Evans}, T. and {Factourovich}, M. and {Fairhurst}, S. and {Fan}, X. and {Fang}, Q. and {Farr}, B. and {Farr}, W.~M. and {Favata}, M. and {Fays}, M. and {Fehrmann}, H. and {Fejer}, M.~M. and {Feldbaum}, D. and {Ferreira}, E.~C. and {Fisher}, R.~P. and {Frei}, Z. and {Freise}, A. and {Frey}, R. and {Fricke}, T.~T. and {Fritschel}, P. and {Frolov}, V.~V. and {Fuentes-Tapia}, S. and {Fulda}, P. and {Fyffe}, M. and {Gair}, J.~R.},
        title = "{Advanced LIGO}",
      journal = {Classical and Quantum Gravity},
         year = 2015,
        month = apr,
       volume = {32},
       number = {7},
          eid = {074001},
        pages = {074001},
          doi = {10.1088/0264-9381/32/7/074001},
archivePrefix = {arXiv},
       eprint = {1411.4547},
 primaryClass = {gr-qc},
       adsurl = {https://ui.adsabs.harvard.edu/abs/2015CQGra..32g4001L}
}

@ARTICLE{2025arXiv250818082T,
       author = {{The LIGO Scientific Collaboration} and {the Virgo Collaboration} and {the KAGRA Collaboration} and {Abac}, A.~G. and {Abouelfettouh}, I. and {Acernese}, F. and {Ackley}, K. and {Adamcewicz}, C. and {Adhicary}, S. and {Adhikari}, D. and {Adhikari}, N. and {Adhikari}, R.~X. and {Adkins}, V.~K. and {Afroz}, S. and {Agapito}, A. and {Agarwal}, D. and {Agathos}, M. and {Aggarwal}, N. and {Aggarwal}, S. and {Aguiar}, O.~D. and {Ahrend}, I.-L. and {Aiello}, L. and {Ain}, A. and {Ajith}, P. and {Akutsu}, T. and {Albanesi}, S. and {Ali}, W. and {Al-Kershi}, S. and {All{\'e}n{\'e}}, C. and {Allocca}, A. and {Al-Shammari}, S. and {Altin}, P.~A. and {Alvarez-Lopez}, S. and {Amar}, W. and {Amarasinghe}, O. and {Amato}, A. and {Amicucci}, F. and {Amra}, C. and {Ananyeva}, A. and {Anderson}, S.~B. and {Anderson}, W.~G. and {Andia}, M. and {Ando}, M. and {Andr{\'e}s-Carcasona}, M. and {Andri{\'c}}, T. and {Anglin}, J. and {Ansoldi}, S. and {Antelis}, J.~M. and {Antier}, S. and {Aoumi}, M. and {Appavuravther}, E.~Z. and {Appert}, S. and {Apple}, S.~K. and {Arai}, K. and {Araya}, A. and {Araya}, M.~C. and {Arca Sedda}, M. and {Areeda}, J.~S. and {Aritomi}, N. and {Armato}, F. and {Armstrong}, S. and {Arnaud}, N. and {Arogeti}, M. and {Aronson}, S.~M. and {Arun}, K.~G. and {Ashton}, G. and {Aso}, Y. and {Asprea}, L. and {Assiduo}, M. and {Assis de Souza Melo}, S. and {Aston}, S.~M. and {Astone}, P. and {Attadio}, F. and {Aubin}, F. and {AultONeal}, K. and {Avallone}, G. and {Avila}, E.~A. and {Babak}, S. and {Badger}, C. and {Bae}, S. and {Bagnasco}, S. and {Baiotti}, L. and {Bajpai}, R. and {Baka}, T. and {Baker}, A.~M. and {Baker}, K.~A. and {Baker}, T. and {Baldi}, G. and {Baldicchi}, N. and {Ball}, M. and {Ballardin}, G. and {Ballmer}, S.~W. and {Banagiri}, S. and {Banerjee}, B. and {Bankar}, D. and {Baptiste}, T.~M. and {Baral}, P. and {Baratti}, M. and {Barayoga}, J.~C. and {Barish}, B.~C. and {Barker}, D. and {Barman}, N. and {Barneo}, P. and {Barone}, F. and {Barr}, B. and {Barsotti}, L. and {Barsuglia}, M. and {Barta}, D. and {Bartoletti}, A.~M. and {Barton}, M.~A. and {Bartos}, I. and {Basalaev}, A. and {Bassiri}, R. and {Basti}, A. and {Bawaj}, M. and {Baxi}, P. and {Bayley}, J.~C. and {Baylor}, A.~C. and {Baynard}, II, P.~A. and {Bazzan}, M. and {Bedakihale}, V.~M. and {Beirnaert}, F. and {Bejger}, M. and {Belardinelli}, D. and {Bell}, A.~S. and {Bellie}, D.~S. and {Bellizzi}, L. and {Benoit}, W. and {Bentara}, I. and {Bentley}, J.~D. and {Ben Yaala}, M. and {Bera}, S. and {Bergamin}, F. and {Berger}, B.~K. and {Bernuzzi}, S. and {Beroiz}, M. and {Berry}, C.~P.~L. and {Bersanetti}, D. and {Bertheas}, T. and {Bertolini}, A. and {Betzwieser}, J. and {Beveridge}, D. and {Bevilacqua}, G. and {Bevins}, N. and {Bhandare}, R. and {Bhatt}, R. and {Bhattacharjee}, D. and {Bhattacharyya}, S. and {Bhaumik}, S. and {Biancalana}, V. and {Bianchi}, A. and {Bilenko}, I.~A. and {Billingsley}, G. and {Binetti}, A. and {Bini}, S. and {Binu}, C. and {Biot}, S. and {Birnholtz}, O. and {Biscoveanu}, S. and {Bisht}, A. and {Bitossi}, M. and {Bizouard}, M.-A. and {Blaber}, S. and {Blackburn}, J.~K. and {Blagg}, L.~A. and {Blair}, C.~D. and {Blair}, D.~G. and {Bode}, N. and {Boettner}, N. and {Boileau}, G. and {Boldrini}, M. and {Bolingbroke}, G.~N. and {Bolliand}, A. and {Bonavena}, L.~D. and {Bondarescu}, R. and {Bondu}, F. and {Bonilla}, E. and {Bonilla}, M.~S. and {Bonino}, A. and {Bonnand}, R. and {Borchers}, A. and {Borhanian}, S. and {Boschi}, V. and {Bose}, S. and {Bossilkov}, V. and {Bothra}, Y. and {Boudon}, A. and {Bourg}, L. and {Boyle}, M. and {Bozzi}, A. and {Bradaschia}, C. and {Brady}, P.~R. and {Branch}, A. and {Branchesi}, M. and {Braun}, I. and {Briant}, T. and {Brillet}, A. and {Brinkmann}, M. and {Brockill}, P. and {Brockmueller}, E.},
        title = "{GWTC-4.0: Updating the Gravitational-Wave Transient Catalog with Observations from the First Part of the Fourth LIGO-Virgo-KAGRA Observing Run}",
      journal = {arXiv e-prints},
         year = 2025,
        month = aug,
          eid = {arXiv:2508.18082},
        pages = {arXiv:2508.18082},
          doi = {10.48550/arXiv.2508.18082},
archivePrefix = {arXiv},
       eprint = {2508.18082},
 primaryClass = {gr-qc},
       adsurl = {https://ui.adsabs.harvard.edu/abs/2025arXiv250818082T}
}

@ARTICLE{2026arXiv260527225T,
       author = {{The LIGO Scientific Collaboration} and {the Virgo Collaboration} and {the KAGRA Collaboration}},
        title = "{GWTC-5.0: Observations from the Second Part of the Fourth LIGO-Virgo-KAGRA Observing Run and Updates to the Gravitational-Wave Transient Catalog}",
      journal = {arXiv e-prints},
         year = 2026,
        month = may,
          eid = {arXiv:2605.27225},
        pages = {arXiv:2605.27225},
          doi = {10.48550/arXiv.2605.27225},
archivePrefix = {arXiv},
       eprint = {2605.27225},
 primaryClass = {gr-qc},
       adsurl = {https://ui.adsabs.harvard.edu/abs/2026arXiv260527225T}
}

@ARTICLE{2017PhRvL.119p1101A,
       author = {{Abbott}, B.~P. and {Abbott}, R. and {Abbott}, T.~D. and {Acernese}, F. and {Ackley}, K. and {Adams}, C. and {Adams}, T. and {Addesso}, P. and {Adhikari}, R.~X. and {Adya}, V.~B. and {Affeldt}, C. and {Afrough}, M. and {Agarwal}, B. and {Agathos}, M. and {Agatsuma}, K. and {Aggarwal}, N. and {Aguiar}, O.~D. and {Aiello}, L. and {Ain}, A. and {Ajith}, P. and {Allen}, B. and {Allen}, G. and {Allocca}, A. and {Altin}, P.~A. and {Amato}, A. and {Ananyeva}, A. and {Anderson}, S.~B. and {Anderson}, W.~G. and {Angelova}, S.~V. and {Antier}, S. and {Appert}, S. and {Arai}, K. and {Araya}, M.~C. and {Areeda}, J.~S. and {Arnaud}, N. and {Arun}, K.~G. and {Ascenzi}, S. and {Ashton}, G. and {Ast}, M. and {Aston}, S.~M. and {Astone}, P. and {Atallah}, D.~V. and {Aufmuth}, P. and {Aulbert}, C. and {AultONeal}, K. and {Austin}, C. and {Avila-Alvarez}, A. and {Babak}, S. and {Bacon}, P. and {Bader}, M.~K.~M. and {Bae}, S. and {Bailes}, M. and {Baker}, P.~T. and {Baldaccini}, F. and {Ballardin}, G. and {Ballmer}, S.~W. and {Banagiri}, S. and {Barayoga}, J.~C. and {Barclay}, S.~E. and {Barish}, B.~C. and {Barker}, D. and {Barkett}, K. and {Barone}, F. and {Barr}, B. and {Barsotti}, L. and {Barsuglia}, M. and {Barta}, D. and {Barthelmy}, S.~D. and {Bartlett}, J. and {Bartos}, I. and {Bassiri}, R. and {Basti}, A. and {Batch}, J.~C. and {Bawaj}, M. and {Bayley}, J.~C. and {Bazzan}, M. and {B{\'e}csy}, B. and {Beer}, C. and {Bejger}, M. and {Belahcene}, I. and {Bell}, A.~S. and {Berger}, B.~K. and {Bergmann}, G. and {Bernuzzi}, S. and {Bero}, J.~J. and {Berry}, C.~P.~L. and {Bersanetti}, D. and {Bertolini}, A. and {Betzwieser}, J. and {Bhagwat}, S. and {Bhandare}, R. and {Bilenko}, I.~A. and {Billingsley}, G. and {Billman}, C.~R. and {Birch}, J. and {Birney}, R. and {Birnholtz}, O. and {Biscans}, S. and {Biscoveanu}, S. and {Bisht}, A. and {Bitossi}, M. and {Biwer}, C. and {Bizouard}, M.~A. and {Blackburn}, J.~K. and {Blackman}, J. and {Blair}, C.~D. and {Blair}, D.~G. and {Blair}, R.~M. and {Bloemen}, S. and {Bock}, O. and {Bode}, N. and {Boer}, M. and {Bogaert}, G. and {Bohe}, A. and {Bondu}, F. and {Bonilla}, E. and {Bonnand}, R. and {Boom}, B.~A. and {Bork}, R. and {Boschi}, V. and {Bose}, S. and {Bossie}, K. and {Bouffanais}, Y. and {Bozzi}, A. and {Bradaschia}, C. and {Brady}, P.~R. and {Branchesi}, M. and {Brau}, J.~E. and {Briant}, T. and {Brillet}, A. and {Brinkmann}, M. and {Brisson}, V. and {Brockill}, P. and {Broida}, J.~E. and {Brooks}, A.~F. and {Brown}, D.~A. and {Brown}, D.~D. and {Brunett}, S. and {Buchanan}, C.~C. and {Buikema}, A. and {Bulik}, T. and {Bulten}, H.~J. and {Buonanno}, A. and {Buskulic}, D. and {Buy}, C. and {Byer}, R.~L. and {Cabero}, M. and {Cadonati}, L. and {Cagnoli}, G. and {Cahillane}, C. and {Calder{\'o}n Bustillo}, J. and {Callister}, T.~A. and {Calloni}, E. and {Camp}, J.~B. and {Canepa}, M. and {Canizares}, P. and {Cannon}, K.~C. and {Cao}, H. and {Cao}, J. and {Capano}, C.~D. and {Capocasa}, E. and {Carbognani}, F. and {Caride}, S. and {Carney}, M.~F. and {Carullo}, G. and {Casanueva Diaz}, J. and {Casentini}, C. and {Caudill}, S. and {Cavagli{\`a}}, M. and {Cavalier}, F. and {Cavalieri}, R. and {Cella}, G. and {Cepeda}, C.~B. and {Cerd{\'a}-Dur{\'a}n}, P. and {Cerretani}, G. and {Cesarini}, E. and {Chamberlin}, S.~J. and {Chan}, M. and {Chao}, S. and {Charlton}, P. and {Chase}, E. and {Chassande-Mottin}, E. and {Chatterjee}, D. and {Chatziioannou}, K. and {Cheeseboro}, B.~D. and {Chen}, H.~Y. and {Chen}, X. and {Chen}, Y. and {Cheng}, H.-P. and {Chia}, H. and {Chincarini}, A. and {Chiummo}, A. and {Chmiel}, T. and {Cho}, H.~S. and {Cho}, M. and {Chow}, J.~H. and {Christensen}, N. and {Chu}, Q. and {Chua}, A.~J.~K. and {Chua}, S.},
        title = "{GW170817: Observation of Gravitational Waves from a Binary Neutron Star Inspiral}",
      journal = {\prl},
         year = 2017,
        month = oct,
       volume = {119},
       number = {16},
          eid = {161101},
        pages = {161101},
          doi = {10.1103/PhysRevLett.119.161101},
archivePrefix = {arXiv},
       eprint = {1710.05832},
 primaryClass = {gr-qc},
       adsurl = {https://ui.adsabs.harvard.edu/abs/2017PhRvL.119p1101A}
}

@ARTICLE{Ascenzi2019,
       author = {{Ascenzi}, Stefano and {Coughlin}, Michael W. and {Dietrich}, Tim and {Foley}, Ryan J. and {Ramirez-Ruiz}, Enrico and {Piranomonte}, Silvia and {Mockler}, Brenna and {Murguia-Berthier}, Ariadna and {Fryer}, Chris L. and {Lloyd-Ronning}, Nicole M. and et al.},
        title = "{A luminosity distribution for kilonovae based on short gamma-ray burst afterglows}",
      journal = {\mnras},
         year = 2019,
        month = jun,
       volume = {486},
       number = {1},
        pages = {672-690},
          doi = {10.1093/mnras/stz891},
archivePrefix = {arXiv},
       eprint = {1811.05506},
 primaryClass = {astro-ph.HE},
       adsurl = {https://ui.adsabs.harvard.edu/abs/2019MNRAS.486..672A}
}

@ARTICLE{Brethauer2026,
       author = {{Brethauer}, D. and {Kasen}, D. and {Margutti}, R. and {Chornock}, R.},
        title = "{Nonthermal Ionization of Kilonova Ejecta: Observable Impacts}",
      journal = {\apj},
         year = 2026,
        month = jan,
       volume = {996},
       number = {1},
          eid = {64},
        pages = {64},
          doi = {10.3847/1538-4357/ae1b8d},
archivePrefix = {arXiv},
       eprint = {2508.18364},
 primaryClass = {astro-ph.HE},
       adsurl = {https://ui.adsabs.harvard.edu/abs/2026ApJ...996...64B}
}

@ARTICLE{Brethauer26b,
       author = {{Brethauer}, Daniel and {Kasen}, Daniel and {Banerjee}, Smaranika and {Margutti}, Raffaella and {Chornock}, Ryan},
        title = "{Free Neutron Decay in Kilonova Ejecta: X-ray/UV Flashes with Non-Thermal Effects}",
      journal = {arXiv e-prints},
         year = 2026,
        month = sep,
          eid = {arXiv:2609.09311},
        pages = {arXiv:2609.09311},
archivePrefix = {arXiv},
       eprint = {2609.09311},
 primaryClass = {astro-ph.HE},
       adsurl = {https://ui.adsabs.harvard.edu/abs/2026arXiv260909311B}
}

@ARTICLE{Ai2025,
       author = {{Ai}, Shunke and {Gao}, He and {Zhang}, Bing},
        title = "{Engine-fed Kilonovae (Mergernovae). II. Radiation}",
      journal = {\apj},
         year = 2025,
        month = jan,
       volume = {978},
       number = {1},
          eid = {52},
        pages = {52},
          doi = {10.3847/1538-4357/ad93b4},
archivePrefix = {arXiv},
       eprint = {2405.00638},
 primaryClass = {astro-ph.HE},
       adsurl = {https://ui.adsabs.harvard.edu/abs/2025ApJ...978...52A}
}

@ARTICLE{Gompertz2018,
       author = {{Gompertz}, B.~P. and {Levan}, A.~J. and {Tanvir}, N.~R. and {Hjorth}, J. and {Covino}, S. and {Evans}, P.~A. and {Fruchter}, A.~S. and {Gonz{\'a}lez-Fern{\'a}ndez}, C. and {Jin}, Z.~P. and {Lyman}, J.~D. and {Oates}, S.~R. and {O'Brien}, P.~T. and {Wiersema}, K.},
        title = "{The Diversity of Kilonova Emission in Short Gamma-Ray Bursts}",
      journal = {\apj},
         year = 2018,
        month = jun,
       volume = {860},
       number = {1},
          eid = {62},
        pages = {62},
          doi = {10.3847/1538-4357/aac206},
archivePrefix = {arXiv},
       eprint = {1710.05442},
 primaryClass = {astro-ph.HE},
       adsurl = {https://ui.adsabs.harvard.edu/abs/2018ApJ...860...62G}
}

@ARTICLE{Gottlieb2020,
       author = {{Gottlieb}, Ore and {Loeb}, Abraham},
        title = "{Electromagnetic signals from the decay of free neutrons in the first hours of neutron star mergers}",
      journal = {\mnras},
         year = 2020,
        month = apr,
       volume = {493},
       number = {2},
        pages = {1753-1760},
          doi = {10.1093/mnras/staa363},
archivePrefix = {arXiv},
       eprint = {1912.04289},
 primaryClass = {astro-ph.HE},
       adsurl = {https://ui.adsabs.harvard.edu/abs/2020MNRAS.493.1753G}
}

@ARTICLE{Metzger2015,
       author = {{Metzger}, Brian D. and {Bauswein}, Andreas and {Goriely}, Stephane and {Kasen}, Daniel},
        title = "{Neutron-powered precursors of kilonovae}",
      journal = {\mnras},
         year = 2015,
        month = jan,
       volume = {446},
       number = {1},
        pages = {1115-1120},
          doi = {10.1093/mnras/stu2225},
archivePrefix = {arXiv},
       eprint = {1409.0544},
 primaryClass = {astro-ph.HE},
       adsurl = {https://ui.adsabs.harvard.edu/abs/2015MNRAS.446.1115M}
}

@ARTICLE{2022ApJ...928..186A,
       author = {{Abbott}, R. and {Abbott}, T.~D. and {Acernese}, F. and {Ackley}, K. and {Adams}, C. and {Adhikari}, N. and {Adhikari}, R.~X. and {Adya}, V.~B. and {Affeldt}, C. and {Agarwal}, D. and {Agathos}, M. and {Agatsuma}, K. and {Aggarwal}, N. and {Aguiar}, O.~D. and {Aiello}, L. and {Ain}, A. and {Ajith}, P. and {Akutsu}, T. and {Albanesi}, S. and {Allocca}, A. and {Altin}, P.~A. and {Amato}, A. and {Anand}, C. and {Anand}, S. and {Ananyeva}, A. and {Anderson}, S.~B. and {Anderson}, W.~G. and {Ando}, M. and {Andrade}, T. and {Andres}, N. and {Andri{\'c}}, T. and {Angelova}, S.~V. and {Ansoldi}, S. and {Antelis}, J.~M. and {Antier}, S. and {Appert}, S. and {Arai}, Koji and {Arai}, Koya and {Arai}, Y. and {Araki}, S. and {Araya}, A. and {Araya}, M.~C. and {Areeda}, J.~S. and {Ar{\`e}ne}, M. and {Aritomi}, N. and {Arnaud}, N. and {Aronson}, S.~M. and {Arun}, K.~G. and {Asada}, H. and {Asali}, Y. and {Ashton}, G. and {Aso}, Y. and {Assiduo}, M. and {Aston}, S.~M. and {Astone}, P. and {Aubin}, F. and {Austin}, C. and {Babak}, S. and {Badaracco}, F. and {Bader}, M.~K.~M. and {Badger}, C. and {Bae}, S. and {Bae}, Y. and {Baer}, A.~M. and {Bagnasco}, S. and {Bai}, Y. and {Baiotti}, L. and {Baird}, J. and {Bajpai}, R. and {Ball}, M. and {Ballardin}, G. and {Ballmer}, S.~W. and {Balsamo}, A. and {Baltus}, G. and {Banagiri}, S. and {Bankar}, D. and {Barayoga}, J.~C. and {Barbieri}, C. and {Barish}, B.~C. and {Barker}, D. and {Barneo}, P. and {Barone}, F. and {Barr}, B. and {Barsotti}, L. and {Barsuglia}, M. and {Barta}, D. and {Bartlett}, J. and {Barton}, M.~A. and {Bartos}, I. and {Bassiri}, R. and {Basti}, A. and {Bawaj}, M. and {Bayley}, J.~C. and {Baylor}, A.~C. and {Bazzan}, M. and {B{\'e}csy}, B. and {Bedakihale}, V.~M. and {Bejger}, M. and {Belahcene}, I. and {Benedetto}, V. and {Beniwal}, D. and {Bennett}, T.~F. and {Bentley}, J.~D. and {Benyaala}, M. and {Bergamin}, F. and {Berger}, B.~K. and {Bernuzzi}, S. and {Berry}, C.~P.~L. and {Bersanetti}, D. and {Bertolini}, A. and {Betzwieser}, J. and {Beveridge}, D. and {Bhandare}, R. and {Bhardwaj}, U. and {Bhattacharjee}, D. and {Bhaumik}, S. and {Bilenko}, I.~A. and {Billingsley}, G. and {Bini}, S. and {Birney}, R. and {Birnholtz}, O. and {Biscans}, S. and {Bischi}, M. and {Biscoveanu}, S. and {Bisht}, A. and {Biswas}, B. and {Bitossi}, M. and {Bizouard}, M.-A. and {Blackburn}, J.~K. and {Blair}, C.~D. and {Blair}, D.~G. and {Blair}, R.~M. and {Bobba}, F. and {Bode}, N. and {Boer}, M. and {Bogaert}, G. and {Boldrini}, M. and {Bonavena}, L.~D. and {Bondu}, F. and {Bonilla}, E. and {Bonnand}, R. and {Booker}, P. and {Boom}, B.~A. and {Bork}, R. and {Boschi}, V. and {Bose}, N. and {Bose}, S. and {Bossilkov}, V. and {Boudart}, V. and {Bouffanais}, Y. and {Bozzi}, A. and {Bradaschia}, C. and {Brady}, P.~R. and {Bramley}, A. and {Branch}, A. and {Branchesi}, M. and {Brau}, J.~E. and {Breschi}, M. and {Briant}, T. and {Briggs}, J.~H. and {Brillet}, A. and {Brinkmann}, M. and {Brockill}, P. and {Brooks}, A.~F. and {Brooks}, J. and {Brown}, D.~D. and {Brunett}, S. and {Bruno}, G. and {Bruntz}, R. and {Bryant}, J. and {Bulik}, T. and {Bulten}, H.~J. and {Buonanno}, A. and {Buscicchio}, R. and {Buskulic}, D. and {Buy}, C. and {Byer}, R.~L. and {Cadonati}, L. and {Cagnoli}, G. and {Cahillane}, C. and {Bustillo}, J. Calder{\'o}n and {Callaghan}, J.~D. and {Callister}, T.~A. and {Calloni}, E. and {Cameron}, J. and {Camp}, J.~B. and {Canepa}, M. and {Canevarolo}, S. and {Cannavacciuolo}, M. and {Cannon}, K.~C. and {Cao}, H. and {Cao}, Z. and {Capocasa}, E. and {Capote}, E. and {Carapella}, G. and {Carbognani}, F. and {Carlin}, J.~B. and {Carney}, M.~F. and {Carpinelli}, M. and {Carrillo}, G.},
        title = "{Search for Gravitational Waves Associated with Gamma-Ray Bursts Detected by Fermi and Swift during the LIGO-Virgo Run O3b}",
      journal = {\apj},
         year = 2022,
        month = apr,
       volume = {928},
       number = {2},
          eid = {186},
        pages = {186},
          doi = {10.3847/1538-4357/ac532b},
archivePrefix = {arXiv},
       eprint = {2111.03608},
 primaryClass = {astro-ph.HE},
       adsurl = {https://ui.adsabs.harvard.edu/abs/2022ApJ...928..186A}
}

@ARTICLE{2017ApJ...848L..12A,
       author = {{Abbott}, B.~P. and {Abbott}, R. and {Abbott}, T.~D. and {Acernese}, F. and {Ackley}, K. and {Adams}, C. and {Adams}, T. and {Addesso}, P. and {Adhikari}, R.~X. and {Adya}, V.~B. and {Affeldt}, C. and {Afrough}, M. and {Agarwal}, B. and {Agathos}, M. and {Agatsuma}, K. and {Aggarwal}, N. and {Aguiar}, O.~D. and {Aiello}, L. and {Ain}, A. and {Ajith}, P. and {Allen}, B. and {Allen}, G. and {Allocca}, A. and {Altin}, P.~A. and {Amato}, A. and {Ananyeva}, A. and {Anderson}, S.~B. and {Anderson}, W.~G. and {Angelova}, S.~V. and {Antier}, S. and {Appert}, S. and {Arai}, K. and {Araya}, M.~C. and {Areeda}, J.~S. and {Arnaud}, N. and {Arun}, K.~G. and {Ascenzi}, S. and {Ashton}, G. and {Ast}, M. and {Aston}, S.~M. and {Astone}, P. and {Atallah}, D.~V. and {Aufmuth}, P. and {Aulbert}, C. and {AultONeal}, K. and {Austin}, C. and {Avila-Alvarez}, A. and {Babak}, S. and {Bacon}, P. and {Bader}, M.~K.~M. and {Bae}, S. and {Baker}, P.~T. and {Baldaccini}, F. and {Ballardin}, G. and {Ballmer}, S.~W. and {Banagiri}, S. and {Barayoga}, J.~C. and {Barclay}, S.~E. and {Barish}, B.~C. and {Barker}, D. and {Barkett}, K. and {Barone}, F. and {Barr}, B. and {Barsotti}, L. and {Barsuglia}, M. and {Barta}, D. and {Barthelmy}, S.~D. and {Bartlett}, J. and {Bartos}, I. and {Bassiri}, R. and {Basti}, A. and {Batch}, J.~C. and {Bawaj}, M. and {Bayley}, J.~C. and {Bazzan}, M. and {B{\'e}csy}, B. and {Beer}, C. and {Bejger}, M. and {Belahcene}, I. and {Bell}, A.~S. and {Berger}, B.~K. and {Bergmann}, G. and {Bero}, J.~J. and {Berry}, C.~P.~L. and {Bersanetti}, D. and {Bertolini}, A. and {Betzwieser}, J. and {Bhagwat}, S. and {Bhandare}, R. and {Bilenko}, I.~A. and {Billingsley}, G. and {Billman}, C.~R. and {Birch}, J. and {Birney}, R. and {Birnholtz}, O. and {Biscans}, S. and {Biscoveanu}, S. and {Bisht}, A. and {Bitossi}, M. and {Biwer}, C. and {Bizouard}, M.~A. and {Blackburn}, J.~K. and {Blackman}, J. and {Blair}, C.~D. and {Blair}, D.~G. and {Blair}, R.~M. and {Bloemen}, S. and {Bock}, O. and {Bode}, N. and {Boer}, M. and {Bogaert}, G. and {Bohe}, A. and {Bondu}, F. and {Bonilla}, E. and {Bonnand}, R. and {Boom}, B.~A. and {Bork}, R. and {Boschi}, V. and {Bose}, S. and {Bossie}, K. and {Bouffanais}, Y. and {Bozzi}, A. and {Bradaschia}, C. and {Brady}, P.~R. and {Branchesi}, M. and {Brau}, J.~E. and {Briant}, T. and {Brillet}, A. and {Brinkmann}, M. and {Brisson}, V. and {Brockill}, P. and {Broida}, J.~E. and {Brooks}, A.~F. and {Brown}, D.~A. and {Brown}, D.~D. and {Brunett}, S. and {Buchanan}, C.~C. and {Buikema}, A. and {Bulik}, T. and {Bulten}, H.~J. and {Buonanno}, A. and {Buskulic}, D. and {Buy}, C. and {Byer}, R.~L. and {Cabero}, M. and {Cadonati}, L. and {Cagnoli}, G. and {Cahillane}, C. and {Calder{\'o}n Bustillo}, J. and {Callister}, T.~A. and {Calloni}, E. and {Camp}, J.~B. and {Canepa}, M. and {Canizares}, P. and {Cannon}, K.~C. and {Cao}, H. and {Cao}, J. and {Capano}, C.~D. and {Capocasa}, E. and {Carbognani}, F. and {Caride}, S. and {Carney}, M.~F. and {Casanueva Diaz}, J. and {Casentini}, C. and {Caudill}, S. and {Cavagli{\`a}}, M. and {Cavalier}, F. and {Cavalieri}, R. and {Cella}, G. and {Cepeda}, C.~B. and {Cerd{\'a}-Dur{\'a}n}, P. and {Cerretani}, G. and {Cesarini}, E. and {Chamberlin}, S.~J. and {Chan}, M. and {Chao}, S. and {Charlton}, P. and {Chase}, E. and {Chassande-Mottin}, E. and {Chatterjee}, D. and {Chatziioannou}, K. and {Cheeseboro}, B.~D. and {Chen}, H.~Y. and {Chen}, X. and {Chen}, Y. and {Cheng}, H.-P. and {Chia}, H. and {Chincarini}, A. and {Chiummo}, A. and {Chmiel}, T. and {Cho}, H.~S. and {Cho}, M. and {Chow}, J.~H. and {Christensen}, N. and {Chu}, Q. and {Chua}, A.~J.~K. and {Chua}, S. and {Chung}, A.~K.~W. and {Chung}, S. and {Ciani}, G.},
        title = "{Multi-messenger Observations of a Binary Neutron Star Merger}",
      journal = {\apjl},
         year = 2017,
        month = oct,
       volume = {848},
       number = {2},
          eid = {L12},
        pages = {L12},
          doi = {10.3847/2041-8213/aa91c9},
archivePrefix = {arXiv},
       eprint = {1710.05833},
 primaryClass = {astro-ph.HE},
       adsurl = {https://ui.adsabs.harvard.edu/abs/2017ApJ...848L..12A}
}

@ARTICLE{2020ApJ...892L...3A,
       author = {{Abbott}, B.~P. and {Abbott}, R. and {Abbott}, T.~D. and {Abraham}, S. and {Acernese}, F. and {Ackley}, K. and {Adams}, C. and {Adhikari}, R.~X. and {Adya}, V.~B. and {Affeldt}, C. and {Agathos}, M. and {Agatsuma}, K. and {Aggarwal}, N. and {Aguiar}, O.~D. and {Aiello}, L. and {Ain}, A. and {Ajith}, P. and {Allen}, G. and {Allocca}, A. and {Aloy}, M.~A. and {Altin}, P.~A. and {Amato}, A. and {Anand}, S. and {Ananyeva}, A. and {Anderson}, S.~B. and {Anderson}, W.~G. and {Angelova}, S.~V. and {Antier}, S. and {Appert}, S. and {Arai}, K. and {Araya}, M.~C. and {Areeda}, J.~S. and {Ar{\`e}ne}, M. and {Arnaud}, N. and {Aronson}, S.~M. and {Arun}, K.~G. and {Ascenzi}, S. and {Ashton}, G. and {Aston}, S.~M. and {Astone}, P. and {Aubin}, F. and {Aufmuth}, P. and {AultONeal}, K. and {Austin}, C. and {Avendano}, V. and {Avila-Alvarez}, A. and {Babak}, S. and {Bacon}, P. and {Badaracco}, F. and {Bader}, M.~K.~M. and {Bae}, S. and {Baird}, J. and {Baker}, P.~T. and {Baldaccini}, F. and {Ballardin}, G. and {Ballmer}, S.~W. and {Bals}, A. and {Banagiri}, S. and {Barayoga}, J.~C. and {Barbieri}, C. and {Barclay}, S.~E. and {Barish}, B.~C. and {Barker}, D. and {Barkett}, K. and {Barnum}, S. and {Barone}, F. and {Barr}, B. and {Barsotti}, L. and {Barsuglia}, M. and {Barta}, D. and {Bartlett}, J. and {Bartos}, I. and {Bassiri}, R. and {Basti}, A. and {Bawaj}, M. and {Bayley}, J.~C. and {Baylor}, A.~C. and {Bazzan}, M. and {B{\'e}csy}, B. and {Bejger}, M. and {Belahcene}, I. and {Bell}, A.~S. and {Beniwal}, D. and {Benjamin}, M.~G. and {Berger}, B.~K. and {Bergmann}, G. and {Bernuzzi}, S. and {Berry}, C.~P.~L. and {Bersanetti}, D. and {Bertolini}, A. and {Betzwieser}, J. and {Bhandare}, R. and {Bidler}, J. and {Biggs}, E. and {Bilenko}, I.~A. and {Bilgili}, S.~A. and {Billingsley}, G. and {Birney}, R. and {Birnholtz}, O. and {Biscans}, S. and {Bischi}, M. and {Biscoveanu}, S. and {Bisht}, A. and {Bitossi}, M. and {Bizouard}, M.~A. and {Blackburn}, J.~K. and {Blackman}, J. and {Blair}, C.~D. and {Blair}, D.~G. and {Blair}, R.~M. and {Bloemen}, S. and {Bobba}, F. and {Bode}, N. and {Boer}, M. and {Boetzel}, Y. and {Bogaert}, G. and {Bondu}, F. and {Bonnand}, R. and {Booker}, P. and {Boom}, B.~A. and {Bork}, R. and {Boschi}, V. and {Bose}, S. and {Bossilkov}, V. and {Bosveld}, J. and {Bouffanais}, Y. and {Bozzi}, A. and {Bradaschia}, C. and {Brady}, P.~R. and {Bramley}, A. and {Branchesi}, M. and {Brau}, J.~E. and {Breschi}, M. and {Briant}, T. and {Briggs}, J.~H. and {Brighenti}, F. and {Brillet}, A. and {Brinkmann}, M. and {Brockill}, P. and {Brooks}, A.~F. and {Brooks}, J. and {Brown}, D.~D. and {Brunett}, S. and {Buikema}, A. and {Bulik}, T. and {Bulten}, H.~J. and {Buonanno}, A. and {Buskulic}, D. and {Buy}, C. and {Byer}, R.~L. and {Cabero}, M. and {Cadonati}, L. and {Cagnoli}, G. and {Cahillane}, C. and {Calder{\'o}n Bustillo}, J. and {Callister}, T.~A. and {Calloni}, E. and {Camp}, J.~B. and {Campbell}, W.~A. and {Canepa}, M. and {Cannon}, K.~C. and {Cao}, H. and {Cao}, J. and {Carapella}, G. and {Carbognani}, F. and {Caride}, S. and {Carney}, M.~F. and {Carullo}, G. and {Casanueva Diaz}, J. and {Casentini}, C. and {Caudill}, S. and {Cavagli{\`a}}, M. and {Cavalier}, F. and {Cavalieri}, R. and {Cella}, G. and {Cerd{\'a}-Dur{\'a}n}, P. and {Cesarini}, E. and {Chaibi}, O. and {Chakravarti}, K. and {Chamberlin}, S.~J. and {Chan}, M. and {Chao}, S. and {Charlton}, P. and {Chase}, E.~A. and {Chassande-Mottin}, E. and {Chatterjee}, D. and {Chaturvedi}, M. and {Chatziioannou}, K. and {Cheeseboro}, B.~D. and {Chen}, H.~Y. and {Chen}, X. and {Chen}, Y. and {Cheng}, H.-P. and {Cheong}, C.~K. and {Chia}, H.~Y. and {Chiadini}, F. and {Chincarini}, A. and {Chiummo}, A. and {Cho}, G. and {Cho}, H.~S.},
        title = "{GW190425: Observation of a Compact Binary Coalescence with Total Mass {\ensuremath{\sim}} 3.4 M$_{☉}$}",
      journal = {\apjl},
         year = 2020,
        month = mar,
       volume = {892},
       number = {1},
          eid = {L3},
        pages = {L3},
          doi = {10.3847/2041-8213/ab75f5},
archivePrefix = {arXiv},
       eprint = {2001.01761},
 primaryClass = {astro-ph.HE},
       adsurl = {https://ui.adsabs.harvard.edu/abs/2020ApJ...892L...3A}
}

@ARTICLE{2015CQGra..32b4001A,
       author = {{Acernese}, F. and {Agathos}, M. and {Agatsuma}, K. and {Aisa}, D. and {Allemandou}, N. and {Allocca}, A. and {Amarni}, J. and {Astone}, P. and {Balestri}, G. and {Ballardin}, G. and {Barone}, F. and {Baronick}, J.-P. and {Barsuglia}, M. and {Basti}, A. and {Basti}, F. and {Bauer}, Th S. and {Bavigadda}, V. and {Bejger}, M. and {Beker}, M.~G. and {Belczynski}, C. and {Bersanetti}, D. and {Bertolini}, A. and {Bitossi}, M. and {Bizouard}, M.~A. and {Bloemen}, S. and {Blom}, M. and {Boer}, M. and {Bogaert}, G. and {Bondi}, D. and {Bondu}, F. and {Bonelli}, L. and {Bonnand}, R. and {Boschi}, V. and {Bosi}, L. and {Bouedo}, T. and {Bradaschia}, C. and {Branchesi}, M. and {Briant}, T. and {Brillet}, A. and {Brisson}, V. and {Bulik}, T. and {Bulten}, H.~J. and {Buskulic}, D. and {Buy}, C. and {Cagnoli}, G. and {Calloni}, E. and {Campeggi}, C. and {Canuel}, B. and {Carbognani}, F. and {Cavalier}, F. and {Cavalieri}, R. and {Cella}, G. and {Cesarini}, E. and {Mottin}, E. Chassande- and {Chincarini}, A. and {Chiummo}, A. and {Chua}, S. and {Cleva}, F. and {Coccia}, E. and {Cohadon}, P.-F. and {Colla}, A. and {Colombini}, M. and {Conte}, A. and {Coulon}, J.-P. and {Cuoco}, E. and {Dalmaz}, A. and {D'Antonio}, S. and {Dattilo}, V. and {Davier}, M. and {Day}, R. and {Debreczeni}, G. and {Degallaix}, J. and {Del{\'e}glise}, S. and {Pozzo}, W. Del and {Dereli}, H. and {Rosa}, R. De and {Fiore}, L. Di and {Lieto}, A. Di and {Virgilio}, A. Di and {Doets}, M. and {Dolique}, V. and {Drago}, M. and {Ducrot}, M. and {Endr{\H{o}}czi}, G. and {Fafone}, V. and {Farinon}, S. and {Ferrante}, I. and {Ferrini}, F. and {Fidecaro}, F. and {Fiori}, I. and {Flaminio}, R. and {Fournier}, J.-D. and {Franco}, S. and {Frasca}, S. and {Frasconi}, F. and {Gammaitoni}, L. and {Garufi}, F. and {Gaspard}, M. and {Gatto}, A. and {Gemme}, G. and {Gendre}, B. and {Genin}, E. and {Gennai}, A. and {Ghosh}, S. and {Giacobone}, L. and {Giazotto}, A. and {Gouaty}, R. and {Granata}, M. and {Greco}, G. and {Groot}, P. and {Guidi}, G.~M. and {Harms}, J. and {Heidmann}, A. and {Heitmann}, H. and {Hello}, P. and {Hemming}, G. and {Hennes}, E. and {Hofman}, D. and {Jaranowski}, P. and {Jonker}, R.~J.~G. and {Kasprzack}, M. and {K{\'e}f{\'e}lian}, F. and {Kowalska}, I. and {Kraan}, M. and {Kr{\'o}lak}, A. and {Kutynia}, A. and {Lazzaro}, C. and {Leonardi}, M. and {Leroy}, N. and {Letendre}, N. and {Li}, T.~G.~F. and {Lieunard}, B. and {Lorenzini}, M. and {Loriette}, V. and {Losurdo}, G. and {Magazz{\`u}}, C. and {Majorana}, E. and {Maksimovic}, I. and {Malvezzi}, V. and {Man}, N. and {Mangano}, V. and {Mantovani}, M. and {Marchesoni}, F. and {Marion}, F. and {Marque}, J. and {Martelli}, F. and {Martellini}, L. and {Masserot}, A. and {Meacher}, D. and {Meidam}, J. and {Mezzani}, F. and {Michel}, C. and {Milano}, L. and {Minenkov}, Y. and {Moggi}, A. and {Mohan}, M. and {Montani}, M. and {Morgado}, N. and {Mours}, B. and {Mul}, F. and {Nagy}, M.~F. and {Nardecchia}, I. and {Naticchioni}, L. and {Nelemans}, G. and {Neri}, I. and {Neri}, M. and {Nocera}, F. and {Pacaud}, E. and {Palomba}, C. and {Paoletti}, F. and {Paoli}, A. and {Pasqualetti}, A. and {Passaquieti}, R. and {Passuello}, D. and {Perciballi}, M. and {Petit}, S. and {Pichot}, M. and {Piergiovanni}, F. and {Pillant}, G. and {Piluso}, A. and {Pinard}, L. and {Poggiani}, R. and {Prijatelj}, M. and {Prodi}, G.~A. and {Punturo}, M. and {Puppo}, P. and {Rabeling}, D.~S. and {R{\'a}cz}, I. and {Rapagnani}, P. and {Razzano}, M. and {Re}, V. and {Regimbau}, T. and {Ricci}, F. and {Robinet}, F. and {Rocchi}, A. and {Rolland}, L. and {Romano}, R. and {Rosi{\'n}ska}, D. and {Ruggi}, P. and {Saracco}, E.},
        title = "{Advanced Virgo: a second-generation interferometric gravitational wave detector}",
      journal = {Classical and Quantum Gravity},
         year = 2015,
        month = jan,
       volume = {32},
       number = {2},
          eid = {024001},
        pages = {024001},
          doi = {10.1088/0264-9381/32/2/024001},
archivePrefix = {arXiv},
       eprint = {1408.3978},
 primaryClass = {gr-qc},
       adsurl = {https://ui.adsabs.harvard.edu/abs/2015CQGra..32b4001A}
}

@ARTICLE{2019CQGra..36p5008A,
       author = {{Akutsu}, T. and {Ando}, M. and {Arai}, K. and {Arai}, Y. and {Araki}, S. and {Araya}, A. and {Aritomi}, N. and {Asada}, H. and {Aso}, Y. and {Atsuta}, S. and {Awai}, K. and {Bae}, S. and {Baiotti}, L. and {Barton}, M.~A. and {Cannon}, K. and {Capocasa}, E. and {Chen}, C.-S. and {Chiu}, T.-W. and {Cho}, K. and {Chu}, Y.-K. and {Craig}, K. and {Creus}, W. and {Doi}, K. and {Eda}, K. and {Enomoto}, Y. and {Flaminio}, R. and {Fujii}, Y. and {Fujimoto}, M.-K. and {Fukunaga}, M. and {Fukushima}, M. and {Furuhata}, T. and {Hagiwara}, A. and {Haino}, S. and {Hasegawa}, K. and {Hashino}, K. and {Hayama}, K. and {Hirobayashi}, S. and {Hirose}, E. and {Hsieh}, B.~H. and {Huang}, C.-Z. and {Ikenoue}, B. and {Inoue}, Y. and {Ioka}, K. and {Itoh}, Y. and {Izumi}, K. and {Kaji}, T. and {Kajita}, T. and {Kakizaki}, M. and {Kamiizumi}, M. and {Kanbara}, S. and {Kanda}, N. and {Kanemura}, S. and {Kaneyama}, M. and {Kang}, G. and {Kasuya}, J. and {Kataoka}, Y. and {Kawai}, N. and {Kawamura}, S. and {Kawasaki}, T. and {Kim}, C. and {Kim}, J. and {Kim}, J.~C. and {Kim}, W.~S. and {Kim}, Y.-M. and {Kimura}, N. and {Kinugawa}, T. and {Kirii}, S. and {Kitaoka}, Y. and {Kitazawa}, H. and {Kojima}, Y. and {Kokeyama}, K. and {Komori}, K. and {Kong}, A.~K.~H. and {Kotake}, K. and {Kozu}, R. and {Kumar}, R. and {Kuo}, H.-S. and {Kuroyanagi}, S. and {Lee}, H.~K. and {Lee}, H.~M. and {Lee}, H.~W. and {Leonardi}, M. and {Lin}, C.-Y. and {Lin}, F.-L. and {Liu}, G.~C. and {Liu}, Y. and {Majorana}, E. and {Mano}, S. and {Marchio}, M. and {Matsui}, T. and {Matsushima}, F. and {Michimura}, Y. and {Mio}, N. and {Miyakawa}, O. and {Miyamoto}, A. and {Miyamoto}, T. and {Miyo}, K. and {Miyoki}, S. and {Morii}, W. and {Morisaki}, S. and {Moriwaki}, Y. and {Morozumi}, T. and {Murakami}, I. and {Musha}, M. and {Nagano}, K. and {Nagano}, S. and {Nakamura}, K. and {Nakamura}, T. and {Nakano}, H. and {Nakano}, M. and {Nakao}, K. and {Namai}, Y. and {Narikawa}, T. and {Naticchioni}, L. and {Nguyen Quynh}, L. and {Ni}, W.-T. and {Nishizawa}, A. and {Obuchi}, Y. and {Ochi}, T. and {Oh}, J.~J. and {Oh}, S.~H. and {Ohashi}, M. and {Ohishi}, N. and {Ohkawa}, M. and {Okutomi}, K. and {Ono}, K. and {Oohara}, K. and {Ooi}, C.~P. and {Pan}, S.-S. and {Park}, J. and {Pe{\~n}a Arellano}, F.~E. and {Pinto}, I. and {Sago}, N. and {Saijo}, M. and {Saito}, Y. and {Saitou}, S. and {Sakai}, K. and {Sakai}, Y. and {Sakai}, Y. and {Sasai}, M. and {Sasaki}, M. and {Sasaki}, Y. and {Sato}, N. and {Sato}, S. and {Sato}, T. and {Sekiguchi}, Y. and {Seto}, N. and {Shibata}, M. and {Shimoda}, T. and {Shinkai}, H. and {Shishido}, T. and {Shoda}, A. and {Somiya}, K. and {Son}, E.~J. and {Suemasa}, A. and {Suzuki}, T. and {Suzuki}, T. and {Tagoshi}, H. and {Tahara}, H. and {Takahashi}, H. and {Takahashi}, R. and {Takamori}, A. and {Takeda}, H. and {Tanaka}, H. and {Tanaka}, K. and {Tanaka}, T. and {Tanioka}, S. and {Tapia San Martin}, E.~N. and {Tatsumi}, D. and {Terashima}, S. and {Tomaru}, T. and {Tomura}, T. and {Travasso}, F. and {Tsubono}, K. and {Tsuchida}, S. and {Uchikata}, N. and {Uchiyama}, T. and {Ueda}, A. and {Uehara}, T. and {Ueki}, S. and {Ueno}, K. and {Uraguchi}, F. and {Ushiba}, T. and {van Putten}, M.~H.~P.~M. and {Vocca}, H. and {Wada}, S. and {Wakamatsu}, T. and {Watanabe}, Y. and {Xu}, W.-R. and {Yamada}, T. and {Yamamoto}, A. and {Yamamoto}, K. and {Yamamoto}, K. and {Yamamoto}, S. and {Yamamoto}, T. and {Yokogawa}, K. and {Yokoyama}, J. and {Yokozawa}, T. and {Yoon}, T.~H. and {Yoshioka}, T.},
        title = "{First cryogenic test operation of underground km-scale gravitational-wave observatory KAGRA}",
      journal = {Classical and Quantum Gravity},
         year = 2019,
        month = aug,
       volume = {36},
       number = {16},
          eid = {165008},
        pages = {165008},
          doi = {10.1088/1361-6382/ab28a9},
archivePrefix = {arXiv},
       eprint = {1901.03569},
 primaryClass = {astro-ph.IM},
       adsurl = {https://ui.adsabs.harvard.edu/abs/2019CQGra..36p5008A}
}

@ARTICLE{2017ApJ...848L..14G,
       author = {{Goldstein}, A. and {Veres}, P. and {Burns}, E. and {Briggs}, M.~S. and {Hamburg}, R. and {Kocevski}, D. and {Wilson-Hodge}, C.~A. and {Preece}, R.~D. and {Poolakkil}, S. and {Roberts}, O.~J. and {Hui}, C.~M. and {Connaughton}, V. and {Racusin}, J. and {von Kienlin}, A. and {Dal Canton}, T. and {Christensen}, N. and {Littenberg}, T. and {Siellez}, K. and {Blackburn}, L. and {Broida}, J. and {Bissaldi}, E. and {Cleveland}, W.~H. and {Gibby}, M.~H. and {Giles}, M.~M. and {Kippen}, R.~M. and {McBreen}, S. and {McEnery}, J. and {Meegan}, C.~A. and {Paciesas}, W.~S. and {Stanbro}, M.},
        title = "{An Ordinary Short Gamma-Ray Burst with Extraordinary Implications: Fermi-GBM Detection of GRB 170817A}",
      journal = {\apjl},
         year = 2017,
        month = oct,
       volume = {848},
       number = {2},
          eid = {L14},
        pages = {L14},
          doi = {10.3847/2041-8213/aa8f41},
archivePrefix = {arXiv},
       eprint = {1710.05446},
 primaryClass = {astro-ph.HE},
       adsurl = {https://ui.adsabs.harvard.edu/abs/2017ApJ...848L..14G}
}

@ARTICLE{2016ApJ...829..110B,
       author = {{Barnes}, Jennifer and {Kasen}, Daniel and {Wu}, Meng-Ru and {Mart{\'\i}nez-Pinedo}, Gabriel},
        title = "{Radioactivity and Thermalization in the Ejecta of Compact Object Mergers and Their Impact on Kilonova Light Curves}",
      journal = {\apj},
         year = 2016,
        month = oct,
       volume = {829},
       number = {2},
          eid = {110},
        pages = {110},
          doi = {10.3847/0004-637X/829/2/110},
archivePrefix = {arXiv},
       eprint = {1605.07218},
 primaryClass = {astro-ph.HE},
       adsurl = {https://ui.adsabs.harvard.edu/abs/2016ApJ...829..110B}
}

@ARTICLE{2011PhRvD..83h4002H,
       author = {{Harry}, I.~W. and {Fairhurst}, S.},
        title = "{Targeted coherent search for gravitational waves from compact binary coalescences}",
      journal = {\prd},
         year = 2011,
        month = apr,
       volume = {83},
       number = {8},
          eid = {084002},
        pages = {084002},
          doi = {10.1103/PhysRevD.83.084002},
archivePrefix = {arXiv},
       eprint = {1012.4939},
 primaryClass = {gr-qc},
       adsurl = {https://ui.adsabs.harvard.edu/abs/2011PhRvD..83h4002H}
}

@ARTICLE{2019ApJ...880L...4H,
       author = {{Hosseinzadeh}, G. and {Cowperthwaite}, P.~S. and {Gomez}, S. and {Villar}, V.~A. and {Nicholl}, M. and {Margutti}, R. and {Berger}, E. and {Chornock}, R. and {Paterson}, K. and {Fong}, W. and {Savchenko}, V. and {Short}, P. and {Alexander}, K.~D. and {Blanchard}, P.~K. and {Braga}, J. and {Calkins}, M.~L. and {Cartier}, R. and {Coppejans}, D.~L. and {Eftekhari}, T. and {Laskar}, T. and {Ly}, C. and {Patton}, L. and {Pelisoli}, I. and {Reichart}, D.~E. and {Terreran}, G. and {Williams}, P.~K.~G.},
        title = "{Follow-up of the Neutron Star Bearing Gravitational-wave Candidate Events S190425z and S190426c with MMT and SOAR}",
      journal = {\apjl},
         year = 2019,
        month = jul,
       volume = {880},
       number = {1},
          eid = {L4},
        pages = {L4},
          doi = {10.3847/2041-8213/ab271c},
archivePrefix = {arXiv},
       eprint = {1905.02186},
 primaryClass = {astro-ph.HE},
       adsurl = {https://ui.adsabs.harvard.edu/abs/2019ApJ...880L...4H}
}

@ARTICLE{2017Natur.551...67P,
       author = {{Pian}, E. and {D'Avanzo}, P. and {Benetti}, S. and {Branchesi}, M. and {Brocato}, E. and {Campana}, S. and {Cappellaro}, E. and {Covino}, S. and {D'Elia}, V. and {Fynbo}, J.~P.~U. and {Getman}, F. and {Ghirlanda}, G. and {Ghisellini}, G. and {Grado}, A. and {Greco}, G. and {Hjorth}, J. and {Kouveliotou}, C. and {Levan}, A. and {Limatola}, L. and {Malesani}, D. and {Mazzali}, P.~A. and {Melandri}, A. and {M{\o}ller}, P. and {Nicastro}, L. and {Palazzi}, E. and {Piranomonte}, S. and {Rossi}, A. and {Salafia}, O.~S. and {Selsing}, J. and {Stratta}, G. and {Tanaka}, M. and {Tanvir}, N.~R. and {Tomasella}, L. and {Watson}, D. and {Yang}, S. and {Amati}, L. and {Antonelli}, L.~A. and {Ascenzi}, S. and {Bernardini}, M.~G. and {Bo{\"e}r}, M. and {Bufano}, F. and {Bulgarelli}, A. and {Capaccioli}, M. and {Casella}, P. and {Castro-Tirado}, A.~J. and {Chassande-Mottin}, E. and {Ciolfi}, R. and {Copperwheat}, C.~M. and {Dadina}, M. and {De Cesare}, G. and {di Paola}, A. and {Fan}, Y.~Z. and {Gendre}, B. and {Giuffrida}, G. and {Giunta}, A. and {Hunt}, L.~K. and {Israel}, G.~L. and {Jin}, Z.-P. and {Kasliwal}, M.~M. and {Klose}, S. and {Lisi}, M. and {Longo}, F. and {Maiorano}, E. and {Mapelli}, M. and {Masetti}, N. and {Nava}, L. and {Patricelli}, B. and {Perley}, D. and {Pescalli}, A. and {Piran}, T. and {Possenti}, A. and {Pulone}, L. and {Razzano}, M. and {Salvaterra}, R. and {Schipani}, P. and {Spera}, M. and {Stamerra}, A. and {Stella}, L. and {Tagliaferri}, G. and {Testa}, V. and {Troja}, E. and {Turatto}, M. and {Vergani}, S.~D. and {Vergani}, D.},
        title = "{Spectroscopic identification of r-process nucleosynthesis in a double neutron-star merger}",
      journal = {\nat},
         year = 2017,
        month = nov,
       volume = {551},
       number = {7678},
        pages = {67-70},
          doi = {10.1038/nature24298},
archivePrefix = {arXiv},
       eprint = {1710.05858},
 primaryClass = {astro-ph.HE},
       adsurl = {https://ui.adsabs.harvard.edu/abs/2017Natur.551...67P}
}

@phdthesis{Piotrzkowski2022,
  title  = {Searching for Gravitational Wave Associations with High-Energy Astrophysical Transients},
  author = {Piotrzkowski, B. J.},
  school = {University of Wisconsin-Milwaukee},
  year   = {2022}
}

@ARTICLE{2017ApJ...848L..15S,
       author = {{Savchenko}, V. and {Ferrigno}, C. and {Kuulkers}, E. and {Bazzano}, A. and {Bozzo}, E. and {Brandt}, S. and {Chenevez}, J. and {Courvoisier}, T.~J.-L. and {Diehl}, R. and {Domingo}, A. and {Hanlon}, L. and {Jourdain}, E. and {von Kienlin}, A. and {Laurent}, P. and {Lebrun}, F. and {Lutovinov}, A. and {Martin-Carrillo}, A. and {Mereghetti}, S. and {Natalucci}, L. and {Rodi}, J. and {Roques}, J.-P. and {Sunyaev}, R. and {Ubertini}, P.},
        title = "{INTEGRAL Detection of the First Prompt Gamma-Ray Signal Coincident with the Gravitational-wave Event GW170817}",
      journal = {\apjl},
         year = 2017,
        month = oct,
       volume = {848},
       number = {2},
          eid = {L15},
        pages = {L15},
          doi = {10.3847/2041-8213/aa8f94},
archivePrefix = {arXiv},
       eprint = {1710.05449},
 primaryClass = {astro-ph.HE},
       adsurl = {https://ui.adsabs.harvard.edu/abs/2017ApJ...848L..15S}
}

@ARTICLE{2017Natur.551...75S,
       author = {{Smartt}, S.~J. and {Chen}, T.-W. and {Jerkstrand}, A. and {Coughlin}, M. and {Kankare}, E. and {Sim}, S.~A. and {Fraser}, M. and {Inserra}, C. and {Maguire}, K. and {Chambers}, K.~C. and {Huber}, M.~E. and {Kr{\"u}hler}, T. and {Leloudas}, G. and {Magee}, M. and {Shingles}, L.~J. and {Smith}, K.~W. and {Young}, D.~R. and {Tonry}, J. and {Kotak}, R. and {Gal-Yam}, A. and {Lyman}, J.~D. and {Homan}, D.~S. and {Agliozzo}, C. and {Anderson}, J.~P. and {Angus}, C.~R. and {Ashall}, C. and {Barbarino}, C. and {Bauer}, F.~E. and {Berton}, M. and {Botticella}, M.~T. and {Bulla}, M. and {Bulger}, J. and {Cannizzaro}, G. and {Cano}, Z. and {Cartier}, R. and {Cikota}, A. and {Clark}, P. and {De Cia}, A. and {Della Valle}, M. and {Denneau}, L. and {Dennefeld}, M. and {Dessart}, L. and {Dimitriadis}, G. and {Elias-Rosa}, N. and {Firth}, R.~E. and {Flewelling}, H. and {Fl{\"o}rs}, A. and {Franckowiak}, A. and {Frohmaier}, C. and {Galbany}, L. and {Gonz{\'a}lez-Gait{\'a}n}, S. and {Greiner}, J. and {Gromadzki}, M. and {Guelbenzu}, A. Nicuesa and {Guti{\'e}rrez}, C.~P. and {Hamanowicz}, A. and {Hanlon}, L. and {Harmanen}, J. and {Heintz}, K.~E. and {Heinze}, A. and {Hernandez}, M.-S. and {Hodgkin}, S.~T. and {Hook}, I.~M. and {Izzo}, L. and {James}, P.~A. and {Jonker}, P.~G. and {Kerzendorf}, W.~E. and {Klose}, S. and {Kostrzewa-Rutkowska}, Z. and {Kowalski}, M. and {Kromer}, M. and {Kuncarayakti}, H. and {Lawrence}, A. and {Lowe}, T.~B. and {Magnier}, E.~A. and {Manulis}, I. and {Martin-Carrillo}, A. and {Mattila}, S. and {McBrien}, O. and {M{\"u}ller}, A. and {Nordin}, J. and {O'Neill}, D. and {Onori}, F. and {Palmerio}, J.~T. and {Pastorello}, A. and {Patat}, F. and {Pignata}, G. and {Podsiadlowski}, Ph. and {Pumo}, M.~L. and {Prentice}, S.~J. and {Rau}, A. and {Razza}, A. and {Rest}, A. and {Reynolds}, T. and {Roy}, R. and {Ruiter}, A.~J. and {Rybicki}, K.~A. and {Salmon}, L. and {Schady}, P. and {Schultz}, A.~S.~B. and {Schweyer}, T. and {Seitenzahl}, I.~R. and {Smith}, M. and {Sollerman}, J. and {Stalder}, B. and {Stubbs}, C.~W. and {Sullivan}, M. and {Szegedi}, H. and {Taddia}, F. and {Taubenberger}, S. and {Terreran}, G. and {van Soelen}, B. and {Vos}, J. and {Wainscoat}, R.~J. and {Walton}, N.~A. and {Waters}, C. and {Weiland}, H. and {Willman}, M. and {Wiseman}, P. and {Wright}, D.~E. and {Wyrzykowski}, {\L}. and {Yaron}, O.},
        title = "{A kilonova as the electromagnetic counterpart to a gravitational-wave source}",
      journal = {\nat},
         year = 2017,
        month = nov,
       volume = {551},
       number = {7678},
        pages = {75-79},
          doi = {10.1038/nature24303},
archivePrefix = {arXiv},
       eprint = {1710.05841},
 primaryClass = {astro-ph.HE},
       adsurl = {https://ui.adsabs.harvard.edu/abs/2017Natur.551...75S}
}

@ARTICLE{2010NJPh...12e3034S,
       author = {{Sutton}, Patrick J. and {Jones}, Gareth and {Chatterji}, Shourov and {Kalmus}, Peter and {Leonor}, Isabel and {Poprocki}, Stephen and {Rollins}, Jameson and {Searle}, Antony and {Stein}, Leo and {Tinto}, Massimo and {Was}, Michal},
        title = "{X-Pipeline: an analysis package for autonomous gravitational-wave burst searches}",
      journal = {New Journal of Physics},
         year = 2010,
        month = may,
       volume = {12},
       number = {5},
          eid = {053034},
        pages = {053034},
          doi = {10.1088/1367-2630/12/5/053034},
archivePrefix = {arXiv},
       eprint = {0908.3665},
 primaryClass = {gr-qc},
       adsurl = {https://ui.adsabs.harvard.edu/abs/2010NJPh...12e3034S}
}

@phdthesis{Urban2016,
  title  = {Monsters in the Dark: High Energy Signatures of Black Hole Formation with Multimessenger Astronomy},
  author = {Urban, A. L.},
  school = {University of Wisconsin-Milwaukee},
  year   = {2016}
}

@ARTICLE{2012PhRvD..86b2003W,
       author = {{W{\k{a}}s}, Micha{\l} and {Sutton}, Patrick J. and {Jones}, Gareth and {Leonor}, Isabel},
        title = "{Performance of an externally triggered gravitational-wave burst search}",
      journal = {\prd},
         year = 2012,
        month = jul,
       volume = {86},
       number = {2},
          eid = {022003},
        pages = {022003},
          doi = {10.1103/PhysRevD.86.022003},
archivePrefix = {arXiv},
       eprint = {1201.5599},
 primaryClass = {gr-qc},
       adsurl = {https://ui.adsabs.harvard.edu/abs/2012PhRvD..86b2003W}
}

@ARTICLE{2014PhRvD..90l2004W,
       author = {{Williamson}, A.~R. and {Biwer}, C. and {Fairhurst}, S. and {Harry}, I.~W. and {Macdonald}, E. and {Macleod}, D. and {Predoi}, V.},
        title = "{Improved methods for detecting gravitational waves associated with short gamma-ray bursts}",
      journal = {\prd},
         year = 2014,
        month = dec,
       volume = {90},
       number = {12},
          eid = {122004},
        pages = {122004},
          doi = {10.1103/PhysRevD.90.122004},
archivePrefix = {arXiv},
       eprint = {1410.6042},
 primaryClass = {gr-qc},
       adsurl = {https://ui.adsabs.harvard.edu/abs/2014PhRvD..90l2004W}
}

@ARTICLE{Rastinejad_2021,
       author = {{Rastinejad}, J.~C. and {Fong}, W. and {Kilpatrick}, C.~D. and {Paterson}, K. and {Tanvir}, N.~R. and {Levan}, A.~J. and {Metzger}, B.~D. and {Berger}, E. and {Chornock}, R. and {Cobb}, B.~E. and {Laskar}, T. and {Milne}, P. and {Nugent}, A.~E. and {Smith}, N.},
        title = "{Probing Kilonova Ejecta Properties Using a Catalog of Short Gamma-Ray Burst Observations}",
      journal = {\apj},
         year = 2021,
        month = aug,
       volume = {916},
       number = {2},
          eid = {89},
        pages = {89},
          doi = {10.3847/1538-4357/ac04b4},
archivePrefix = {arXiv},
       eprint = {2101.03175},
 primaryClass = {astro-ph.HE},
       adsurl = {https://ui.adsabs.harvard.edu/abs/2021ApJ...916...89R}
}

@article{Rastinejad_2025,
doi = {10.3847/1538-4357/ad9c77},
url = {https://doi.org/10.3847/1538-4357/ad9c77},
year = {2025},
month = {jan},
publisher = {The American Astronomical Society},
volume = {979},
number = {2},
pages = {190},
author = {Rastinejad, J. C. and Fong, W. and Kilpatrick, C. D. and Nicholl, M. and Metzger, B. D.},
title = {Uniform Modeling of Observed Kilonovae: Implications for Diversity and the Progenitors of Merger-driven Long Gamma-Ray Bursts},
journal = {The Astrophysical Journal}
}

@ARTICLE{Ashton:2019,
       author = {{Ashton}, Gregory and {H{\"u}bner}, Moritz and {Lasky}, Paul D. and others},
        title = "{BILBY: A User-friendly Bayesian Inference Library for Gravitational-wave Astronomy}",
      journal = {\apjs},
         year = 2019,
        month = apr,
       volume = {241},
       number = {2},
        pages = {27},
          doi = {10.3847/1538-4365/ab06fc},
archivePrefix = {arXiv},
       eprint = {1811.02042},
 primaryClass = {astro-ph.IM}
}

@ARTICLE{Berger2013,
       author = {{Berger}, E. and {Fong}, W. and {Chornock}, R.},
        title = "{An r-process Kilonova Associated with the Short-hard GRB 130603B}",
      journal = {\apjl},
         year = 2013,
        month = sep,
       volume = {774},
       number = {2},
          eid = {L23},
        pages = {L23},
          doi = {10.1088/2041-8205/774/2/L23},
archivePrefix = {arXiv},
       eprint = {1306.3960},
 primaryClass = {astro-ph.HE},
       adsurl = {https://ui.adsabs.harvard.edu/abs/2013ApJ...774L..23B}
}

@ARTICLE{Bulla:2019,
       author = {{Bulla}, M.},
        title = "{POSSIS: predicting spectra, light curves, and polarization for multidimensional models of supernovae and kilonovae}",
      journal = {\mnras},
         year = 2019,
        month = nov,
       volume = {489},
       number = {4},
        pages = {5037-5045},
          doi = {10.1093/mnras/stz2495},
archivePrefix = {arXiv},
       eprint = {1906.04205},
 primaryClass = {astro-ph.HE}
}

@ARTICLE{Dietrich:2020,
       author = {{Dietrich}, Tim and {Coughlin}, Michael W. and {Pang}, Peter T.~H. and {Bulla}, Mattia and {Heinzel}, Jack and {Issa}, Lina and {Tews}, Ingo and {Antier}, Sarah},
        title = "{Multimessenger constraints on the neutron-star equation of state and the Hubble constant}",
      journal = {Science},
         year = 2020,
        month = dec,
       volume = {370},
       number = {6523},
        pages = {1450-1453},
          doi = {10.1126/science.abb4317},
archivePrefix = {arXiv},
       eprint = {2002.11355},
 primaryClass = {astro-ph.HE}
}

@ARTICLE{Jin2016,
       author = {{Jin}, Zhi-Ping and {Hotokezaka}, Kenta and {Li}, Xiang and {Tanaka}, Masaomi and {D'Avanzo}, Paolo and {Fan}, Yi-Zhong and {Covino}, Stefano and {Wei}, Da-Ming and {Piran}, Tsvi},
        title = "{The Macronova in GRB 050709 and the GRB-macronova connection}",
      journal = {Nature Communications},
         year = 2016,
        month = sep,
       volume = {7},
          eid = {12898},
        pages = {12898},
          doi = {10.1038/ncomms12898},
archivePrefix = {arXiv},
       eprint = {1603.07869},
 primaryClass = {astro-ph.HE},
       adsurl = {https://ui.adsabs.harvard.edu/abs/2016NatCo...712898J}
}

@ARTICLE{Koehn:2025b,
       author = {{Koehn}, H. and {Wouters}, T. and {Pang}, P.~T.~H. and {Bulla}, M. and {Rose}, H. and {Wichern}, H. and {Dietrich}, T.},
        title = "{Efficient Bayesian analysis of kilonovae and gamma ray burst afterglows with FIESTA}",
      journal = {\aap},
         year = 2025,
        month = dec,
       volume = {704},
        pages = {A55},
          doi = {10.1051/0004-6361/202556626},
archivePrefix = {arXiv},
       eprint = {2507.13807},
 primaryClass = {astro-ph.HE}
}

@ARTICLE{Pang:2023,
       author = {{Pang}, Peter T.~H. and {Dietrich}, Tim and {Coughlin}, Michael W. and others},
        title = "{An updated nuclear-physics and multi-messenger astrophysics framework for binary neutron star mergers}",
      journal = {Nature Communications},
         year = 2023,
        month = dec,
       volume = {14},
        pages = {8352},
          doi = {10.1038/s41467-023-43932-6},
archivePrefix = {arXiv},
       eprint = {2205.08513},
 primaryClass = {astro-ph.HE}
}

@ARTICLE{Rose:2026,
       author = {{Rose}, Henrik and {Koehn}, Hauke and {Wouters}, Thibeau and {Pang}, Peter T.~H. and {Bulla}, Mattia and {Coughlin}, Michael W. and {Dietrich}, Tim},
        title = "{nmma: An extended Bayesian framework for Nuclear Multimessenger Astronomy in the Era of Next-Generation Detectors}",
      journal = {arXiv e-prints},
         year = 2026,
        month = jul,
          eid = {arXiv:2607.03045},
        pages = {arXiv:2607.03045},
          doi = {10.48550/arXiv.2607.03045},
archivePrefix = {arXiv},
       eprint = {2607.03045},
 primaryClass = {astro-ph.IM},
}

@ARTICLE{Tanaka2013,
       author = {{Tanaka}, Masaomi and {Hotokezaka}, Kenta},
        title = "{Radiative Transfer Simulations of Neutron Star Merger Ejecta}",
      journal = {\apj},
         year = 2013,
        month = oct,
       volume = {775},
       number = {2},
          eid = {113},
        pages = {113},
          doi = {10.1088/0004-637X/775/2/113},
archivePrefix = {arXiv},
       eprint = {1306.3742},
 primaryClass = {astro-ph.HE},
       adsurl = {https://ui.adsabs.harvard.edu/abs/2013ApJ...775..113T}
}

@ARTICLE{Tanvir2013,
       author = {{Tanvir}, N.~R. and {Levan}, A.~J. and {Fruchter}, A.~S. and {Hjorth}, J. and {Hounsell}, R.~A. and {Wiersema}, K. and {Tunnicliffe}, R.~L.},
        title = "{A `kilonova' associated with the short-duration {\ensuremath{\gamma}}-ray burst GRB 130603B}",
      journal = {\nat},
         year = 2013,
        month = aug,
       volume = {500},
       number = {7464},
        pages = {547-549},
          doi = {10.1038/nature12505},
archivePrefix = {arXiv},
       eprint = {1306.4971},
 primaryClass = {astro-ph.HE},
       adsurl = {https://ui.adsabs.harvard.edu/abs/2013Natur.500..547T}
}

@ARTICLE{Kasen2013,
       author = {{Kasen}, Daniel and {Badnell}, N.~R. and {Barnes}, Jennifer},
        title = "{Opacities and Spectra of the r-process Ejecta from Neutron Star Mergers}",
      journal = {\apj},
         year = 2013,
        month = sep,
       volume = {774},
       number = {1},
          eid = {25},
        pages = {25},
          doi = {10.1088/0004-637X/774/1/25},
archivePrefix = {arXiv},
       eprint = {1303.5788},
 primaryClass = {astro-ph.HE},
       adsurl = {https://ui.adsabs.harvard.edu/abs/2013ApJ...774...25K}
}

@ARTICLE{Wollaeger2018,
       author = {{Wollaeger}, Ryan T. and {Korobkin}, Oleg and {Fontes}, Christopher J. and {Rosswog}, Stephan K. and {Even}, Wesley P. and {Fryer}, Christopher L. and {Sollerman}, Jesper and {Hungerford}, Aimee L. and {van Rossum}, Daniel R. and {Wollaber}, Allan B.},
        title = "{Impact of ejecta morphology and composition on the electromagnetic signatures of neutron star mergers}",
      journal = {\mnras},
         year = 2018,
        month = aug,
       volume = {478},
       number = {3},
        pages = {3298-3334},
          doi = {10.1093/mnras/sty1018},
archivePrefix = {arXiv},
       eprint = {1705.07084},
 primaryClass = {astro-ph.HE},
       adsurl = {https://ui.adsabs.harvard.edu/abs/2018MNRAS.478.3298W}
}

@ARTICLE{Yang2015,
       author = {{Yang}, Bin and {Jin}, Zhi-Ping and {Li}, Xiang and {Covino}, Stefano and {Zheng}, Xian-Zhong and {Hotokezaka}, Kenta and {Fan}, Yi-Zhong and {Piran}, Tsvi and {Wei}, Da-Ming},
        title = "{A possible macronova in the late afterglow of the long-short burst GRB 060614}",
      journal = {Nature Communications},
         year = 2015,
        month = jun,
       volume = {6},
          eid = {7323},
        pages = {7323},
          doi = {10.1038/ncomms8323},
archivePrefix = {arXiv},
       eprint = {1503.07761},
 primaryClass = {astro-ph.HE},
       adsurl = {https://ui.adsabs.harvard.edu/abs/2015NatCo...6.7323Y}
}

@ARTICLE{Shingles2023,
       author = {{Shingles}, Luke J. and {Collins}, Christine E. and {Vijayan}, Vimal and {Fl{\"o}rs}, Andreas and {Just}, Oliver and {Leck}, Gerrit and {Xiong}, Zewei and {Bauswein}, Andreas and {Mart{\'\i}nez-Pinedo}, Gabriel and {Sim}, Stuart A.},
        title = "{Self-consistent 3D Radiative Transfer for Kilonovae: Directional Spectra from Merger Simulations}",
      journal = {\apjl},
         year = 2023,
        month = sep,
       volume = {954},
       number = {2},
          eid = {L41},
        pages = {L41},
          doi = {10.3847/2041-8213/acf29a},
archivePrefix = {arXiv},
       eprint = {2306.17612},
 primaryClass = {astro-ph.HE},
       adsurl = {https://ui.adsabs.harvard.edu/abs/2023ApJ...954L..41S}
}

@software{jones_2026_20607766,
  author       = {Jones, Lynne and
                  Bianco, Federica Bettina and
                  Yoachim, Peter and
                  Neilsen, Eric},
  title        = {https://survey-strategy.lsst.io},
  month        = jun,
  year         = 2026,
  publisher    = {Zenodo},
  version      = {v0.2.0},
  doi          = {10.5281/zenodo.20607766},
  url          = {https://doi.org/10.5281/zenodo.20607766},
}

@ARTICLE{lsst_mag,
       author = {{Ivezi{\'c}}, {\v{Z}}eljko and {Kahn}, Steven M. and {Tyson}, J. Anthony and {Abel}, Bob and {Acosta}, Emily and {Allsman}, Robyn and {Alonso}, David and {AlSayyad}, Yusra and {Anderson}, Scott F. and {Andrew}, John and {Angel}, James Roger P. and {Angeli}, George Z. and {Ansari}, Reza and {Antilogus}, Pierre and {Araujo}, Constanza and {Armstrong}, Robert and {Arndt}, Kirk T. and {Astier}, Pierre and {Aubourg}, {\'E}ric and {Auza}, Nicole and {Axelrod}, Tim S. and {Bard}, Deborah J. and {Barr}, Jeff D. and {Barrau}, Aurelian and {Bartlett}, James G. and {Bauer}, Amanda E. and {Bauman}, Brian J. and {Baumont}, Sylvain and {Bechtol}, Ellen and {Bechtol}, Keith and {Becker}, Andrew C. and {Becla}, Jacek and {Beldica}, Cristina and {Bellavia}, Steve and {Bianco}, Federica B. and {Biswas}, Rahul and {Blanc}, Guillaume and {Blazek}, Jonathan and {Blandford}, Roger D. and {Bloom}, Josh S. and {Bogart}, Joanne and {Bond}, Tim W. and {Booth}, Michael T. and {Borgland}, Anders W. and {Borne}, Kirk and {Bosch}, James F. and {Boutigny}, Dominique and {Brackett}, Craig A. and {Bradshaw}, Andrew and {Brandt}, William Nielsen and {Brown}, Michael E. and {Bullock}, James S. and {Burchat}, Patricia and {Burke}, David L. and {Cagnoli}, Gianpietro and {Calabrese}, Daniel and {Callahan}, Shawn and {Callen}, Alice L. and {Carlin}, Jeffrey L. and {Carlson}, Erin L. and {Chandrasekharan}, Srinivasan and {Charles-Emerson}, Glenaver and {Chesley}, Steve and {Cheu}, Elliott C. and {Chiang}, Hsin-Fang and {Chiang}, James and {Chirino}, Carol and {Chow}, Derek and {Ciardi}, David R. and {Claver}, Charles F. and {Cohen-Tanugi}, Johann and {Cockrum}, Joseph J. and {Coles}, Rebecca and {Connolly}, Andrew J. and {Cook}, Kem H. and {Cooray}, Asantha and {Covey}, Kevin R. and {Cribbs}, Chris and {Cui}, Wei and {Cutri}, Roc and {Daly}, Philip N. and {Daniel}, Scott F. and {Daruich}, Felipe and {Daubard}, Guillaume and {Daues}, Greg and {Dawson}, William and {Delgado}, Francisco and {Dellapenna}, Alfred and {de Peyster}, Robert and {de Val-Borro}, Miguel and {Digel}, Seth W. and {Doherty}, Peter and {Dubois}, Richard and {Dubois-Felsmann}, Gregory P. and {Durech}, Josef and {Economou}, Frossie and {Eifler}, Tim and {Eracleous}, Michael and {Emmons}, Benjamin L. and {Fausti Neto}, Angelo and {Ferguson}, Henry and {Figueroa}, Enrique and {Fisher-Levine}, Merlin and {Focke}, Warren and {Foss}, Michael D. and {Frank}, James and {Freemon}, Michael D. and {Gangler}, Emmanuel and {Gawiser}, Eric and {Geary}, John C. and {Gee}, Perry and {Geha}, Marla and {Gessner}, Charles J.~B. and {Gibson}, Robert R. and {Gilmore}, D. Kirk and {Glanzman}, Thomas and {Glick}, William and {Goldina}, Tatiana and {Goldstein}, Daniel A. and {Goodenow}, Iain and {Graham}, Melissa L. and {Gressler}, William J. and {Gris}, Philippe and {Guy}, Leanne P. and {Guyonnet}, Augustin and {Haller}, Gunther and {Harris}, Ron and {Hascall}, Patrick A. and {Haupt}, Justine and {Hernandez}, Fabio and {Herrmann}, Sven and {Hileman}, Edward and {Hoblitt}, Joshua and {Hodgson}, John A. and {Hogan}, Craig and {Howard}, James D. and {Huang}, Dajun and {Huffer}, Michael E. and {Ingraham}, Patrick and {Innes}, Walter R. and {Jacoby}, Suzanne H. and {Jain}, Bhuvnesh and {Jammes}, Fabrice and {Jee}, M. James and {Jenness}, Tim and {Jernigan}, Garrett and {Jevremovi{\'c}}, Darko and {Johns}, Kenneth and {Johnson}, Anthony S. and {Johnson}, Margaret W.~G. and {Jones}, R. Lynne and {Juramy-Gilles}, Claire and {Juri{\'c}}, Mario and {Kalirai}, Jason S. and {Kallivayalil}, Nitya J. and {Kalmbach}, Bryce and {Kantor}, Jeffrey P. and {Karst}, Pierre and {Kasliwal}, Mansi M. and {Kelly}, Heather and {Kessler}, Richard and {Kinnison}, Veronica and {Kirkby}, David and {Knox}, Lloyd and {Kotov}, Ivan V. and {Krabbendam}, Victor L. and {Krughoff}, K. Simon and {Kub{\'a}nek}, Petr and {Kuczewski}, John and {Kulkarni}, Shri and {Ku}, John and {Kurita}, Nadine R. and {Lage}, Craig S. and {Lambert}, Ron and {Lange}, Travis and {Langton}, J. Brian and {Le Guillou}, Laurent and {Levine}, Deborah and {Liang}, Ming and {Lim}, Kian-Tat and {Lintott}, Chris J. and {Long}, Kevin E. and {Lopez}, Margaux and {Lotz}, Paul J. and {Lupton}, Robert H. and {Lust}, Nate B. and {MacArthur}, Lauren A. and {Mahabal}, Ashish and {Mandelbaum}, Rachel and {Markiewicz}, Thomas W. and {Marsh}, Darren S. and {Marshall}, Philip J. and {Marshall}, Stuart and {May}, Morgan and {McKercher}, Robert and {McQueen}, Michelle and {Meyers}, Joshua and {Migliore}, Myriam and {Miller}, Michelle and {Mills}, David J.},
        title = "{LSST: From Science Drivers to Reference Design and Anticipated Data Products}",
      journal = {\apj},
         year = 2019,
        month = mar,
       volume = {873},
       number = {2},
          eid = {111},
        pages = {111},
          doi = {10.3847/1538-4357/ab042c},
archivePrefix = {arXiv},
       eprint = {0805.2366},
 primaryClass = {astro-ph},
       adsurl = {https://ui.adsabs.harvard.edu/abs/2019ApJ...873..111I}
}

@ARTICLE{RIFT,
       author = {{Lange}, Jacob and {O'Shaughnessy}, Richard and {Rizzo}, Monica},
        title = "{Rapid and accurate parameter inference for coalescing, precessing compact binaries}",
      journal = {arXiv e-prints},
         year = 2018,
        month = may,
          eid = {arXiv:1805.10457},
        pages = {arXiv:1805.10457},
          doi = {10.48550/arXiv.1805.10457},
archivePrefix = {arXiv},
       eprint = {1805.10457},
 primaryClass = {gr-qc},
       adsurl = {https://ui.adsabs.harvard.edu/abs/2018arXiv180510457L}
}

@ARTICLE{RIFT-GPU,
       author = {{Wysocki}, D. and {O'Shaughnessy}, R. and {Lange}, Jacob and {Fang}, Yao-Lung L.},
        title = "{Accelerating parameter inference with graphics processing units}",
      journal = {\prd},
         year = 2019,
        month = apr,
       volume = {99},
       number = {8},
          eid = {084026},
        pages = {084026},
          doi = {10.1103/PhysRevD.99.084026},
archivePrefix = {arXiv},
       eprint = {1902.04934},
 primaryClass = {astro-ph.IM},
       adsurl = {https://ui.adsabs.harvard.edu/abs/2019PhRvD..99h4026W}
}

@ARTICLE{Jan2020,
       author = {{Jan}, A.~Z. and {Yelikar}, A.~B. and {Lange}, J. and {O'Shaughnessy}, R.},
        title = "{Assessing and marginalizing over compact binary coalescence waveform systematics with RIFT}",
      journal = {\prd},
         year = 2020,
        month = dec,
       volume = {102},
       number = {12},
          eid = {124069},
        pages = {124069},
          doi = {10.1103/PhysRevD.102.124069},
archivePrefix = {arXiv},
       eprint = {2011.03571},
 primaryClass = {gr-qc},
       adsurl = {https://ui.adsabs.harvard.edu/abs/2020PhRvD.102l4069J}
}

@ARTICLE{Wagner2024,
       author = {{Wagner}, Katelyn J. and {O'Shaughnessy}, R.},
        title = "{Parameter estimation for low-mass eccentric black hole binaries}",
      journal = {\prd},
         year = 2024,
        month = dec,
       volume = {110},
       number = {12},
          eid = {124024},
        pages = {124024},
          doi = {10.1103/PhysRevD.110.124024},
archivePrefix = {arXiv},
       eprint = {2402.08039},
 primaryClass = {gr-qc},
       adsurl = {https://ui.adsabs.harvard.edu/abs/2024PhRvD.110l4024W}
}

@ARTICLE{RuizRocha2025,
       author = {{Ruiz-Rocha}, Krystal and {Yelikar}, Anjali B. and {Lange}, Jacob and {Gabella}, William and {Weller}, Robert A. and {O'Shaughnessy}, Richard and {Holley-Bockelmann}, Kelly and {Jani}, Karan},
        title = "{Properties of ``Lite'' Intermediate-mass Black Hole Candidates in LIGO-Virgo's Third Observing Run}",
      journal = {\apjl},
         year = 2025,
        month = jun,
       volume = {985},
       number = {2},
          eid = {L37},
        pages = {L37},
          doi = {10.3847/2041-8213/adc5f8},
archivePrefix = {arXiv},
       eprint = {2502.17681},
 primaryClass = {astro-ph.HE},
       adsurl = {https://ui.adsabs.harvard.edu/abs/2025ApJ...985L..37R}
}

@ARTICLE{Henshaw2025,
       author = {{Henshaw}, Chad and {Lange}, Jacob and {Lott}, Peter and {O'Shaughnessy}, Richard and {Cadonati}, Laura},
        title = "{Parameter estimation of gravitational waves from hyperbolic black hole encounters}",
      journal = {Classical and Quantum Gravity},
         year = 2025,
        month = dec,
       volume = {42},
       number = {24},
          eid = {24LT01},
        pages = {24LT01},
          doi = {10.1088/1361-6382/ae2415},
archivePrefix = {arXiv},
       eprint = {2412.04329},
 primaryClass = {gr-qc},
       adsurl = {https://ui.adsabs.harvard.edu/abs/2025CQGra..42L.T01H}
}

@ARTICLE{Ristic-LC,
       author = {{Risti\'c}, M. and {Champion}, E. and {O'Shaughnessy}, R. and {Wollaeger}, R. and {Korobkin}, O. and {Chase}, E.~A. and {Fryer}, C.~L. and {Hungerford}, A.~L. and {Fontes}, C.~J.},
        title = "{Interpolating detailed simulations of kilonovae: Adaptive learning and parameter inference applications}",
      journal = {Physical Review Research},
         year = 2022,
        month = jan,
       volume = {4},
       number = {1},
          eid = {013046},
        pages = {013046},
          doi = {10.1103/PhysRevResearch.4.013046},
archivePrefix = {arXiv},
       eprint = {2105.07013},
 primaryClass = {astro-ph.HE},
       adsurl = {https://ui.adsabs.harvard.edu/abs/2022PhRvR...4a3046R}
}

@ARTICLE{Kedia2023,
       author = {{Kedia}, Atul and {Risti\'c}, Marko and {O'Shaughnessy}, Richard and {Yelikar}, Anjali B. and {Wollaeger}, Ryan T. and {Korobkin}, Oleg and {Chase}, Eve A. and {Fryer}, Christopher L. and {Fontes}, Christopher J.},
        title = "{Surrogate light curve models for kilonovae with comprehensive wind ejecta outflows and parameter estimation for AT2017gfo}",
      journal = {Physical Review Research},
         year = 2023,
        month = mar,
       volume = {5},
       number = {1},
          eid = {013168},
        pages = {013168},
          doi = {10.1103/PhysRevResearch.5.013168},
archivePrefix = {arXiv},
       eprint = {2211.04363},
 primaryClass = {astro-ph.HE},
       adsurl = {https://ui.adsabs.harvard.edu/abs/2023PhRvR...5a3168K}
}

@ARTICLE{Ristic-nucleosynthesis,
       author = {{Risti{\'c}}, Marko and {Holmbeck}, Erika M. and {Wollaeger}, Ryan T. and {Korobkin}, Oleg and {Champion}, Elizabeth and {O'Shaughnessy}, Richard and {Fryer}, Chris L. and {Fontes}, Christopher J. and {Mumpower}, Matthew R. and {Sprouse}, Trevor M.},
        title = "{Constraining Inputs to Realistic Kilonova Simulations through Comparison to Observed r-process Abundances}",
      journal = {\apj},
         year = 2023,
        month = oct,
       volume = {956},
       number = {1},
          eid = {64},
        pages = {64},
          doi = {10.3847/1538-4357/acf3e0},
archivePrefix = {arXiv},
       eprint = {2206.02273},
 primaryClass = {astro-ph.HE},
       adsurl = {https://ui.adsabs.harvard.edu/abs/2023ApJ...956...64R}
}

@ARTICLE{Ristic-spectra,
       author = {{Risti{\'c}}, Marko and {O'Shaughnessy}, Richard and {Villar}, V. Ashley and {Wollaeger}, Ryan T. and {Korobkin}, Oleg and {Fryer}, Chris L. and {Fontes}, Christopher J. and {Kedia}, Atul},
        title = "{Interpolated kilonova spectra models: Examining the effects of a phenomenological, blue component in the fitting of AT2017gfo spectra}",
      journal = {Physical Review Research},
         year = 2023,
        month = nov,
       volume = {5},
       number = {4},
          eid = {043106},
        pages = {043106},
          doi = {10.1103/PhysRevResearch.5.043106},
archivePrefix = {arXiv},
       eprint = {2304.06699},
 primaryClass = {astro-ph.HE},
       adsurl = {https://ui.adsabs.harvard.edu/abs/2023PhRvR...5d3106R}
}

@ARTICLE{Ristic-LGRB,
       author = {{Risti{\'c}}, Marko and {Barker}, Brandon L. and {Cupp}, Samuel and {Gross}, Axel and {Lloyd-Ronning}, Nicole and {Korobkin}, Oleg and {Miller}, Jonah M. and {Mumpower}, Matthew R.},
        title = "{Kilonovae and Long-duration Gamma-Ray Bursts}",
      journal = {\apjl},
         year = 2026,
        month = may,
       volume = {1002},
       number = {1},
          eid = {L24},
        pages = {L24},
          doi = {10.3847/2041-8213/ae5e53},
archivePrefix = {arXiv},
       eprint = {2509.03003},
 primaryClass = {astro-ph.HE},
       adsurl = {https://ui.adsabs.harvard.edu/abs/2026ApJ..1002L..24R}
}

@ARTICLE{Ristic-GWEM,
       author = {{Risti{\'c}}, Marko and {O'Shaughnessy}, Richard and {Wagner}, Kate and {Fontes}, Christopher J. and {Fryer}, Chris L. and {Korobkin}, Oleg and {Mumpower}, Matthew R. and {Wollaeger}, Ryan T.},
        title = "{Joint electromagnetic and gravitational wave inference of binary neutron star merger GW170817 using forward-modeling ejecta predictions}",
      journal = {Physical Review Research},
         year = 2026,
        month = may,
       volume = {8},
       number = {2},
          eid = {023117},
        pages = {023117},
          doi = {10.1103/gffk-xjtg},
archivePrefix = {arXiv},
       eprint = {2503.12320},
 primaryClass = {astro-ph.HE},
       adsurl = {https://ui.adsabs.harvard.edu/abs/2026PhRvR...8b3117R}
}

@ARTICLE{Korobkin2021,
       author = {{Korobkin}, Oleg and {Wollaeger}, Ryan T. and {Fryer}, Christopher L. and {Hungerford}, Aimee L. and {Rosswog}, Stephan and {Fontes}, Christopher J. and {Mumpower}, Matthew R. and {Chase}, Eve A. and {Even}, Wesley P. and {Miller}, Jonah and {Misch}, G. Wendell and {Lippuner}, Jonas},
        title = "{Axisymmetric Radiative Transfer Models of Kilonovae}",
      journal = {\apj},
         year = 2021,
        month = apr,
       volume = {910},
       number = {2},
          eid = {116},
        pages = {116},
          doi = {10.3847/1538-4357/abe1b5},
archivePrefix = {arXiv},
       eprint = {2004.00102},
 primaryClass = {astro-ph.HE},
       adsurl = {https://ui.adsabs.harvard.edu/abs/2021ApJ...910..116K}
}

@ARTICLE{SuperNu,
       author = {{Wollaeger}, Ryan T. and {van Rossum}, Daniel R.},
        title = "{Radiation Transport for Explosive Outflows: Opacity Regrouping}",
      journal = {\apjs},
         year = 2014,
        month = oct,
       volume = {214},
       number = {2},
          eid = {28},
        pages = {28},
          doi = {10.1088/0067-0049/214/2/28},
archivePrefix = {arXiv},
       eprint = {1407.3833},
 primaryClass = {astro-ph.HE},
       adsurl = {https://ui.adsabs.harvard.edu/abs/2014ApJS..214...28W}
}

@ARTICLE{Pankow2015,
       author = {{Pankow}, C. and {Brady}, P. and {Ochsner}, E. and {O'Shaughnessy}, R.},
        title = "{Novel scheme for rapid parallel parameter estimation of gravitational waves from compact binary coalescences}",
      journal = {\prd},
         year = 2015,
        month = jul,
       volume = {92},
       number = {2},
          eid = {023002},
        pages = {023002},
          doi = {10.1103/PhysRevD.92.023002},
archivePrefix = {arXiv},
       eprint = {1502.04370},
 primaryClass = {gr-qc},
       adsurl = {https://ui.adsabs.harvard.edu/abs/2015PhRvD..92b3002P}
}

@ARTICLE{Mooley2018,
       author = {{Mooley}, K.~P. and {Deller}, A.~T. and {Gottlieb}, O. and {Nakar}, E. and {Hallinan}, G. and {Bourke}, S. and {Frail}, D.~A. and {Horesh}, A. and {Corsi}, A. and {Hotokezaka}, K.},
        title = "{Superluminal motion of a relativistic jet in the neutron-star merger GW170817}",
      journal = {\nat},
         year = 2018,
        month = sep,
       volume = {561},
       number = {7723},
        pages = {355-359},
          doi = {10.1038/s41586-018-0486-3},
archivePrefix = {arXiv},
       eprint = {1806.09693},
 primaryClass = {astro-ph.HE},
       adsurl = {https://ui.adsabs.harvard.edu/abs/2018Natur.561..355M}
}

@book{kl-div,
	address = {Hoboken, NJ},
	title = {Elements of information theory},
	isbn = {978-0-471-06259-2 978-0-471-20061-1},
	language = {en},
	publisher = {Wiley-Interscience},
	author = {Cover, Thomas M. and Thomas, Joy A.},
	year = {2001},
}

@ARTICLE{Bulla2023,
       author = {{Bulla}, Mattia},
        title = "{The critical role of nuclear heating rates, thermalization efficiencies, and opacities for kilonova modelling and parameter inference}",
      journal = {\mnras},
         year = 2023,
        month = apr,
       volume = {520},
       number = {2},
        pages = {2558-2570},
          doi = {10.1093/mnras/stad232},
archivePrefix = {arXiv},
       eprint = {2211.14348},
 primaryClass = {astro-ph.HE},
       adsurl = {https://ui.adsabs.harvard.edu/abs/2023MNRAS.520.2558B}
}

@ARTICLE{Sarin2022,
       author = {{Sarin}, Nikhil and {Omand}, Conor M.~B. and {Margalit}, Ben and {Jones}, David I.},
        title = "{On the diversity of magnetar-driven kilonovae}",
      journal = {\mnras},
         year = 2022,
        month = nov,
       volume = {516},
       number = {4},
        pages = {4949-4962},
          doi = {10.1093/mnras/stac2609},
archivePrefix = {arXiv},
       eprint = {2205.14159},
 primaryClass = {astro-ph.HE},
       adsurl = {https://ui.adsabs.harvard.edu/abs/2022MNRAS.516.4949S}
}

@ARTICLE{2024arXiv240210797C,
       author = {{Cabezas}, Alberto and {Corenflos}, Adrien and {Lao}, Junpeng and {Louf}, R{\'e}mi and {Carnec}, Antoine and {Chaudhari}, Kaustubh and {Cohn-Gordon}, Reuben and {Coullon}, Jeremie and {Deng}, Wei and {Duffield}, Sam and {Dur{\'a}n-Mart{\'\i}n}, Gerardo and {Elantkowski}, Marcin and {Foreman-Mackey}, Dan and {Gregori}, Michele and {Iguaran}, Carlos and {Kumar}, Ravin and {Lysy}, Martin and {Murphy}, Kevin and {Orduz}, Juan Camilo and {Patel}, Karm and {Wang}, Xi and {Zinkov}, Rob},
        title = "{BlackJAX: Composable Bayesian inference in JAX}",
      journal = {arXiv e-prints},
         year = 2024,
        month = feb,
          eid = {arXiv:2402.10797},
        pages = {arXiv:2402.10797},
          doi = {10.48550/arXiv.2402.10797},
archivePrefix = {arXiv},
       eprint = {2402.10797},
 primaryClass = {cs.MS},
       adsurl = {https://ui.adsabs.harvard.edu/abs/2024arXiv240210797C}
}

@ARTICLE{Sarin2024,
       author = {{Sarin}, Nikhil and {H{\"u}bner}, Moritz and {Omand}, Conor M.~B. and {Setzer}, Christian N. and {Schulze}, Steve and {Adhikari}, Naresh and {Sagu{\'e}s-Carracedo}, Ana and {Galaudage}, Shanika and {Wallace}, Wendy F. and {Lamb}, Gavin P. and {Lin}, En-Tzu},
        title = "{REDBACK: a Bayesian inference software package for electromagnetic transients}",
      journal = {\mnras},
         year = 2024,
        month = jun,
       volume = {531},
       number = {1},
        pages = {1203-1227},
          doi = {10.1093/mnras/stae1238},
archivePrefix = {arXiv},
       eprint = {2308.12806},
 primaryClass = {astro-ph.HE},
       adsurl = {https://ui.adsabs.harvard.edu/abs/2024MNRAS.531.1203S}
}

@ARTICLE{2018ApJ...855..103P,
       author = {{Piro}, Anthony L. and {Kollmeier}, Juna A.},
        title = "{Evidence for Cocoon Emission from the Early Light Curve of SSS17a}",
      journal = {\apj},
         year = 2018,
        month = mar,
       volume = {855},
       number = {2},
          eid = {103},
        pages = {103},
          doi = {10.3847/1538-4357/aaaab3},
archivePrefix = {arXiv},
       eprint = {1710.05822},
 primaryClass = {astro-ph.HE},
       adsurl = {https://ui.adsabs.harvard.edu/abs/2018ApJ...855..103P}
}

@ARTICLE{Lukosiute2022,
       author = {{Luko{\v{s}}iute}, K. and {Raaijmakers}, G. and {Doctor}, Z. and {Soares-Santos}, M. and {Nord}, B.},
        title = "{KilonovaNet: Surrogate models of kilonova spectra with conditional variational autoencoders}",
      journal = {\mnras},
         year = 2022,
        month = oct,
       volume = {516},
       number = {1},
        pages = {1137-1148},
          doi = {10.1093/mnras/stac2342},
archivePrefix = {arXiv},
       eprint = {2204.00285},
 primaryClass = {astro-ph.IM},
       adsurl = {https://ui.adsabs.harvard.edu/abs/2022MNRAS.516.1137L}
}

@ARTICLE{Kasen2006,
       author = {{Kasen}, Daniel and {Thomas}, R.~C. and {Nugent}, P.},
        title = "{Time-dependent Monte Carlo Radiative Transfer Calculations for Three-dimensional Supernova Spectra, Light Curves, and Polarization}",
      journal = {\apj},
         year = 2006,
        month = nov,
       volume = {651},
       number = {1},
        pages = {366-380},
          doi = {10.1086/506190},
archivePrefix = {arXiv},
       eprint = {astro-ph/0606111},
 primaryClass = {astro-ph},
       adsurl = {https://ui.adsabs.harvard.edu/abs/2006ApJ...651..366K}
}

@ARTICLE{2015ApJ...815...82L,
       author = {{Lippuner}, Jonas and {Roberts}, Luke F.},
        title = "{r-process Lanthanide Production and Heating Rates in Kilonovae}",
      journal = {\apj},
         year = 2015,
        month = dec,
       volume = {815},
       number = {2},
          eid = {82},
        pages = {82},
          doi = {10.1088/0004-637X/815/2/82},
archivePrefix = {arXiv},
       eprint = {1508.03133},
 primaryClass = {astro-ph.HE},
       adsurl = {https://ui.adsabs.harvard.edu/abs/2015ApJ...815...82L}
}

@ARTICLE{Sarin2024b,
       author = {{Sarin}, Nikhil and {Rosswog}, Stephan},
        title = "{Cautionary Tales on Heating-rate Prescriptions in Kilonovae}",
      journal = {\apjl},
         year = 2024,
        month = sep,
       volume = {973},
       number = {1},
          eid = {L24},
        pages = {L24},
          doi = {10.3847/2041-8213/ad739d},
archivePrefix = {arXiv},
       eprint = {2404.07271},
 primaryClass = {astro-ph.HE},
       adsurl = {https://ui.adsabs.harvard.edu/abs/2024ApJ...973L..24S}
}

@ARTICLE{Brethauer2024,
       author = {{Brethauer}, D. and {Kasen}, D. and {Margutti}, R. and {Chornock}, R.},
        title = "{Impact of Systematic Modeling Uncertainties on Kilonova Property Estimation}",
      journal = {\apj},
         year = 2024,
        month = nov,
       volume = {975},
       number = {2},
          eid = {213},
        pages = {213},
          doi = {10.3847/1538-4357/ad7d83},
archivePrefix = {arXiv},
       eprint = {2408.02731},
 primaryClass = {astro-ph.HE},
       adsurl = {https://ui.adsabs.harvard.edu/abs/2024ApJ...975..213B}
}

@article{Palmese_2017,
doi = {10.3847/2041-8213/aa9660},
url = {https://doi.org/10.3847/2041-8213/aa9660},
year = {2017},
month = {nov},
publisher = {The American Astronomical Society},
volume = {849},
number = {2},
pages = {L34},
author = {Palmese, A. and Hartley, W. and Tarsitano, F. and Conselice, C. and Lahav, O. and Allam, S. and Annis, J. and Lin, H. and Soares-Santos, M. and Tucker, D. and Brout, D. and Banerji, M. and Bechtol, K. and Diehl, H. T. and Fruchter, A. and García-Bellido, J. and Herner, K. and Levan, A. J. and Li, T. S. and Lidman, C. and Misra, K. and Sako, M. and Scolnic, D. and Smith, M. and Abbott, T. M. C. and Abdalla, F. B. and Benoit-Lévy, A. and Bertin, E. and Brooks, D. and Buckley-Geer, E. and Rosell, A. Carnero and Kind, M. Carrasco and Carretero, J. and Castander, F. J. and Cunha, C. E. and D’Andrea, C. B. and da Costa, L. N. and Davis, C. and DePoy, D. L. and Desai, S. and Dietrich, J. P. and Doel, P. and Drlica-Wagner, A. and Eifler, T. F. and Evrard, A. E. and Flaugher, B. and Fosalba, P. and Frieman, J. and Gaztanaga, E. and Gerdes, D. W. and Giannantonio, T. and Gruen, D. and Gruendl, R. A. and Gschwend, J. and Gutierrez, G. and Honscheid, K. and Jain, B. and James, D. J. and Jeltema, T. and Johnson, M. W. G. and Johnson, M. D. and Krause, E. and Kron, R. and Kuehn, K. and Kuhlmann, S. and Kuropatkin, N. and Lima, M. and Maia, M. A. G. and March, M. and Marshall, J. L. and McMahon, R. G. and Menanteau, F. and Miller, C. J. and Miquel, R. and Neilsen, E. and Ogando, R. L. C. and Plazas, A. A. and Reil, K. and Romer, A. K. and Sanchez, E. and Schindler, R. and Smith, R. C. and Sobreira, F. and Suchyta, E. and Swanson, M. E. C. and Tarle, G. and Thomas, D. and Thomas, R. C. and Walker, A. R. and Weller, J. and Zhang, Y. and Zuntz, J.},
title = {Evidence for Dynamically Driven Formation of the GW170817 Neutron Star Binary in NGC 4993},
journal = {The Astrophysical Journal Letters}}

@ARTICLE{2025A&A...703A.101G,
       author = {{Guidorzi}, C. and {Maccary}, R. and {Maistrello}, M. and {Kobayashi}, S. and {Bulla}, M. and {Frontera}, F.},
        title = "{Gamma-ray burst X-ray plateaus as evidence of pre-prompt afterglow}",
      journal = {\aap},
         year = 2025,
        month = nov,
       volume = {703},
          eid = {A101},
        pages = {A101},
          doi = {10.1051/0004-6361/202556663},
archivePrefix = {arXiv},
       eprint = {2509.18996},
 primaryClass = {astro-ph.HE},
       adsurl = {https://ui.adsabs.harvard.edu/abs/2025A&A...703A.101G}
}

@ARTICLE{Koehn2026,
       author = {{Koehn}, Hauke and {Wouters}, Thibeau and {Sadeh}, Gilad and {Pang}, Peter T.~H. and {Bulla}, Mattia and {Van Den Broeck}, Chris and {Coughlin}, Michael W. and {Dietrich}, Tim},
        title = "{Binary neutron stars in the next-generation era: Multi-messenger detection prospects and constraints on the equation of state, mass distribution, and cosmology}",
      journal = {arXiv e-prints},
         year = 2026,
        month = jul,
          eid = {arXiv:2607.28438},
        pages = {arXiv:2607.28438},
archivePrefix = {arXiv},
       eprint = {2607.28438},
 primaryClass = {astro-ph.HE},
       adsurl = {https://ui.adsabs.harvard.edu/abs/2026arXiv260728438K}
}

@article{Schlafly_2011,
doi = {10.1088/0004-637X/737/2/103},
url = {https://doi.org/10.1088/0004-637X/737/2/103},
year = {2011},
month = {aug},
publisher = {The American Astronomical Society},
volume = {737},
number = {2},
pages = {103},
author = {Schlafly, Edward F. and Finkbeiner, Douglas P.},
title = {MEASURING REDDENING WITH SLOAN DIGITAL SKY SURVEY STELLAR SPECTRA AND RECALIBRATING SFD},
journal = {The Astrophysical Journal}
}

@article{Planck2018,
   title={<i>Planck</i>
                    2018 results: VI. Cosmological parameters},
   volume={641},
   ISSN={1432-0746},
   url={http://dx.doi.org/10.1051/0004-6361/201833910},
   DOI={10.1051/0004-6361/201833910},
   journal={Astronomy \&amp; Astrophysics},
   publisher={EDP Sciences},
   author={ Aghanim, N. and Akrami, Y. and Ashdown, M. and Aumont, J. and Baccigalupi, C. and Ballardini, M. and Banday, A. J. and Barreiro, R. B. and Bartolo, N. and Basak, S. and Battye, R. and Benabed, K. and Bernard, J.-P. and Bersanelli, M. and Bielewicz, P. and Bock, J. J. and Bond, J. R. and Borrill, J. and Bouchet, F. R. and Boulanger, F. and Bucher, M. and Burigana, C. and Butler, R. C. and Calabrese, E. and Cardoso, J.-F. and Carron, J. and Challinor, A. and Chiang, H. C. and Chluba, J. and Colombo, L. P. L. and Combet, C. and Contreras, D. and Crill, B. P. and Cuttaia, F. and de Bernardis, P. and de Zotti, G. and Delabrouille, J. and Delouis, J.-M. and Di Valentino, E. and Diego, J. M. and Doré, O. and Douspis, M. and Ducout, A. and Dupac, X. and Dusini, S. and Efstathiou, G. and Elsner, F. and Enßlin, T. A. and Eriksen, H. K. and Fantaye, Y. and Farhang, M. and Fergusson, J. and Fernandez-Cobos, R. and Finelli, F. and Forastieri, F. and Frailis, M. and Fraisse, A. A. and Franceschi, E. and Frolov, A. and Galeotta, S. and Galli, S. and Ganga, K. and Génova-Santos, R. T. and Gerbino, M. and Ghosh, T. and González-Nuevo, J. and Górski, K. M. and Gratton, S. and Gruppuso, A. and Gudmundsson, J. E. and Hamann, J. and Handley, W. and Hansen, F. K. and Herranz, D. and Hildebrandt, S. R. and Hivon, E. and Huang, Z. and Jaffe, A. H. and Jones, W. C. and Karakci, A. and Keihänen, E. and Keskitalo, R. and Kiiveri, K. and Kim, J. and Kisner, T. S. and Knox, L. and Krachmalnicoff, N. and Kunz, M. and Kurki-Suonio, H. and Lagache, G. and Lamarre, J.-M. and Lasenby, A. and Lattanzi, M. and Lawrence, C. R. and Le Jeune, M. and Lemos, P. and Lesgourgues, J. and Levrier, F. and Lewis, A. and Liguori, M. and Lilje, P. B. and Lilley, M. and Lindholm, V. and López-Caniego, M. and Lubin, P. M. and Ma, Y.-Z. and Macías-Pérez, J. F. and Maggio, G. and Maino, D. and Mandolesi, N. and Mangilli, A. and Marcos-Caballero, A. and Maris, M. and Martin, P. G. and Martinelli, M. and Martínez-González, E. and Matarrese, S. and Mauri, N. and McEwen, J. D. and Meinhold, P. R. and Melchiorri, A. and Mennella, A. and Migliaccio, M. and Millea, M. and Mitra, S. and Miville-Deschênes, M.-A. and Molinari, D. and Montier, L. and Morgante, G. and Moss, A. and Natoli, P. and Nørgaard-Nielsen, H. U. and Pagano, L. and Paoletti, D. and Partridge, B. and Patanchon, G. and Peiris, H. V. and Perrotta, F. and Pettorino, V. and Piacentini, F. and Polastri, L. and Polenta, G. and Puget, J.-L. and Rachen, J. P. and Reinecke, M. and Remazeilles, M. and Renzi, A. and Rocha, G. and Rosset, C. and Roudier, G. and Rubiño-Martín, J. A. and Ruiz-Granados, B. and Salvati, L. and Sandri, M. and Savelainen, M. and Scott, D. and Shellard, E. P. S. and Sirignano, C. and Sirri, G. and Spencer, L. D. and Sunyaev, R. and Suur-Uski, A.-S. and Tauber, J. A. and Tavagnacco, D. and Tenti, M. and Toffolatti, L. and Tomasi, M. and Trombetti, T. and Valenziano, L. and Valiviita, J. and Van Tent, B. and Vibert, L. and Vielva, P. and Villa, F. and Vittorio, N. and Wandelt, B. D. and Wehus, I. K. and White, M. and White, S. D. M. and Zacchei, A. and Zonca, A.},
   year={2020},
   month=Sept, pages={A6} }

@article{Skilling:2004pqw,
    author = "Skilling, John",
    title = "{Nested Sampling}",
    doi = "10.1063/1.1835238",
    journal = "AIP Conf. Proc.",
    volume = "735",
    number = "1",
    pages = "395",
    year = "2004"
}

@article{Skilling:2006gxv,
    author = "Skilling, John",
    title = "{Nested sampling for general Bayesian computation}",
    doi = "10.1214/06-BA127",
    journal = "Bayesian Analysis",
    volume = "1",
    number = "4",
    pages = "833--859",
    year = "2006"
}

@article{Feroz:2008xx,
    author = "Feroz, F. and Hobson, M. P. and Bridges, M.",
    title = "{MultiNest: an efficient and robust Bayesian inference tool for cosmology and particle physics}",
    eprint = "0809.3437",
    archivePrefix = "arXiv",
    primaryClass = "astro-ph",
    doi = "10.1111/j.1365-2966.2009.14548.x",
    journal = "Mon. Not. Roy. Astron. Soc.",
    volume = "398",
    pages = "1601--1614",
    year = "2009"
}

@ARTICLE{PyMultinest,
       author = {{Buchner}, J. and {Georgakakis}, A. and {Nandra}, K. and {Hsu}, L. and {Rangel}, C. and {Brightman}, M. and {Merloni}, A. and {Salvato}, M. and {Donley}, J. and {Kocevski}, D.},
        title = "{X-ray spectral modelling of the AGN obscuring region in the CDFS: Bayesian model selection and catalogue}",
      journal = {\aap},
         year = 2014,
        month = apr,
       volume = {564},
          eid = {A125},
        pages = {A125},
          doi = {10.1051/0004-6361/201322971},
archivePrefix = {arXiv},
       eprint = {1402.0004},
 primaryClass = {astro-ph.HE},
       adsurl = {https://ui.adsabs.harvard.edu/abs/2014A&A...564A.125B}
}

@article{Speagle:2019ivv,
    author = "Speagle, Joshua S.",
    title = "{dynesty: a dynamic nested sampling package for estimating Bayesian posteriors and evidences}",
    eprint = "1904.02180",
    archivePrefix = "arXiv",
    primaryClass = "astro-ph.IM",
    doi = "10.1093/mnras/staa278",
    journal = "Mon. Not. Roy. Astron. Soc.",
    volume = "493",
    number = "3",
    pages = "3132--3158",
    year = "2020"
}

@article{Romero-Shaw:2020owr,
    author = "Romero-Shaw, I. M. and others",
    title = "{Bayesian inference for compact binary coalescences with bilby: validation and application to the first LIGO{\textendash}Virgo gravitational-wave transient catalogue}",
    eprint = "2006.00714",
    archivePrefix = "arXiv",
    primaryClass = "astro-ph.IM",
    doi = "10.1093/mnras/staa2850",
    journal = "Mon. Not. Roy. Astron. Soc.",
    volume = "499",
    number = "3",
    pages = "3295--3319",
    year = "2020"
}

@article{Karamanis:2022ksp,
    author = "Karamanis, Minas and Nabergoj, David and Beutler, Florian and Peacock, John A. and Seljak, Uros",
    title = "{pocoMC: A Python package for accelerated Bayesian inference in astronomy and cosmology}",
    eprint = "2207.05660",
    archivePrefix = "arXiv",
    primaryClass = "astro-ph.IM",
    doi = "10.21105/joss.04634",
    journal = "J. Open Source Softw.",
    volume = "7",
    number = "79",
    pages = "4634",
    year = "2022"
}

@article{del2006sequential,
  title={Sequential monte carlo samplers},
  author={Del Moral, Pierre and Doucet, Arnaud and Jasra, Ajay},
  journal={Journal of the Royal Statistical Society Series B: Statistical Methodology},
  volume={68},
  number={3},
  pages={411--436},
  year={2006},
  publisher={Oxford University Press}
}

@book{chopin2020introduction,
  title     = {An Introduction to Sequential Monte Carlo},
  author    = {Chopin, Nicolas and Papaspiliopoulos, Omiros},
  year      = {2020},
  publisher = {Springer Cham},
  series    = {Springer Series in Statistics},
  isbn      = {978-3-030-47844-5},
  doi       = {10.1007/978-3-030-47845-2},
  url       = {https://doi.org/10.1007/978-3-030-47845-2},
}

@article{Karamanis:2022alw,
    author = "Karamanis, Minas and Beutler, Florian and Peacock, John A. and Nabergoj, David and Seljak, Uros",
    title = "{Accelerating astronomical and cosmological inference with preconditioned Monte Carlo}",
    eprint = "2207.05652",
    archivePrefix = "arXiv",
    primaryClass = "astro-ph.IM",
    doi = "10.1093/mnras/stac2272",
    journal = "Mon. Not. Roy. Astron. Soc.",
    volume = "516",
    number = "2",
    pages = "1644--1653",
    year = "2022"
}

@article{Williams:2025szm,
    author = "Williams, Michael J. and Karamanis, Minas and Luo, Yilin and Seljak, Uro{\v{s}}",
    title = "{Validating sequential Monte Carlo for gravitational-wave inference}",
    eprint = "2506.18977",
    archivePrefix = "arXiv",
    primaryClass = "astro-ph.IM",
    reportNumber = "LIGO-P2500231",
    doi = "10.1093/mnras/staf1458",
    journal = "Mon. Not. Roy. Astron. Soc.",
    volume = "543",
    number = "2",
    pages = "1479--1493",
    year = "2025"
}

@ARTICLE{EOS_paper,
  title     = "Constraining neutron-star matter with microscopic and
               macroscopic collisions",
  author    = "Huth, Sabrina and Pang, Peter T H and Tews, Ingo and Dietrich,
               Tim and Le F{\`e}vre, Arnaud and Schwenk, Achim and Trautmann,
               Wolfgang and Agarwal, Kshitij and Bulla, Mattia and Coughlin,
               Michael W and Van Den Broeck, Chris",
  journal   = "Nature",
  publisher = "Springer Science and Business Media LLC",
  volume    =  606,
  number    =  7913,
  pages     = "276--280",
  month     =  jun,
  year      =  2022,
  copyright = "https://creativecommons.org/licenses/by/4.0",
  language  = "en"
}

@article{Wollaeger_2021,
doi = {10.3847/1538-4357/ac0d03},
url = {https://doi.org/10.3847/1538-4357/ac0d03},
year = {2021},
month = {aug},
publisher = {The American Astronomical Society},
volume = {918},
number = {1},
pages = {10},
author = {Wollaeger, R. T. and Fryer, C. L. and Chase, E. A. and Fontes, C. J. and Ristic, M. and Hungerford, A. L. and Korobkin, O. and O’Shaughnessy, R. and Herring, A. M.},
title = {A Broad Grid of 2D Kilonova Emission Models},
journal = {The Astrophysical Journal}
}

@article{formuladyn,
  title = {Estimates for disk and ejecta masses produced in compact binary mergers},
  author = {Kr\"uger, Christian J. and Foucart, Francois},
  journal = {Phys. Rev. D},
  volume = {101},
  issue = {10},
  pages = {103002},
  numpages = {11},
  year = {2020},
  month = {May},
  publisher = {American Physical Society},
  doi = {10.1103/PhysRevD.101.103002},
  url = {https://link.aps.org/doi/10.1103/PhysRevD.101.103002}
}

@ARTICLE{2020SciPy-NMeth,
  author  = {Virtanen, Pauli and Gommers, Ralf and Oliphant, Travis E. and
            Haberland, Matt and Reddy, Tyler and Cournapeau, David and
            Burovski, Evgeni and Peterson, Pearu and Weckesser, Warren and
            Bright, Jonathan and {van der Walt}, St{\'e}fan J. and
            Brett, Matthew and Wilson, Joshua and Millman, K. Jarrod and
            Mayorov, Nikolay and Nelson, Andrew R. J. and Jones, Eric and
            Kern, Robert and Larson, Eric and Carey, C J and
            Polat, {\.I}lhan and Feng, Yu and Moore, Eric W. and
            {VanderPlas}, Jake and Laxalde, Denis and Perktold, Josef and
            Cimrman, Robert and Henriksen, Ian and Quintero, E. A. and
            Harris, Charles R. and Archibald, Anne M. and
            Ribeiro, Ant{\^o}nio H. and Pedregosa, Fabian and
            {van Mulbregt}, Paul and {SciPy 1.0 Contributors}},
  title   = {{{SciPy} 1.0: Fundamental Algorithms for Scientific
            Computing in Python}},
  journal = {Nature Methods},
  year    = {2020},
  volume  = {17},
  pages   = {261--272},
  adsurl  = {https://rdcu.be/b08Wh},
  doi     = {10.1038/s41592-019-0686-2},
}

@ARTICLE{2013PASP..125..306F,
       author = {{Foreman-Mackey}, Daniel and {Hogg}, David W. and {Lang}, Dustin and {Goodman}, Jonathan},
        title = "{emcee: The MCMC Hammer}",
      journal = {\pasp},
         year = 2013,
        month = mar,
       volume = {125},
       number = {925},
        pages = {306},
          doi = {10.1086/670067},
archivePrefix = {arXiv},
       eprint = {1202.3665},
 primaryClass = {astro-ph.IM},
       adsurl = {https://ui.adsabs.harvard.edu/abs/2013PASP..125..306F}
}

@ARTICLE{Heinzel21,
       author = {{Heinzel}, J. and {Coughlin}, M.~W. and {Dietrich}, T. and {Bulla}, M. and {Antier}, S. and {Christensen}, N. and {Coulter}, D.~A. and {Foley}, R.~J. and {Issa}, L. and {Khetan}, N.},
        title = "{Comparing inclination-dependent analyses of kilonova transients}",
      journal = {\mnras},
         year = 2021,
        month = apr,
       volume = {502},
       number = {2},
        pages = {3057-3065},
          doi = {10.1093/mnras/stab221},
archivePrefix = {arXiv},
       eprint = {2010.10746},
 primaryClass = {astro-ph.HE},
       adsurl = {https://ui.adsabs.harvard.edu/abs/2021MNRAS.502.3057H}
}

@ARTICLE{Arcavi2018,
       author = {{Arcavi}, Iair},
        title = "{The First Hours of the GW170817 Kilonova and the Importance of Early Optical and Ultraviolet Observations for Constraining Emission Models}",
      journal = {\apjl},
         year = 2018,
        month = mar,
       volume = {855},
       number = {2},
          eid = {L23},
        pages = {L23},
          doi = {10.3847/2041-8213/aab267},
archivePrefix = {arXiv},
       eprint = {1802.02164},
 primaryClass = {astro-ph.HE},
       adsurl = {https://ui.adsabs.harvard.edu/abs/2018ApJ...855L..23A}
}

@ARTICLE{Kasliwal2017,
       author = {{Kasliwal}, M.~M. and {Nakar}, E. and {Singer}, L.~P. and {Kaplan}, D.~L. and {Cook}, D.~O. and {Van Sistine}, A. and {Lau}, R.~M. and {Fremling}, C. and {Gottlieb}, O. and {Jencson}, J.~E. and {Adams}, S.~M. and {Feindt}, U. and {Hotokezaka}, K. and {Ghosh}, S. and {Perley}, D.~A. and {Yu}, P.-C. and {Piran}, T. and {Allison}, J.~R. and {Anupama}, G.~C. and {Balasubramanian}, A. and {Bannister}, K.~W. and {Bally}, J. and {Barnes}, J. and {Barway}, S. and {Bellm}, E. and {Bhalerao}, V. and {Bhattacharya}, D. and {Blagorodnova}, N. and {Bloom}, J.~S. and {Brady}, P.~R. and {Cannella}, C. and {Chatterjee}, D. and {Cenko}, S.~B. and {Cobb}, B.~E. and {Copperwheat}, C. and {Corsi}, A. and {De}, K. and {Dobie}, D. and {Emery}, S.~W.~K. and {Evans}, P.~A. and {Fox}, O.~D. and {Frail}, D.~A. and {Frohmaier}, C. and {Goobar}, A. and {Hallinan}, G. and {Harrison}, F. and {Helou}, G. and {Hinderer}, T. and {Ho}, A.~Y.~Q. and {Horesh}, A. and {Ip}, W.-H. and {Itoh}, R. and {Kasen}, D. and {Kim}, H. and {Kuin}, N.~P.~M. and {Kupfer}, T. and {Lynch}, C. and {Madsen}, K. and {Mazzali}, P.~A. and {Miller}, A.~A. and {Mooley}, K. and {Murphy}, T. and {Ngeow}, C.-C. and {Nichols}, D. and {Nissanke}, S. and {Nugent}, P. and {Ofek}, E.~O. and {Qi}, H. and {Quimby}, R.~M. and {Rosswog}, S. and {Rusu}, F. and {Sadler}, E.~M. and {Schmidt}, P. and {Sollerman}, J. and {Steele}, I. and {Williamson}, A.~R. and {Xu}, Y. and {Yan}, L. and {Yatsu}, Y. and {Zhang}, C. and {Zhao}, W.},
        title = "{Illuminating gravitational waves: A concordant picture of photons from a neutron star merger}",
      journal = {Science},
         year = 2017,
        month = dec,
       volume = {358},
       number = {6370},
        pages = {1559-1565},
          doi = {10.1126/science.aap9455},
archivePrefix = {arXiv},
       eprint = {1710.05436},
 primaryClass = {astro-ph.HE},
       adsurl = {https://ui.adsabs.harvard.edu/abs/2017Sci...358.1559K}
}

@article{stevenson_strategy_2026,
	title = {Strategy for {Identifying} {Vera} {C}. {Rubin} {Observatory} {Kilonova} {Candidates} for {Targeted} {Gravitational}-wave {Searches}},
	volume = {998},
	issn = {0004-637X},
	url = {https://doi.org/10.3847/1538-4357/ae244e},
	doi = {10.3847/1538-4357/ae244e},
	language = {en},
	number = {1},
	urldate = {2026-10-01},
	journal = {The Astrophysical Journal},
	publisher = {The American Astronomical Society},
	author = {Stevenson, Simon and Möller, Anais and Powell, Jade},
	month = feb,
	year = {2026},
	pages = {8},
}

@article{rochiniTDR,
	author = {{Ronchini, S.} and {Chopra, A.} and {Dal Canton, T.} and {Banerjee, B.} and {De Santis, A. L.} and {Branchesi, M.}},
	title = {Gravitational-wave detectability range informed by external messengers},
	DOI= "10.1051/0004-6361/202661039",
	url= "https://doi.org/10.1051/0004-6361/202661039",
	journal = {A\&A},
	year = 2026,
	volume = 714,
	pages = "A25",
}

\begin{appendix} 

\section{Priors used for the AT2017gfo analyses}
\label{sec:priors}

This section includes tables with parameter priors for the different kilonova models used in the AT2017gfo analyzes of Sec.~\ref{sec:170817}. Table~\ref{tab:NMMA2017prior} refers to the \texttt{Bu2019lm} and \texttt{Bu2026\_MLP} grids, Table~\ref{tab:Bret26_priors} to the \texttt{Brethauer26} grid, Table~\ref{tab:Ka2017_priors} to the \texttt{Ka2017} grid and Table~\ref{tab:LANL_priors} to the \texttt{LANL} grid.

\begin{table}[ht]
\centering
\begin{tabular}{llcl}
\hline
\hline
\textbf{Parameter} & \textbf{Distribution} & \textbf{Range / Bounds} & \textbf{Model} \\ \hline
\hline
$\Phi$ & Uniform & ${15 - 75}$ [$^{\circ}$] & \verb|Bu2019lm|\\
$\log_{10}M^{\rm{dyn}}_{\rm{ej}}$ & Uniform & ${-3, -1}$ & \verb|Bu2019lm|\\
$\log_{10}M^{\rm{wind}}_{\rm{ej}}$ & Uniform & ${-3, -0.5}$ & \verb|Bu2019lm|\\

$v_\mathrm{ej}^\mathrm{dyn}$ & Uniform & ${0.12 - 0.35}\,\,[c]$ & \verb|Bu2026_MLP| \\
$v_\mathrm{ej}^\mathrm{wind}$ & Uniform  & ${0.05 - 0.15}\,\,[c]$ & \verb|Bu2026_MLP| \\ $Y_e^\mathrm{dyn}$ & Uniform  & ${0.15 - 0.35}$ & \verb|Bu2026_MLP| \\
$Y_e^\mathrm{wind}$ & Uniform & ${0.2 - 0.4}$ & \verb|Bu2026_MLP| \\
$\log_{10}M^{\rm{dyn}}_{\rm{ej}}$ & Uniform  & ${-4, -1.302}$ & \verb|Bu2026_MLP|\\
$\log_{10}M^{\rm{wind}}_{\rm{ej}}$ & Uniform  & ${-4, -0.56}$ & \verb|Bu2026_MLP|\\
$\theta$ & Sine & ${0 - \displaystyle \pi/2} \,\, [\text{rad}]$ & \rm{both} \\
$D_L$ & Fixed & $40.7\,\,[\text{Mpc}]$ & \rm{both}\\
$t_0$ & Uniform & ${-2 - 0}\,\,[\text{days]}$ & \rm{both}\\ \hline
\hline
\end{tabular}
\caption{Priors used for the different parameters of \texttt{Bu2019lm} and \texttt{Bu2026\_MLP} 
to analyse AT2017gfo with {\sc nmma}. $M^{\rm{dyn,wind}}_{\rm{ej}}$,  $v^{\rm{dyn,wind}}_{\rm{ej}}$ and $Y_e^\mathrm{dyn, wind}$ are the dynamical and disk wind ejecta mass, velocity and electron fraction respectively. $\theta$ is the viewing angle between the source plan and the observer line of sight. $\phi$ is the half opening angle or the dynamical ejecta. $D_L$ and $t_0$ are the luminosity distance and the timeshift between the merger time and the first detection.}
\label{tab:NMMA2017prior}
\end{table}

\begin{table}[ht]
\centering
\begin{tabular}{llcl}
\hline
\hline
\textbf{Parameter} & \textbf{Distribution} & \textbf{Range / Bounds}\\ \hline
\hline
$m_{\rm{ej}}$ & Uniform & $0.001,0.1\,\,[M_{\odot}]$ \\
$v_{\rm{ej}}$ & Uniform & $0.05, 0.3\,\,[c]$  \\
$X_{\rm{lan}}$ & Uniform & $10^{-9},10^{-2}$\\
$n_{\rm outer}$ & Uniform & 6, 10 \\
$D_L$ & Fixed & $40.7\,\,[\text{Mpc}]$ & \\
$t_0$ & Uniform & ${-2, 0}\,\,[\text{days]}$ \\ \hline
\hline
\end{tabular}
\caption{Priors used for the different parameters of \texttt{Brethauer26}}
\label{tab:Bret26_priors}
\end{table}

\begin{table}[ht]
\centering
\begin{tabular}{llcl}
\hline
\hline
\textbf{Parameter} & \textbf{Distribution} & \textbf{Range / Bounds}\\ \hline
\hline
$m_{\rm{ej}}(\rm red)$ & Uniform & $0.001,0.1\,\,[M_{\odot}]$ \\
$v_{\rm{ej}}(\rm red)$ & Uniform & $0.05, 0.3\,\,[c]$  \\
$X_{\rm{lan}}(\rm red)$ & Uniform & $10^{-2},10^{-1}$\\
$m_{\rm{ej}}(\rm blue)$ & Uniform & $0.001,0.1\,\,[M_{\odot}]$ \\
$v_{\rm{ej}}(\rm blue)$ & Uniform &   $0.05,0.3\,\,[c]$\\
$X_{\rm{lan}}(\rm blue)$ & Uniform & $10^{-5},10^{-2}$ \\
$v_{\rm{ej}}(\rm{blue})-v_{\rm{ej}}(\rm red)$ & Constraint & $0,0.29\,\,[c]$ \\
$D_L$ & Fixed & $40.7\,\,[\text{Mpc}]$ & \\
$t_0$ & Uniform & ${-2, 0}\,\,[\text{days]}$ \\ \hline
\hline
\end{tabular}
\caption{Priors used for the different parameters of \texttt{Ka2017}}
\label{tab:Ka2017_priors}
\end{table}

\begin{table}[ht]
\centering
\begin{tabular}{llcl}
\hline
\hline
\textbf{Parameter} & \textbf{Distribution} & \textbf{Range / Bounds}\\ \hline
\hline
$m_d$ & Uniform & $0.001,0.1\,\,[M_{\odot}]$ \\
$v_d$ & Uniform & $0.05, 0.3\,\,[c]$  \\
$m_w$ & Uniform & $0.001,0.1\,\,[M_{\odot}]$ \\
$v_w$ & Uniform &   $0.05,0.3\,\,[c]$\\
$\theta$ & $\mathcal{N}(\mu=20, \sigma=5)$ & $0, 90\,\,(\text{deg})$ \\
$\sigma_{\rm{sys}}$ & Uniform & $0, 2\,\,(\text{mag})$ \\
$t_0$ & Uniform & ${-2, 0}\,\,[\text{days]}$ \\
$D_L$ & Fixed & $40\,\,[\text{Mpc}]$ & \\ \hline
\hline
\end{tabular}
\caption{Priors used for the different parameters of the \texttt{LANL} models.}
\label{tab:LANL_priors}
\end{table}

\section{Best-fit light curves to AT2017gfo}
\label{sec:bestfitlc}

\begin{figure}
    \centering
    \includegraphics[width=0.9\linewidth]{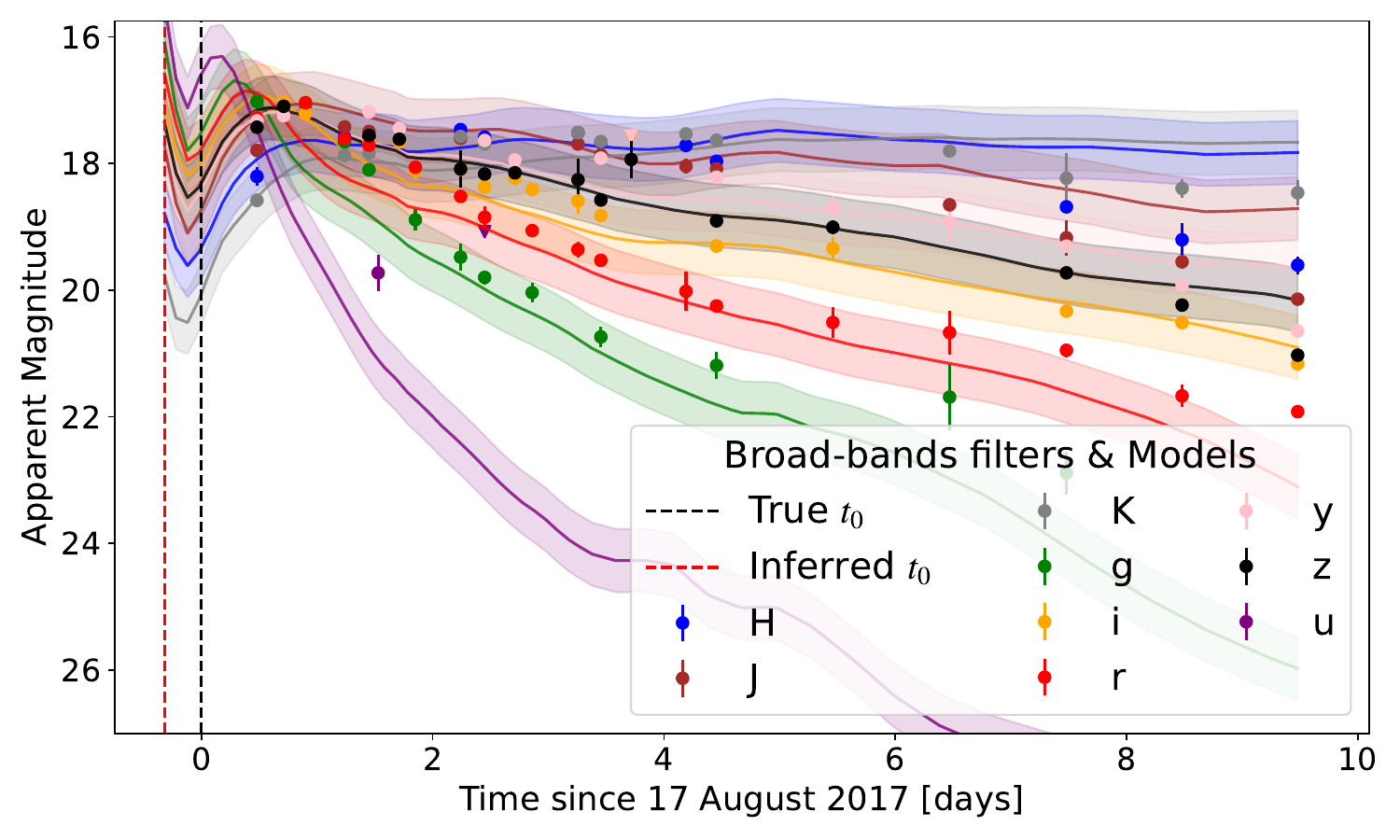}
    \caption{Best-fit light curves in $ugrizyJHK$ obtained with {\sc nmma} using \texttt{Bu2026\_MLP}. The black dotted line indicates the true $t_0$ while the red dotted line shows the inferred $t_0$.}
    \label{fig:nmma_lc}
\end{figure}

\begin{figure}
    \centering
    \includegraphics[width=\linewidth]{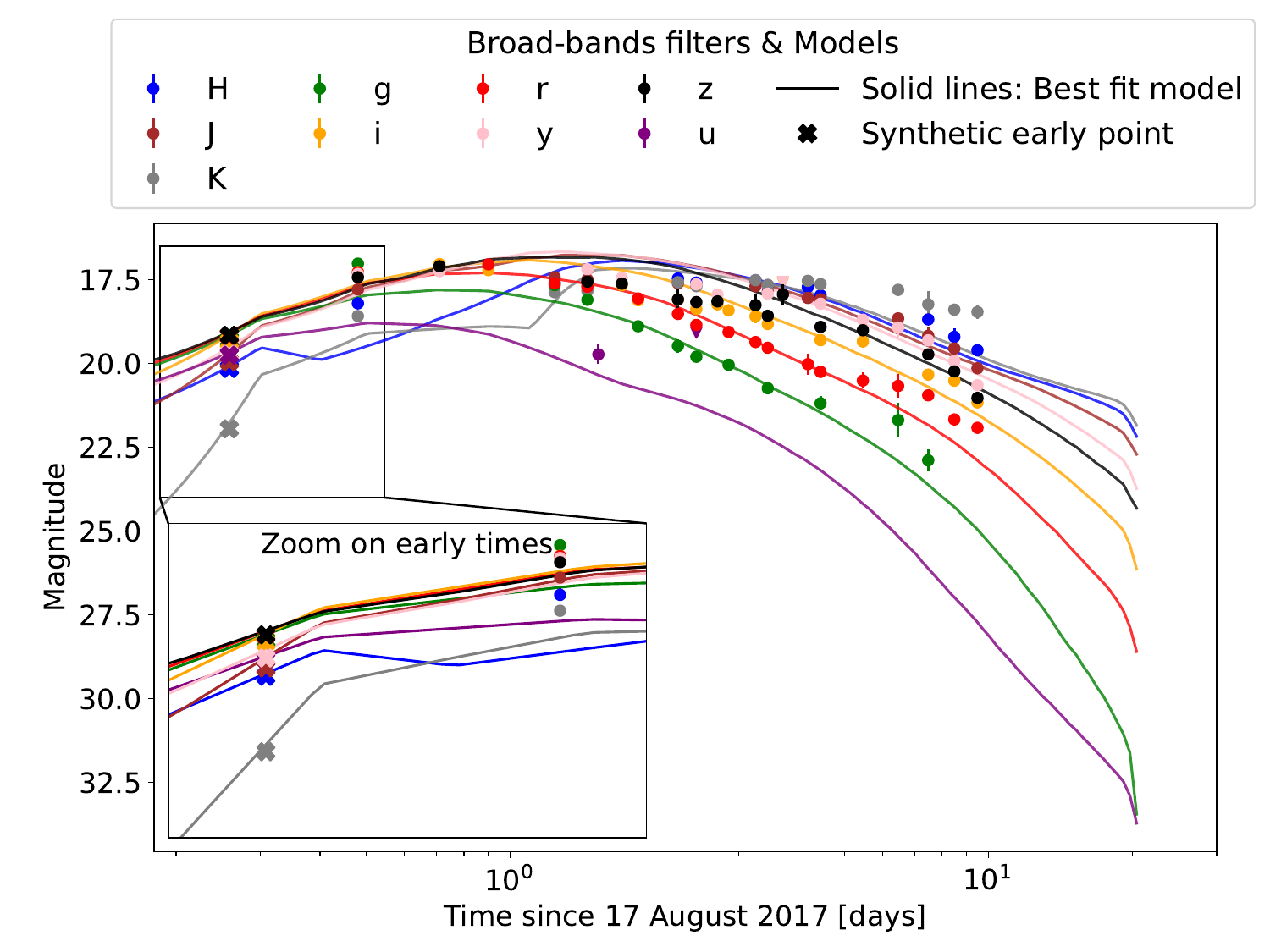}
    \caption{Photometric measurements of AT2017gfo in $ugrizyJHK$ bands with \texttt{Bu2019lm} best-fit lightcurves, corrected for galactic extinctions, fixing the start time to GW170817 merger time. The crosses indicate the fake data point chosen in different filters.}
    \label{fig:2017-fake-detection}
\end{figure}

\begin{figure}
    \centering
    \includegraphics[width=0.98\linewidth]{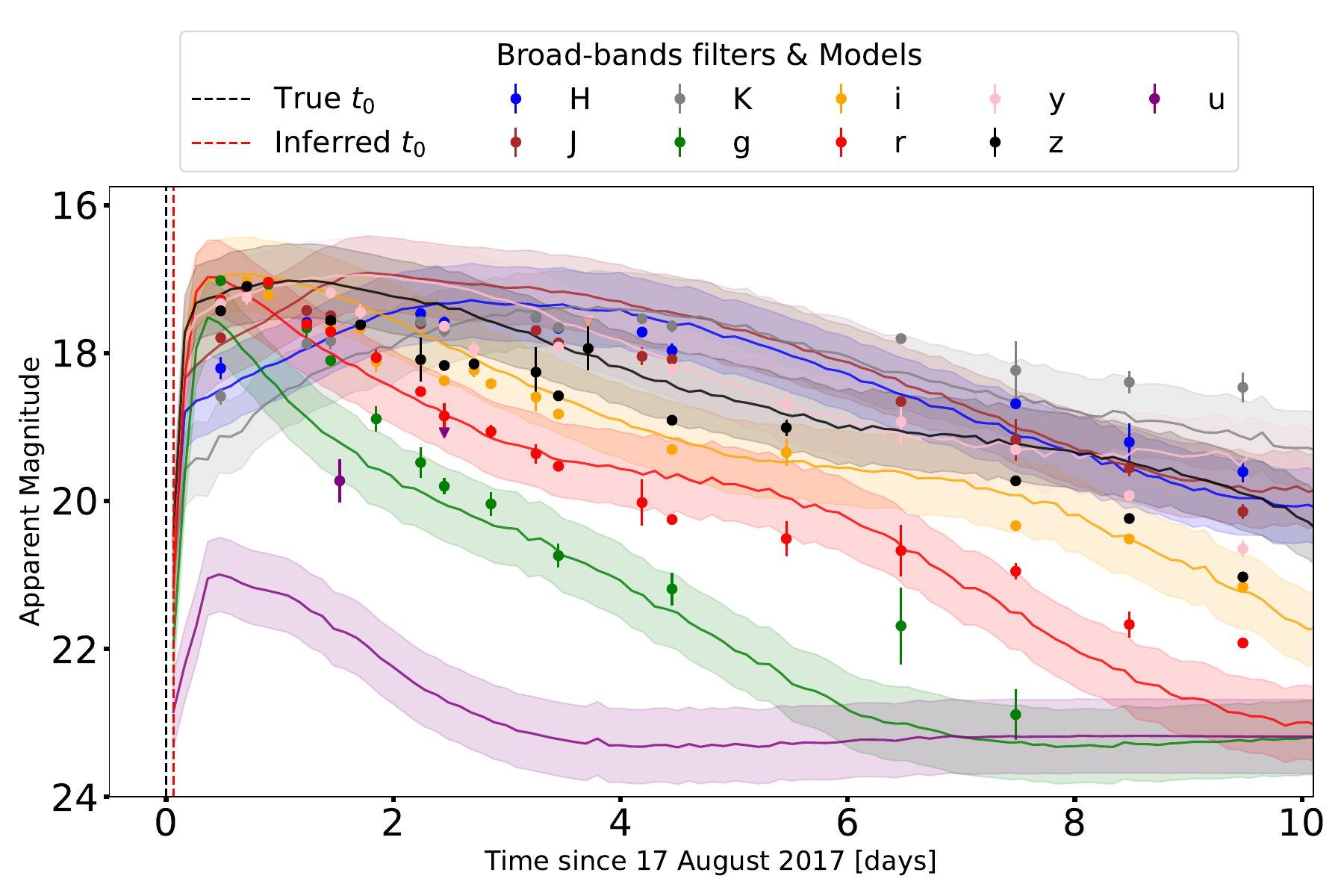}
    \caption{Best-fit light curves in $ugrizyJHK$ obtained with {\sc nmma} using \texttt{Brethauer2026}. The black dotted line indicates the true $t_0$ while the red dotted line shows the inferred $t_0$.}
    \label{fig:nmma_lc_Bret26}
\end{figure}

\begin{figure}
    \centering
    \includegraphics[width=\linewidth]{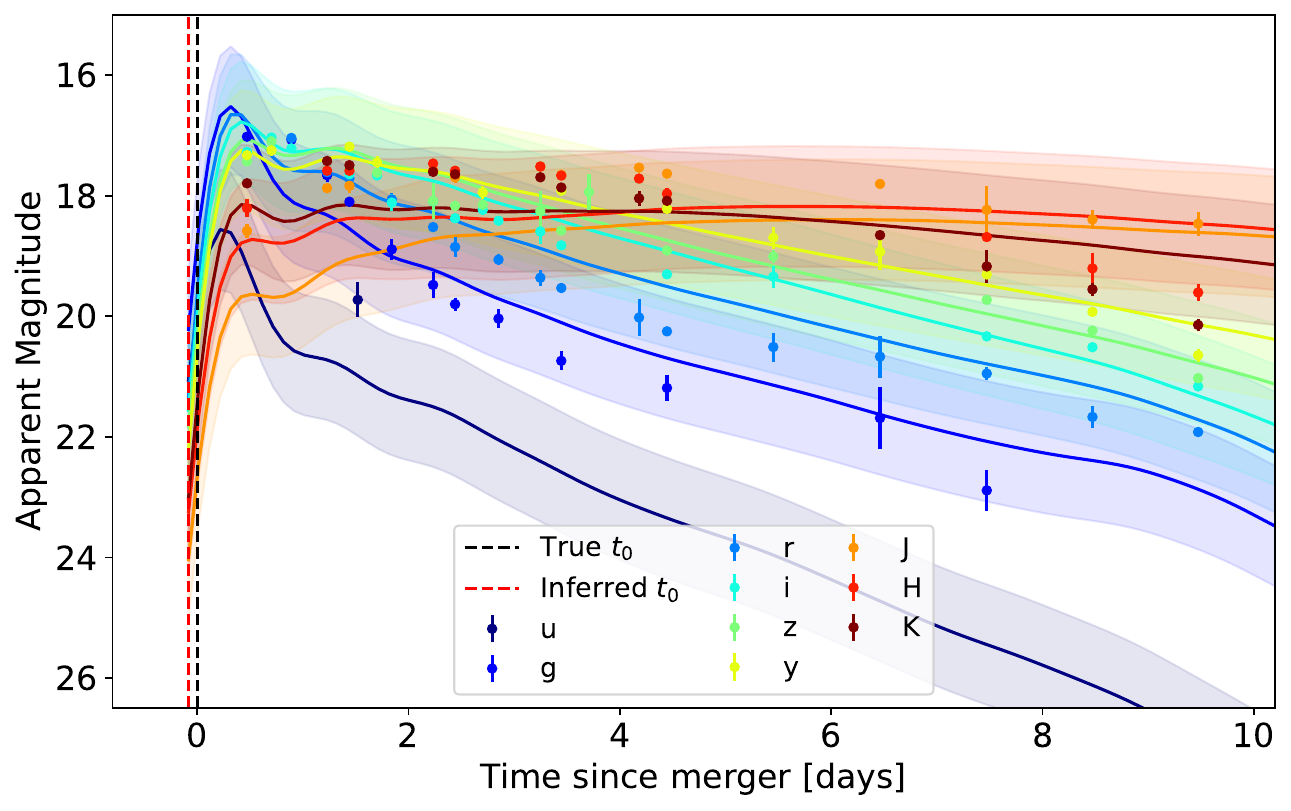}
    \caption{Photometric measurements of AT2017gfo in $ugrizyJHK$ bands with \texttt{Ka2017} best-fit lightcurves, corrected for galactic extinctions, fixing the start time to GW170817 merger time and with $\sigma_{\rm sys} = 1.0$. The dashed lines show the real (black) and inferred (red) $t_0$.}
    \label{fig:bestfit_Ka17_george}
\end{figure}

\begin{figure}
\centering
\includegraphics[width=\linewidth]{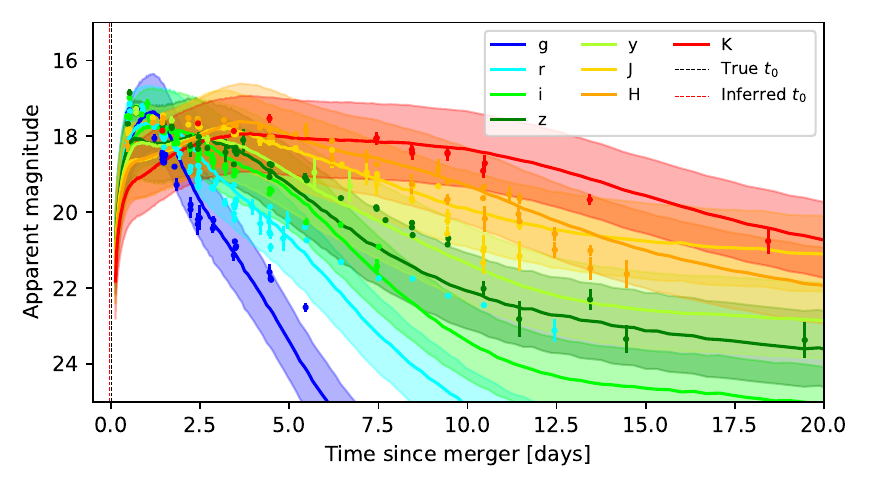}
\caption{Best-fit light curve fit using the \textsc{RIFT} parameter inference engine and the \texttt{LANL} kilonova models presented in \cite{Ristic-LC}, represented by the neural network emulator developed in \cite{Peng2024}. Emulated light curves begin at $t = 0.125$ days.}
\label{fig:RIFT_LANL_AT2017gfo}
\end{figure}

\begin{figure}
    \centering
    \includegraphics[width=\linewidth]{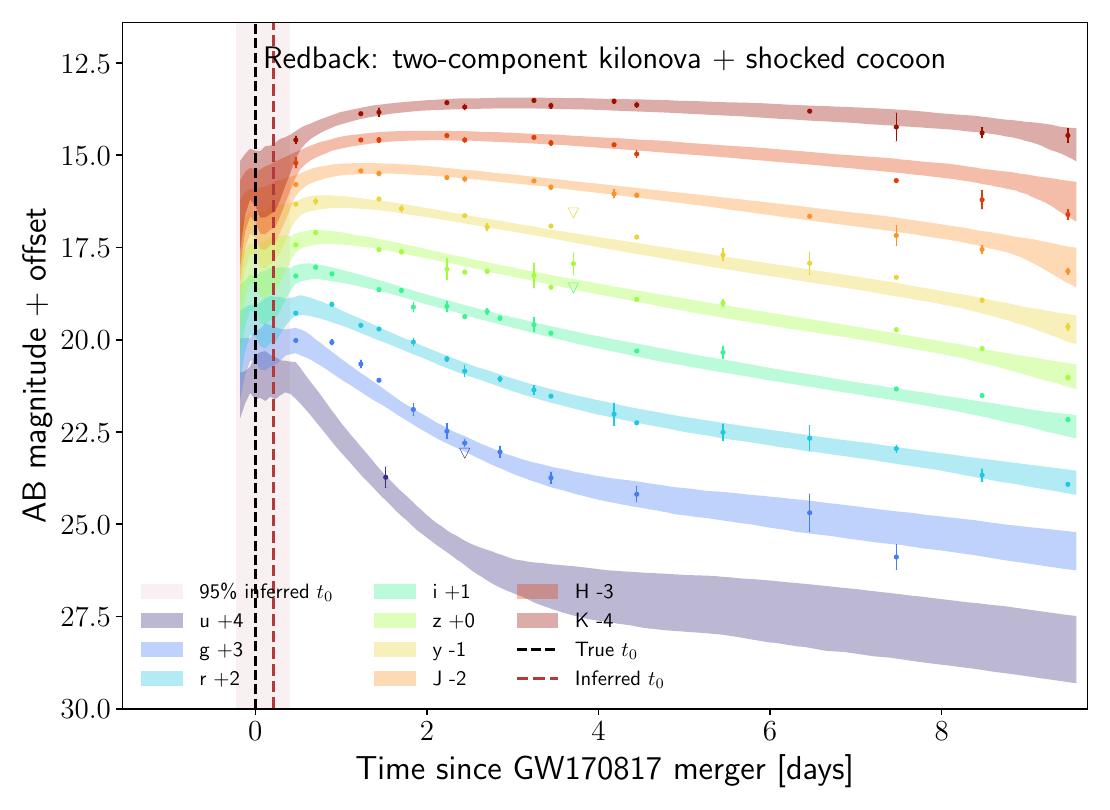}
    \caption{AT2017gfo photometry and posterior-predictive posterior-predictive light-curve $68\%$ credible interval for the $\sigma_{\rm sys}=1$ mag analysis. The pale red region indicates the central 95\% interval for $t_0$, while the black dashed and red dotted lines mark the GW170817 merger time and the posterior-median inferred merger time, respectively.}
    \label{fig:redback-lightcurve}
\end{figure}

This section includes best-fit light curves to AT2017gfo photometry using different kilonova grids: \texttt{Bu2026\_MLP} (Fig.~\ref{fig:nmma_lc}), \texttt{Bu2019lm} with a fake detection on the rise (Fig.~\ref{fig:2017-fake-detection}), \texttt{Brethauer2026} (Fig.~\ref{fig:nmma_lc_Bret26}), \texttt{Ka2017} (Fig.~\ref{fig:bestfit_Ka17_george}), \texttt{LANL} (Fig.~\ref{fig:RIFT_LANL_AT2017gfo}) and \texttt{two\_component\_kilonova\_with\_cocoon} (Fig.~\ref{fig:redback-lightcurve}).

\section{AT2017gfo posterior distribution with other $\sigma_{\rm sys}$ configurations or samplers}
\label{appendix-AT2017gfo}

\begin{figure}
    \centering
    \includegraphics[width=\linewidth]{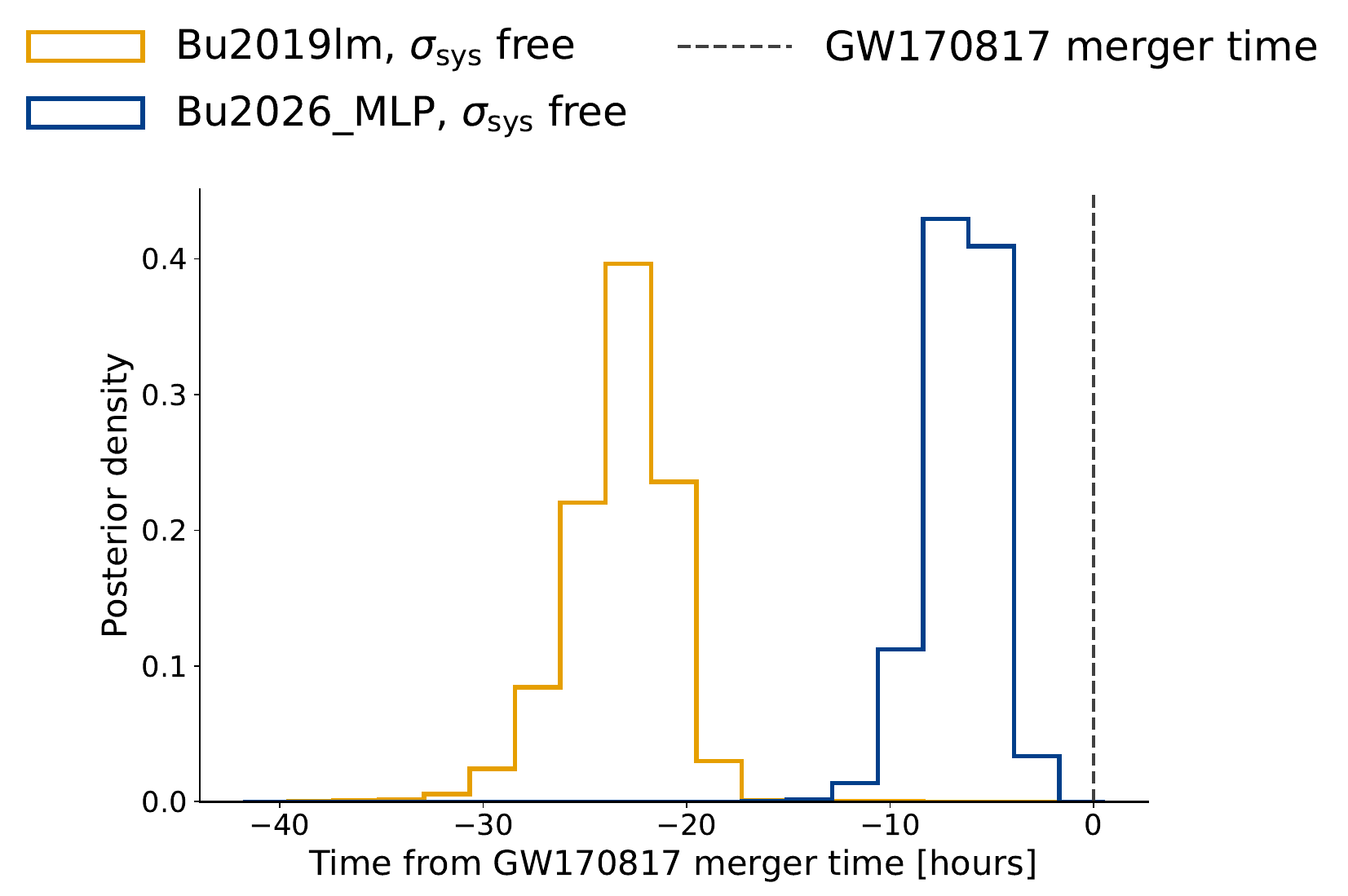}
    \caption{AT2017gfo merger time posterior distribution inferred with {\sc nmma} using \texttt{Bu2019lm} (yellow) and \texttt{Bu2026\_MLP} (blue) leaving $\sigma_{\rm sys}$ as a free parameter.}
    \label{fig:Bulla-nmma-freesigmasys}
\end{figure}

\begin{figure}
    \centering
    \includegraphics[width=0.9\linewidth]{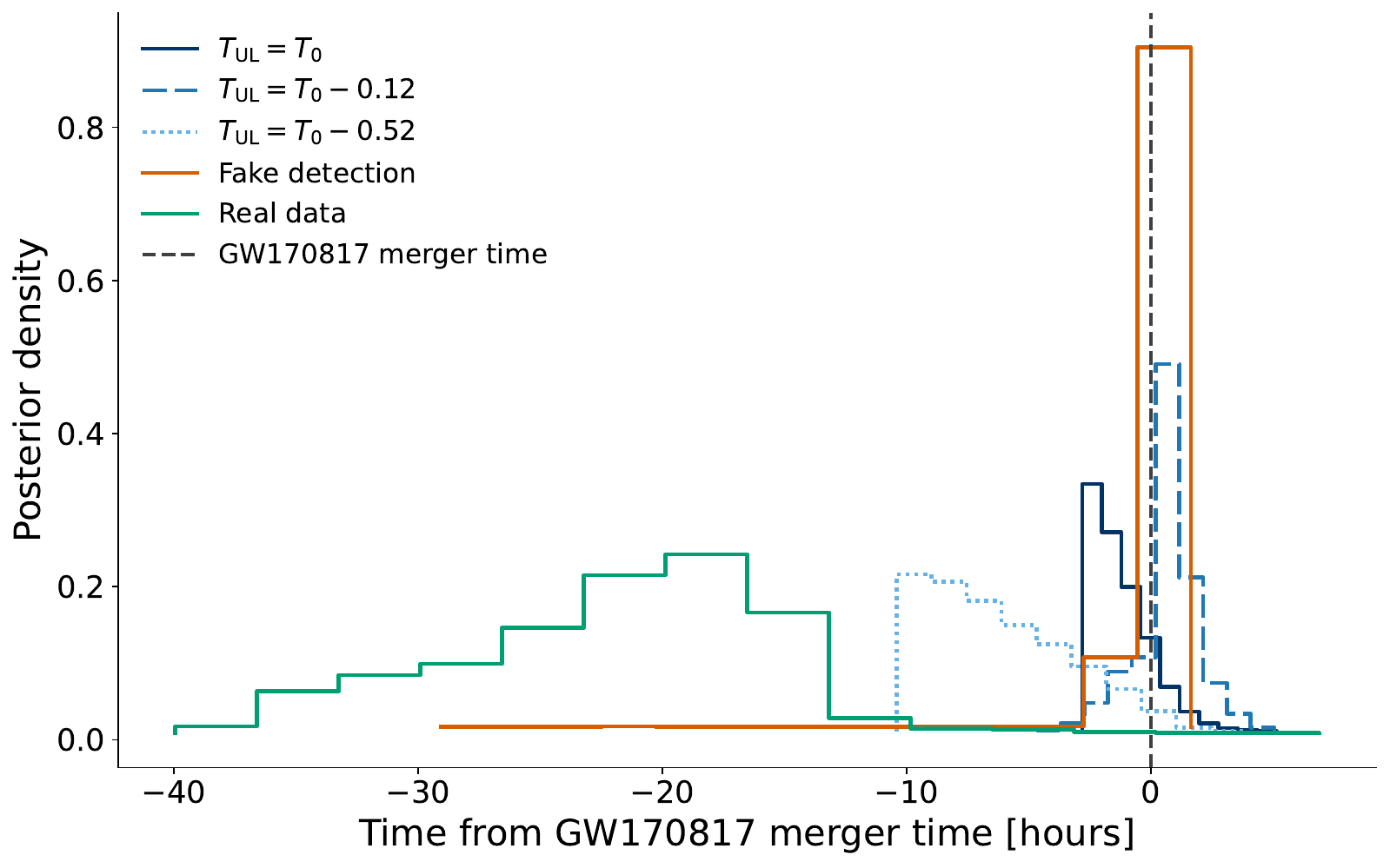}
    \caption{AT2017gfo merger time posterior distribution inferred with {\sc nmma} using \texttt{Bu2019lm} and \texttt{Pymultinest} sampler.}
    \label{fig:Bu19lm-AT2017gfo-pymultinest}
\end{figure}

\begin{figure}
    \centering
    \includegraphics[width=0.85\linewidth]{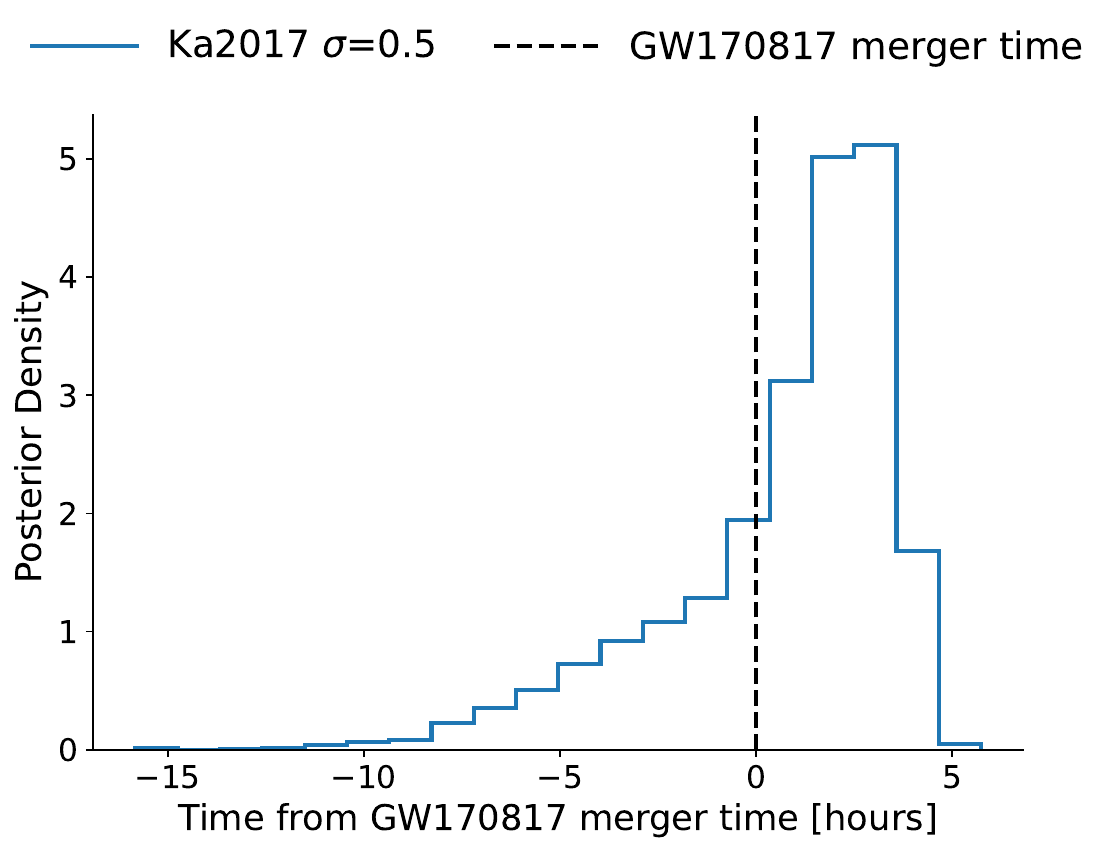}
    \caption{AT2017gfo merger time posterior distribution inferred with george and nessai using \texttt{Ka2017} with $\sigma_{\rm sys} = 0.5 $. GW170817 merger time is displayed in black.}
    \label{fig:Ka2017-sigmasys0.5}
\end{figure}

\begin{figure}
    \centering
    \includegraphics[width=0.95\linewidth]{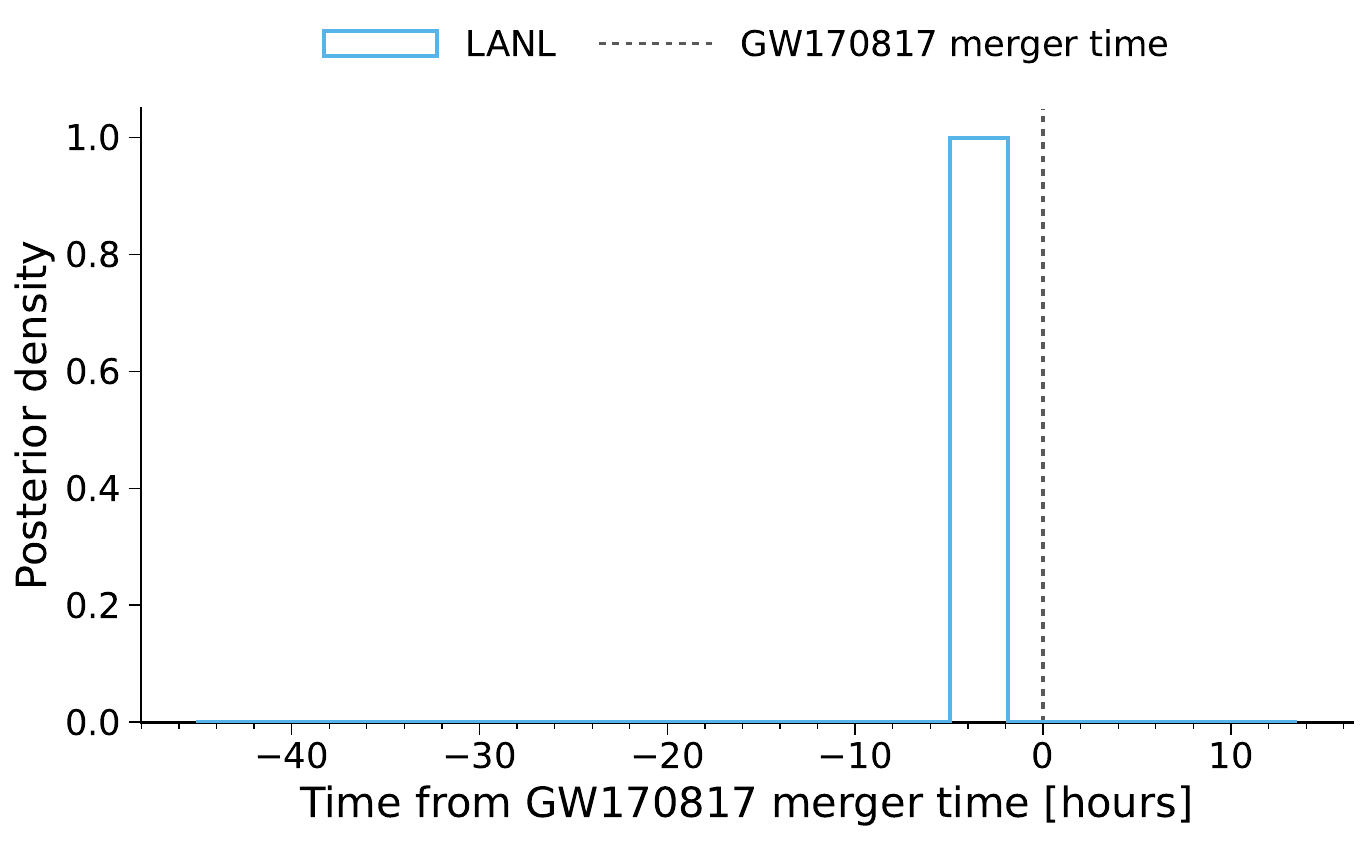}
    \caption{AT2017gfo merger time posterior distribution inferred with \textsc{rift} using \texttt{LANL} with a free $\sigma_{\rm sys}$. GW170817 merger time is displayed in black.}
    \label{fig:LANL-freesigma-sys}
\end{figure}

This section includes additional tests in which fits to AT2017gfo are computed with different assumptions for the systematic uncertainty parameter $\sigma_{\rm sys}$ and for the sampler.
Fig.~\ref{fig:Bulla-nmma-freesigmasys} displays the $t_0$ posterior distribution using \textsc{nmma} leaving $\sigma_{\rm sys}$ as a free parameter, while Fig.~\ref{fig:Bu19lm-AT2017gfo-pymultinest} shows results when using \texttt{Pymultinest} instead of \texttt{pocoMC} as sampler for the \texttt{Bu2019lm} fit. Fig.~\ref{fig:Ka2017-sigmasys0.5} shows similar results with george and nessai using \texttt{Ka2017} and a $\sigma_{\rm sys}=0.5$ mag, while Fig.~\ref{fig:LANL-freesigma-sys} results for the \textsc{rift}/\texttt{LANL} analyses with $\sigma_{\rm sys}$ as a free parameter.

\end{appendix}

\end{document}